\documentclass[
aps,
prd,
floats,
floatfix,
twocolumn,
superscriptaddress,
nofootinbib,
noshowpacs]{revtex4-2}

\usepackage{hyperref}
\usepackage{amssymb}
\usepackage{amsmath}
\usepackage{physics}
\usepackage{latexsym}

\usepackage{verbatim}
\usepackage{mathrsfs}
\usepackage{amsfonts}

\usepackage{graphicx,subfigure}
\usepackage{epsfig}

\usepackage{units}
\usepackage{overpic}
\usepackage[utf8]{inputenc}
\usepackage{rotating}
\usepackage{enumerate}

\usepackage{soul}

\usepackage{xfrac}
\usepackage{xcolor}
\usepackage{multirow}

\newcommand{\Real}{\mathbb{R}}
\newcommand{\Complex}{\mathbb{C}}

\begin{document}

\title{Nonrelativistic Proca stars: Spherical stationary and multi-frequency states}

\author{Emmanuel Ch\'avez Nambo}
\affiliation{Instituto de F\'isica y Matem\'aticas,
Universidad Michoacana de San Nicol\'as de Hidalgo,
Edificio C-3, Ciudad Universitaria, 58040 Morelia, Michoac\'an, M\'exico}
\author{Alberto Diez-Tejedor}
\affiliation{Departamento de Física, División de Ciencias e Ingenierías, Campus León, Universidad de Guanajuato, 37150, León, México}
\author{Edgar~Preciado-Govea}
\affiliation{Departamento de Física, División de Ciencias e Ingenierías, Campus León, Universidad de Guanajuato, 37150, León, México}
\author{Armando A. Roque}
\affiliation{Unidad Acad\'emica de F\'isica, Universidad Aut\'onoma de Zacatecas, 98060 Zacatecas, M\'exico}
\affiliation{Departamento de Física, División de Ciencias e Ingenierías, Campus León, Universidad de Guanajuato, 37150, León, México}
\author{Olivier Sarbach}
\affiliation{Instituto de F\'isica y Matem\'aticas,
Universidad Michoacana de San Nicol\'as de Hidalgo,
Edificio C-3, Ciudad Universitaria, 58040 Morelia, Michoac\'an, M\'exico}
\affiliation{Departamento de Matem\'aticas Aplicadas y Sistemas, Universidad Aut\'onoma Metropolitana-Cuajimalpa, 05348 Cuajimalpa de Morelos, Ciudad de M\'exico, M\'exico}

\date{\today}

\begin{abstract}
In this paper we follow an effective theory approach to study the nonrelativistic limit of a selfgravitating and selfinteracting massive vector field. Our effective theory is characterized by three parameters: the field's mass $m_0$ and the selfinteraction constants $\lambda_n$ and $\lambda_s$.  For definiteness, we focus on a systematic study of the equilibrium configurations, commonly referred to as Proca stars when they have finite energy. We identify two different types of Proca stars, depending on the specific sector of the effective theory that we are exploring. In the generic sector, defined by $\lambda_s\neq 0$, all equilibrium configurations are stationary states described by wave functions that evolve harmonically in time. However, in the symmetry-enhanced sector, for which $\lambda_s=0$, there exist multi-frequency states whose wave functions oscillate with two or three distinct frequencies in addition to the stationary states. We determine the conditions under which a ground state configuration with fixed particle number exists. When these conditions are met, we prove that the lowest energy is reached by a stationary spherically symmetric configuration of constant polarization that is linear or circular depending on the sign of $\lambda_s$. We numerically construct some illustrative examples of spherical stationary and multi-frequency solutions, analyze their properties, and compare them with our analytical predictions. Unlike stationary states and other soliton configurations, which form a discrete set in the solution space associated with fixed particle number, the symmetry-enhanced sector exhibits a continuum of solutions with multi-frequency states connecting stationary states of constant polarization.
\end{abstract}

\maketitle

\section{Introduction}

\begin{figure*}
	\centering
    \includegraphics[width=14.cm]{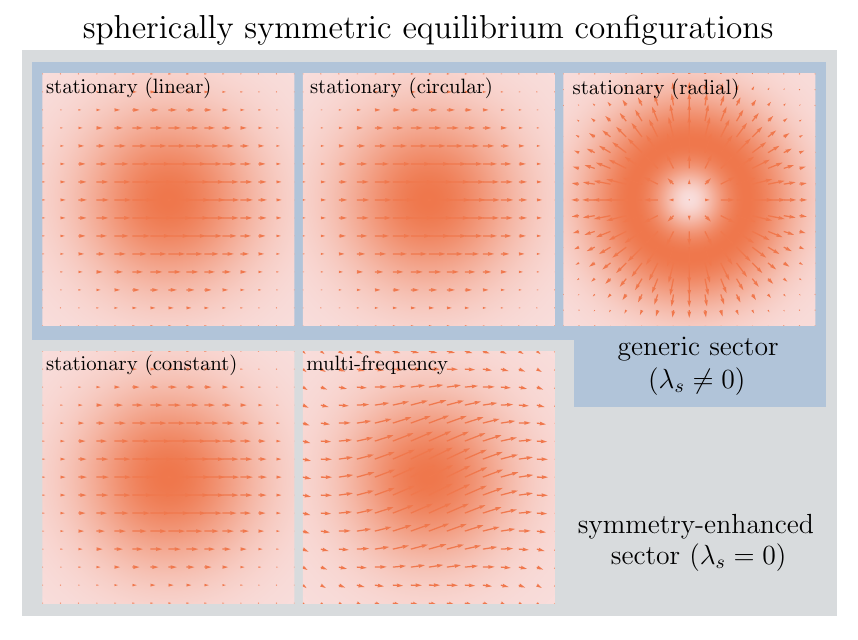}
\caption{{\bf Spherical Proca stars' inventory:}
The normalized real part of the vector field, $\vec{\psi}_R(t,\vec{x})$, and the normalized particle number density, $n(t,\vec{x})$, in color gradient, of some representative equilibrium configurations at time $t=0$. When $\lambda_s\neq 0$, all equilibrium configurations are stationary states, with only linear, circular, and radial polarizations allowed in spherical symmetry. When $\lambda_s = 0$, in addition to the aforementioned stationary states, spherical symmetry permits general constantly polarized stationary states and multi-frequency states. At $t=0$, all constant polarization states look the same, although they differ in their time evolution (see Fig.~\ref{fig.polarizations} for details).}
\label{fig.outline}
\end{figure*}

Boson stars are regular, finite energy configurations that do not disperse in time and are encountered in massive, selfgravitating scalar field theories~\cite{Kaup:1968zz, Ruffini:1969qy, Jetzer:1991jr, Liddle:1992fmk, Schunck:2003kk, Liebling:2012fv, Zhang:2018slz, Visinelli:2021uve}. Furthermore, similar solutions arise in theories with higher rank fields, such as Proca stars in vector field theories. In this paper, we take an effective theory approach to investigate the nonrelativistic limit of a selfgravitating and selfinteracting massive vector field, focusing on equilibrium configurations and on spherically symmetric states in particular. The linear stability of these solutions will be analyzed in a follow-up paper~\cite{Nambo:preparation}.

Proca stars were first introduced by Brito, Cardoso, Herdeiro, and Radu in Ref.~\cite{Brito:2015pxa}, where they constructed solutions of the Einstein-Proca equations with both static spherically symmetric and stationary axially symmetric spacetimes. This pioneering work triggered a surge in research on such stars that includes theoretical investigations~\cite{SalazarLandea:2016bys, Brihaye:2017inn, Minamitsuji:2018kof, Herdeiro:2020kba, Herdeiro:2020jzx, Herdeiro:2023wqf,Joaquin:2024quo}, numerical simulations~\cite{Sanchis-Gual:2017bhw, Sanchis-Gual:2018oui, Wang:2023tly}, and astrophysical applications~\cite{CalderonBustillo:2020fyi, Herdeiro:2021lwl, Sanchis-Gual:2022mkk, Rosa:2022tfv, Sengo:2024pwk}. As for this latter possibility, massive vector fields may be especially relevant to ultralight dark matter models~\cite{Suarez:2013iw, Marsh:2015xka, Urena-Lopez:2019kud, Ferreira:2020fam}, exhibiting a richer phenomenology compared to spin $s=0$ axion-like particles~\cite{Antypas:2022asj}. In galaxies, these particles could form dark matter halos, whose global structure is inherently Newtonian, and this motivates our focus on the nonrelativistic theory in this paper.

In the nonrelativistic regime, Proca stars have been explored by Amin, Jain, and collaborators  in~\cite{Jain:2021pnk, Zhang:2021xxa, Amin:2022pzv, Jain:2022agt, Jain:2023tsr} (see also Refs.~\cite{Adshead:2021kvl, Gorghetto:2022sue, Chen:2024vgh, Zhang:2023ktk, Zhang:2024bjo}). We can think of these objects as selfgravitating condensates of spin $s=1$ particles, where matter is described in terms of a vector-valued wave function $\vec{\psi}(t,\vec{x})$ satisfying the Schr\"odinger equation and gravity by the Newtonian gravitational potential $\mathcal{U}(t,\vec{x})$ obeying Poisson's equation. When selfinteractions are included, the Schr\"odinger equation needs to be replaced by a Gross-Pitaevskii type equation with two coupling constants $\lambda_n$ and $\lambda_s$. We refer to these equations as the {\it $s=1$ Schr\"odinger-Poisson system}  when $\lambda_n=\lambda_s=0$, and as the {\it $s=1$ Gross-Pitaevskii-Poisson system} otherwise, and to the resulting finite energy equilibrium configurations as {\it nonrelativistic  Proca stars}.

The spectrum of Proca star solutions depends on the spin-spin selfinteraction parameter $\lambda_s$. When $\lambda_s\neq 0$, which we henceforth call the {\it generic sector} of the theory, the Proca star's wave function evolves in time harmonically. As in standard quantum mechanics, we shall refer to these equilibrium configurations as {\it stationary} (or single-frequency) states. However, when $\lambda_s=0$, the effective theory acquires an additional (accidental) symmetry, resulting in the {\it symmetry-enhanced sector}. In this sector, new types of equilibrium configurations appear besides the stationary states in which the wave function oscillates with two or three distinct frequencies. We shall call these configurations {\it multi-frequency} states.

By definition, equilibrium configurations are critical points of the total energy functional keeping fixed suitable constants of motion. It is relevant to determine whether these points correspond to local or global minima or maxima, or to saddle points, since this provides information regarding their stability. We prove that, under certain conditions on the parameters $\lambda_n$ and $\lambda_s$, ground state configurations (i.e. lowest energy solutions for fixed particle number) exist. Moreover, when these conditions are satisfied, there exits a spherically symmetric stationary state of constant polarization which has lowest possible energy, regardless what sector of the theory we are exploring. In the free theory, defined by $\lambda_n=\lambda_s=0$, we show that the ground state is unique (up to translations and rigid unitary transformations).
Otherwise, even if there could in principle exist additional states which minimize the energy, they must also be stationary, spherically symmetric, and exhibit constant polarization.
 
\begin{figure*}
	\centering
    \includegraphics[width=18.cm]{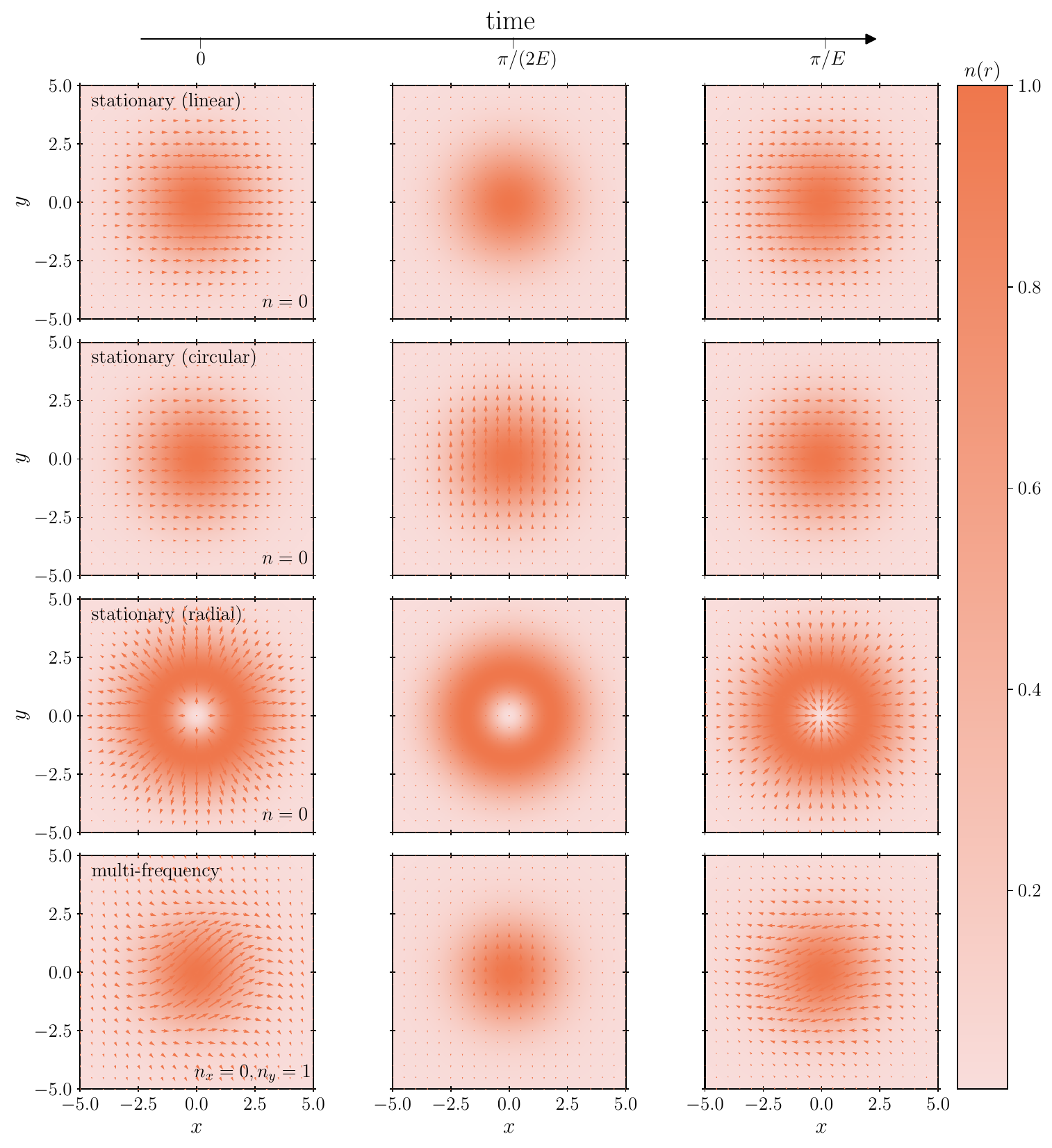}
\caption{{\bf Time evolution of spherical Proca stars:} The normalized real part of the vector field, $\vec{\psi}_R(t,\vec{x})$, and the normalized particle number density, $n(t,\vec{x})$, of some representative equilibrium configurations at different moments of time. In the first three rows, we consider stationary Proca stars of unit central ``amplitude'', $\sigma_0=1$, no nodes, $n=0$, and different polarization vectors $\hat{\epsilon}$ at times $t_1=0$, $t_2=\frac{\pi}{2E}$ and $t_3=\frac{\pi}{E}$. {\it First row:} Linear polarization along the $x$ axis, $\hat{\epsilon}=\hat{e}_x$. {\it Second row:} Circular polarization along the $z$ axis, $\hat{\epsilon} = \hat{\epsilon}_z^{(+)}:=\frac{1}{\sqrt{2}}\left(\hat{e}_x + i\hat{e}_y\right)$. {\it Third row:} Radial polarization, $\hat{\epsilon}=\hat{e}_r$. {\it Fourth row:} A multi-frequency Proca star of central amplitude $(\sigma_{x0},\sigma_{y0},\sigma_{z0})=(1,1,0)$ and nodes $(n_x,n_y,n_z)=(0,1,0)$ at times $t_1=0$, $t_2=\frac{\pi}{2E_x}$ and $t_3=\frac{\pi}{E_x}$. Apart from radially polarized states, the spherical symmetry is only manifest in the gravitational field, since in other cases the wave function selects a preferred direction in space. This occurs because, in these cases, the field transforms under a non-standard representation of the $SO(3)$ group. Radially polarized Proca stars are characterized by a ``hole'' in their center, which is due to the regularity conditions at the origin. In all these cases we have assumed repulsive selfinteractions. To better visualize the time evolution of these objects, we refer the reader to the movies provided in~\cite{youtube}.}\label{fig.polarizations}
\end{figure*}

In this article, we further concentrate on spherical configurations. In this case, stationary Proca stars can be classified according to the node number of their radial profile and their polarization vector, which can be constant or radial, although in the former case the polarization has to be linear or circular when $\lambda_s\neq 0$. In contrast, multi-frequency Proca stars are classified according to the node number of each component of the wave function, the polarization vector being less useful than in the stationary case given that it is time-dependent. Figure~\ref{fig.outline} illustrates the classification of spherical equilibrium configurations that appear in the different sectors of our effective theory, whereas in Fig.~\ref{fig.polarizations} we exhibit some features of their time evolution which will be discussed later. 

It is worth noting that the equilibrium, spherically symmetric  $s=1$ Gross-Pitaevskii-Poisson system is equivalent to other systems studied in the framework of multi-scalar field theories. The free theory possesses an internal global $U(3)$ symmetry and is identical to the $N$-particle $s=0$ Schr\"odinger-Poisson system studied in Ref.~\cite{Roque:2023sjl}, specifically in the case where no more than three orthogonal states are occupied. This system admits nonrelativistic $\ell=0$ and $\ell=1$ boson star solutions~\cite{Roque:2023sjl,NamboThesis}, which in the context of Proca stars lead to stationary states of constant and radial polarization, respectively, and multi-state boson stars solutions~\cite{Matos:2007zza}, which in the present context correspond to multi-frequency states.

When $\lambda_n\neq 0$ and $\lambda_s = 0$, the theory retains the $U(3)$ symmetry and generalizes the theories discussed in Refs.~\cite{Roque:2023sjl} and~\cite{Nambo:2024gvs} in several ways. On the one hand, in the present work we include selfinteractions, which were not considered in Ref.~\cite{Roque:2023sjl}. On the other hand, the internal structure of the vector field allows new configurations, such as the stationary states of radial polarization and the multi-frequency solutions, which do not exist in the theory with a single scalar field considered in Ref.~\cite{Nambo:2024gvs}. In contrast, when $\lambda_s\neq 0$, the spin-spin selfinteraction term breaks the $U(3)$ symmetry which forbids the existence of multi-frequency states and removes the degeneracy of the constant polarization equilibrium configurations.

Conversely, in the framework of relativistic vector field theories, the static and spherically symmetric solutions reported in Ref.~\cite{Brito:2015pxa} correspond to stationary configurations of radial polarization in our classification of the equilibrium solutions. These configurations were subsequently recognized as excited states of the Einstein-Proca theory~\cite{Herdeiro:2023wqf}, indicating the potential existence of lower-energy solutions. In the nonrelativistic regime, the works~\cite{Jain:2021pnk, Zhang:2021xxa} discuss the linearly and circularly polarized states. These studies were further extended in  Ref.~\cite{Wang:2023tly}, where the stability of linearly, circularly, as well as radially polarized Proca stars was investigated by means of $3+1$ dimensional numerical simulations in general relativity, and in Ref.~\cite{Adshead:2021kvl}, where stationary non-spherical solutions were also considered. However, as far as we are aware, multi-frequency states have not been reported in previous studies.

The remainder of this paper is organized as follows. In Sec.~\ref{sec.EFT}, we introduce our effective theory and present the $s=1$ Gross-Pitaevskii-Poisson system, which describes the low energy dynamics of a selfgravitating and selfinteracting massive vector field. Next, in Sec.~\ref{sec.equilibrium}, we introduce the equilibrium configurations as critical points of the total energy functional under suitable constraints, and discuss some general properties that they satisfy. In Sec.~\ref{Sec:.spherically}, we additionally assume spherical symmetry, classify the resulting equilibrium solutions and show that they give rise to a radial set of nonlinear eigenvalue problems. Spherical equilibrium configurations of the $s=1$ Gross-Pitaevskii-Poisson system are constructed numerically in Sec.~\ref{sec.numerical}, where we discuss some of their most important properties. Finally, we conclude in Sec.~\ref{Sec:Conclussions} with an overview of our results. Some complementary material is presented in the appendixes.

In this paper, we work in natural units, for which $c=\hbar=1$.

\section{The \texorpdfstring{$s=1$}{s=1} Gross-Pitaevskii-Poisson system}\label{sec.EFT}

Our starting point is the nonrelativistic, low energy effective theory that describes a selfinteracting vector field $\vec{\psi}(t,\vec{x})$ coupled to the Newtonian gravitational potential $\mathcal{U}(t,\vec{x})$. This theory is expressed in terms of the action
\begin{eqnarray}\label{eq.action.nonrel}
S[\mathcal{U}, \vec{\psi}] &=& \int dt\int dV \bigg[\frac{1}{8\pi G}\mathcal{U}\Delta\mathcal{U}-m_0\mathcal{U} n\nonumber\\ 
&&\hspace{-1.5cm} +\vec{\psi}^*\cdot\left(i\frac{\partial}{\partial t}+\frac{1}{2m_{0}}\Delta \right)\vec{\psi}-\frac{\lambda_n}{4m_{0}^{2}}n^{2}-\frac{\lambda_{s}}{4m_{0}^{2}}\vec{s}^{\,2}\bigg],
\end{eqnarray}
which consists of all operators of mass dimension 6 or lower that can be constructed from the field $\vec{\psi}(t,\vec{x})$ and are scalars under the Galilei group (see App.~\ref{App:EFT} for details). In this expression, $i=\sqrt{-1}$ is the unit imaginary number, $G$ is Newton’s constant, the star denotes complex conjugation, and $dV$ and $\Delta$ refer to the volume element and Laplace operator, respectively, associated with three-dimensional Euclidean space.

The theory described by the action $S[\mathcal{U}, \vec{\psi}]$ is characterized in terms of three parameters: the positive and non-vanishing field's mass $m_0$ and the dimensionless coupling constants $\lambda_n$ and $\lambda_s$.\footnote{ In the effective theory, the parameters $\lambda_n$ and $\lambda_s$ are in principle arbitrary and unrelated. Compare this with e.g. Ref.~\cite{Zhang:2021xxa}, where taking the nonrelativistic limit of a generic massive real-valued vector field the authors obtain $\lambda_s=-\frac{1}{3}\lambda_n$.} The first term in the first line of Eq.~(\ref{eq.action.nonrel}) describes Newtonian gravity, the second line the matter sector and the last term in the first line the interaction between gravity and matter. Note that the matter sector consists of the free Schr\"odinger action plus two short-range selfinteraction terms, one that depends on the particle number density, $n:=\vec{\psi}^*\cdot\vec{\psi}$, and the other on the spin density, $\vec{s}:=-i\vec{\psi}^{*}\times\vec{\psi}$, which by definition are real-valued. As an illustration, in App.~\ref{app.nonrelativistic.Proca} we derive the nonrelativistic limit of the complex Einstein-Proca theory with quartic selfinteraction potential $\lambda_1 (A_\mu^* A^\mu)^2 + \lambda_2 (A_\mu A^\mu) (A_\nu^* A^{\nu*})$, and we show that it leads to our effective theory with all the terms in Eq.~(\ref{eq.action.nonrel}).

Varying Eq.~(\ref{eq.action.nonrel}) with respect to $\vec{\psi}$ we obtain a Gross-Pitaevskii type equation of the form
\begin{subequations}\label{eq.GPPs=1}
\begin{align}\label{eq.GPs=1} i\frac{\partial\vec{\psi}}{\partial t} = -\frac{1}{2m_0}\Delta\vec{\psi}+\frac{\lambda_n}{2m_0^2}n\vec{\psi}+i\frac{\lambda_s}{2m_0^2}\vec{s}\times\vec{\psi}+m_{0}\mathcal{U}\vec{\psi},
\end{align}
whereas varying it with respect to $\mathcal{U}$ leads to the Poisson equation,
\begin{equation}\label{eq.Poisson}
\Delta\mathcal{U} = 4\pi G m_0n.
\end{equation}
\end{subequations}
We will refer to Eqs.~(\ref{eq.GPPs=1}) as the $s=1$ Gross-Pitaevskii-Poisson system. Setting $\lambda_n=\lambda_s=0$ the selfinteractions vanish and Eqs.~(\ref{eq.GPPs=1}) reduce to the $s=1$ Schr\"odinger-Poisson system. 

\subsection{Reformulation of the dynamical equations}

In this section, we reformulate the system~(\ref{eq.GPPs=1}) in terms of a non-linear Hamilton operator, which will bring more transparency for several of the discussions in this paper.

The $s=1$ Gross-Pitaevskii~(\ref{eq.GPs=1}) and Poisson~(\ref{eq.Poisson}) equations can be expressed more compactly as: 
\begin{subequations}\label{eq.dimensionless.GPP.1}
\begin{eqnarray}
i\frac{\partial \vec{\psi}}{\partial t} &=& \hat{{\mathcal{H}}} [\vec{\psi}, \mathcal{U}] \vec{\psi}, \label{eq.dimensionless.GPP.1.GP}\\
\Delta \mathcal{U} &=& 4\pi G m_0 n, \label{eq.dimensionless.GPP.1.P}
\end{eqnarray}
\end{subequations}
where, for fixed $\vec{\psi}$ and $\mathcal{U}$, we have defined the Hamilton operator
\begin{align}
\hat{\mathcal{H}}[\vec{\psi}, \mathcal{U}] &:= -\frac{1}{2m_0}\Delta + \frac{\lambda_n}{2m_0^2} n + i\frac{\lambda_s}{2m_0^2} \vec{s} \times + m_0\mathcal{U}.
\end{align}
In this equation, $\vec{s}\times$ represents the cross product of the spin density with the vector-valued function that the operator $\hat{\mathcal{H}}[\vec{\psi},\mathcal{U}]$ is acting on.

Furthermore, by inverting the Laplacian in Eq.~(\ref{eq.dimensionless.GPP.1.P}), one can eliminate the gravitational potential $\mathcal{U}$ from Eq.~(\ref{eq.dimensionless.GPP.1.GP}), which yields a nonlinear integro-differential equation for the wave function $\vec{\psi}$,\footnote{While we refer to $\vec{\psi}(t,\vec{x})$ as the {\it wave function}, it should be noted that this is a classical field. In particular, $n=\vec{\psi}^*\cdot\vec{\psi}$ describes the particle and not the probability density.}
\begin{align}\label{integro-differential_equation}
i\frac{\partial \vec{\psi}}{\partial t} &= \hat{{\mathcal{H}}} [\vec{\psi}] \vec{\psi},
\end{align}
with
\begin{align}\label{eq.def.H}
\hat{\mathcal{H}}[\vec{\psi}] &:= \hat{\mathcal{H}}[\vec{\psi},\mathcal{U}=\Delta^{-1}(n)]\\
 &= -\frac{1}{2m_0}\Delta + \frac{\lambda_n }{2m_0^2}n + i\frac{\lambda_s}{2m_0^2} \vec{s} \times + 4\pi Gm_0^2\Delta^{-1}(n)\nonumber,
\end{align}
and where for a generic function $f(\vec{x})$ we have introduced the inverse Laplacian as
\begin{equation}
\Delta^{-1}(f)(\vec{x}) := -\frac{1}{4\pi}\int \frac{f(\vec{x}')}{|\vec{x} - \vec{x}'|}dV'.
\end{equation}
It is worth noting that, for fixed $\vec{\psi}$, the Hamilton operator $\hat{\mathcal{H}}[\vec{\psi}]$ is Hermitian, i.e. $(\vec{\psi}_1,\hat{\mathcal{H}}[\vec{\psi}]\vec{\psi}_2)=(\hat{\mathcal{H}}[\vec{\psi}]\vec{\psi}_1,\vec{\psi}_2)$, where $(\vec{\psi}_1,\vec{\psi}_2)=\int (\vec{\psi}_1^*\cdot\vec{\psi}_2) dV$ denotes the standard $L^2$-scalar product between $\vec{\psi}_1$ and $\vec{\psi}_2$.  
However, despite the apparent simplicity of Eq.(\ref{integro-differential_equation}), the operator $\hat{\mathcal{H}}[\vec{\psi}]$ is non-linear, as all terms beyond the first one in Eq.(\ref{eq.def.H}) are quadratic in the field $\vec{\psi}$.

\subsection{Symmetries and conserved quantities}\label{sec.conservation}

Next, we identify some quantities that are conserved during the time evolution of the $s=1$ Gross-Pitaevskii-Poisson system and that will be relevant for the characterization of our solutions and their stability properties.

These quantities are associated with continuous global symmetries. For example, our theory is invariant under time translations, which means that the action~(\ref{eq.action.nonrel}) is not affected by the transformation $\vec{\psi}(t,\vec{x})\mapsto\vec{\psi}(t-t_0,\vec{x})$, with $t_0$ a real constant. Associated to this symmetry is the conserved total energy
\begin{align}\label{eq.energy}
\mathcal{E} = \int \left[\frac{1}{2 m_0}|{\nabla}\vec{\psi}|^2+\frac{\lambda_n}{4m_0^2}n^{2}+\frac{\lambda_s}{4m_0^2}\vec{s}{\,}^2+\frac{m_{0}}{2}n\mathcal{U}\right]dV.
\end{align}

In addition, one can also prove that the action~(\ref{eq.action.nonrel}) remains invariant under rotations, \mbox{$\vec{\psi}(t,\vec{x})\mapsto \hat{U}(R)\vec{\psi}(t,R^{-1}\vec{x})$,} with $\hat{U}(R) = R$ a rotation matrix or $\hat{U}(R) = I$ the identity matrix. Consequently, one has conservation of the internal (or spin),
\begin{subequations}
\begin{equation}\label{eq.spin.density}
\vec{S} = - i\int (\vec{\psi}^{*}\times\vec{\psi}) dV,
\end{equation}
the orbital,\footnote{In the term $\vec{\psi}^*\cdot\nabla \vec{\psi}$ the contraction is over the internal indices of $\vec{\psi}^*$ and $\vec{\psi}$.}
\begin{equation}\label{eq.orbitalL}
\vec{L} = -i\int [\vec{x}\times(\vec{\psi}^*\cdot\nabla \vec{\psi}) ] dV,
\end{equation}
and the total, 
\begin{equation}\label{eq.angular.momentum}
\vec{J} = \vec{S} + \vec{L},
\end{equation}
\end{subequations}
angular momentum.

Other conserved quantities associated with the Galilei group lead to the fact that the center of mass follows a free-particle trajectory. 
However, we will mostly limit ourselves to the study of spherically symmetric equilibrium configurations which are centered at the origin, where these quantities do not play a significant role.

In addition to spacetime symmetries, our effective theory possesses internal symmetries. In general, Eq.~(\ref{eq.action.nonrel}) is invariant under continuous shifts in the  phase of the wave function, \mbox{$\vec{\psi}(t,\vec{x})\mapsto e^{i\alpha}\vec{\psi}(t,\vec{x})$,} with $\alpha$ a real constant, leading to the conservation of the particle number
\begin{equation}\label{eq.number.particles}
N = \int (\vec{\psi}^*\cdot\vec{\psi}) dV.
\end{equation}
From now on we will assume $N>0$, since for $N=0$ only the trivial solution $\vec{\psi}(t,\vec{x})=0$ is possible.

Moreover, our theory features an ``accidental'' symmetry: in absence of spin-spin selfinteractions ($\lambda_s=0$), the action~(\ref{eq.action.nonrel}) is also invariant under unitary transformations, $\vec{\psi}(t,\vec{x})\mapsto \hat{U}\vec{\psi}(t,\vec{x})$, where $\hat{U}$ is a constant unitary $3\times 3$ matrix.
This symmetry induces the conserved self-adjoint second-rank tensor
\begin{equation}\label{eq.globalQ}
\hat{Q}=\int (\vec{\psi}^*\otimes \vec{\psi}) dV,
\end{equation}
with the following properties: $N= \textrm{Tr}(\hat{Q})$ and $\vec{S}=-i\textrm{Tr}(\hat{\varepsilon} \hat{Q})$, where $\hat{\varepsilon}$ is the third-rank Levi-Civita tensor and in the last expression the trace denotes the contraction of the last two indices of $\hat{\varepsilon}$ with the two indices of $\hat{Q}$. The reason why the particle number $N$ and the spin angular momentum $\vec{S}$ are codified in the tensor $\hat{Q}$ is because global phase factors and rotations are elements belonging to the $U(3)$ group. On the other hand, the other components of $\hat{Q}$ are only conserved when $\lambda_s=0$.\footnote{ The tensor $\hat{Q}$ evolves in time according to
\begin{equation}
\frac{d \hat{Q}}{dt} = \frac{\lambda_s}{m_0^2}\Im\int (\vec{\psi}\cdot\vec{\psi})^*(\vec{\psi}\otimes\vec{\psi}) dV.
\end{equation}
If $\lambda_s=0$, this tensor is conserved. If $\lambda_s\neq 0$, however, only the anti-symmetric part of this tensor and its trace are conserved.}

In the following, we will distinguish between two scenarios: the {\it symmetry-enhanced sector} of the effective theory, where $\lambda_s=0$ and the accidental $U(3)$ symmetry is manifest, and the {\it generic sector} of the effective theory, where the two coupling constants $\lambda_n$ and $\lambda_s$ can take arbitrary values except $\lambda_s = 0$. In addition, we will consider two configurations, $\psi_1(t,\vec{x})$ and $\psi_2(t,\vec{x})$, as equivalent if they are related by a symmetry transformation.\footnote{ Note that the notion of equivalent configurations depends on the specific sector of the theory that we are exploring.} In this paper, we will not distinguish between equivalent configurations and will exploit symmetry transformations to simplify the presentation. For example, if $\lambda_s=0$, we can use the $U(3)$ symmetry to diagonalize the operator $\hat{Q}$, leaving only the diagonal elements non-zero.

\subsection{Constant polarization states}\label{sec.constantpolariztion}

In this section, we present some simple properties of constant polarization states which will be of interest later. Constant polarization states are defined to have the form
\begin{equation}\label{Eq.Const.Polariz}
\vec{\psi}(t,\vec{x}) = f(t,\vec{x})\hat{\epsilon},
\end{equation}
where $f(t,\vec{x})$ is an arbitrary complex-valued function and $\hat{\epsilon}$ is a polarization vector that is independent of the space-time coordinates and that, for convenience, we normalize to one, $\hat{\epsilon}^*\cdot\hat{\epsilon}=1$. 

Given that the wave function $\vec{\psi}(t,\vec{x})$ contains an arbitrary global phase, we can parametrize a constant polarization vector as follows:
\begin{equation}\label{eq.general.pol}
\hat{\epsilon} = \sin\theta\cos\phi\hat{e}_x +
e^{i\gamma_1}\sin\theta\sin\phi\hat{e}_y+
e^{i\gamma_2}\cos\theta\hat{e}_z,
\end{equation}
where $\theta$, $\phi$, $\gamma_1$ and $\gamma_2$ are four real constants and $\hat{e}_x$, $\hat{e}_y$, and $\hat{e}_z$ are the Cartesian unit vectors. Furthermore, given the symmetries of our effective theory, one can always perform a rotation such that the polarization vector is contained in the $xy$ plane, and obtain
\begin{equation}\label{eq.pol.elliptical}
\hat{\epsilon} = \cos\phi\hat{e}_x + e^{i\gamma_1}\sin\phi\hat{e}_y,
\end{equation}
which is the expression for a general elliptical polarization vector.\footnote{ This name takes on particular significance when the state is stationary, i.e. of the form $f(t,\vec{x})=e^{-iEt}\bar{f}(\vec{x})$, and the real $\vec{\psi}_R$ and the imaginary $\vec{\psi}_I$ parts of the vector $\vec{\psi}$ evaluated at a fixed point $\vec{x}$ describe an ellipse as time progresses.}

On the other hand, from the definition of the spin density, one has
\begin{equation}
\vec{s}{\,}^2 = |f|^4 \sin^2(2\phi)\sin^2\gamma_1,
\end{equation}
which implies that $|\vec{s}|\leq |f|^2 = n$.\footnote{ Note that, in general, $\vec{s}{\,}^2=n^2-|\vec{\psi}\cdot\vec{\psi}|^2$, so the inequality $|\vec{s}|\le n$ holds true for any wave vector $\vec{\psi}(t,\vec{x})$.} Configurations with zero spin density are (up to a global phase factor) of the form $\hat{\epsilon} = \cos\phi\hat{e}_x \pm\sin\phi\hat{e}_y$, which after a rotation can be further reduced to
\begin{subequations}
\begin{equation}
\hat{\epsilon} = \hat{\epsilon}_x := \hat{e}_x. 
\label{Eq:LinPolarization}
\end{equation}
In contrast, configurations which saturate the spin density, $|\vec{s}| = n$, are of the form $\hat{\epsilon} = \frac{1}{\sqrt{2}}\left( \hat{e}_x \pm i\hat{e}_y \right)$, which using a rotation can be reduced to
\begin{equation}
\hat{\epsilon} = \hat{\epsilon}_z^{(+)} := \frac{1}{\sqrt{2}}\left( \hat{e}_x + i\hat{e}_y \right).
\label{Eq:CircularPolarization}
\end{equation}
\end{subequations}
We call constant polarization states having $\hat{\epsilon}$ as in Eq.~(\ref{Eq:LinPolarization}) {\it linearly polarized} and states as in~(\ref{Eq:CircularPolarization}) {\it circularly polarized}.

When the spin-spin selfinteraction is absent ($\lambda_s=0$) the theory is $U(3)$ invariant, and all states with constant polarization are equivalent to each other.\footnote{Any elliptical polarization vector~(\ref{eq.pol.elliptical}) can be expressed in the form $\hat{\epsilon}=\hat{U}\hat{e}_x$, where
\begin{equation}
\hat{U} = \left(\begin{matrix}
\cos\phi & -e^{-i\gamma_1}
\sin\phi & 0\\
e^{i\gamma_1}\sin\phi & \cos\phi & 0\\
0 & 0 & 1
\end{matrix}
\right)
\end{equation}
is a unitary matrix.} This means that, when exploring constant polarization states in the symmetry-enhanced sector of the effective theory, we can restrict ourselves to the simplest case in which $\hat{\epsilon}=\hat{\epsilon}_x$. In contrast, when $\lambda_s\neq 0$, one can prove (see App.~\ref{app.ConstantPol}) that any solution of Eq.~(\ref{eq.GPs=1}) with constant polarization has either linear  ($\vec{s}{\,}^2 = 0$) or circular ($\vec{s}{\,}^2 = n^2$) polarization, which in this case are inequivalent to each other.

Summarizing, if we restrict ourselves to constant polarization states, it is sufficient to consider the linear and circular ones, which degenerate when $\lambda_s=0$.

\section{Equilibrium configurations}\label{sec.equilibrium}

In this section, we identify the different types of equilibrium configurations that can exist in our effective theory. For that purpose, we define an equilibrium configuration as a critical point of the energy functional. In practice,  we restrict ourselves to variations that keep conserved quantities fixed. As we will demonstrate, the choice of which quantities are fixed might affect the location of the critical points, and, ultimately, the equilibrium states that can exist in the different sectors of the theory.

To proceed, let us first concentrate on the generic sector,  where the two coupling constants $\lambda_n$ and $\lambda_s$ can take arbitrary values (except $\lambda_s = 0$) and the particle number $N$ is conserved. In this case, the relevant critical points are obtained from varying the modified energy functional
\begin{equation}\label{eq.functionalF}
\mathcal{E}_E[\vec{\psi}]:=\mathcal{E}[\vec{\psi}]+\frac{1}{2}E\left(N- \int (\vec{\psi}^*\cdot\vec{\psi}) dV \right),
\end{equation}
where $E$ is a Lagrange multiplier associated with the constraint that guarantees that the particle number remains fixed in the variations. Remember that the functionals $\mathcal{E}[\vec{\psi}]$ and $N[\vec{\psi}]$ defined in Eqs.~(\ref{eq.energy}) and~(\ref{eq.number.particles}) depend on the field $\vec{\psi}$ and its spatial gradients, but not on its time derivatives, and for that reason we will treat $\mathcal{E}_E[\vec{\psi}]$ as a functional of $\vec{\psi}(\vec{x})$ alone, ignoring any time evolution. Correspondingly, we assume that the Lagrange multiplier $E$ is time-independent.

The first variation of $\mathcal{E}_E[\vec{\psi}]$ with respect to $\vec{\psi}$ yields
\begin{equation}
\delta \mathcal{E}_E = \textrm{Re}(\hat{\mathcal{H}}[\vec{\psi}]\vec{\psi}-E\vec{\psi},\delta\vec{\psi}).
\end{equation}
A critical point is characterized by the condition that $\delta \mathcal{E}_E=0$ for all  $\delta\vec{\psi}$; hence, equilibrium configurations $\vec{\psi}(\vec{x})$ must fulfill the nonlinear equation
\begin{equation}\label{eq.eigenvalue.stationary}
E\vec{\psi}=\hat{\mathcal{H}}[\vec{\psi}]\vec{\psi},
\end{equation}
which must be solved subject to appropriate boundary conditions. Note that if $\vec{\psi}(\vec{x})$ satisfies Eq.~(\ref{eq.eigenvalue.stationary}), then $e^{i\alpha}\vec{\psi}(\vec{x})$, where $\alpha$ is an arbitrary phase independent of $\vec{x}$, is also a solution to this equation. If we make this phase time-dependent and introduce this expression into the dynamical Eq.~(\ref{integro-differential_equation}), we obtain 
\begin{equation}\label{eq.single.frequency}
 \vec{\psi}(t,\vec{x})=e^{-i E t} \vec{\psi}(t=0,\vec{x}) ,
\end{equation}
with $\vec{\psi}(t=0,\vec{x})=\vec{\psi}(\vec{x})$. In the context of quantum mechanics, these solutions are usually referred to as {\it stationary states}.  They have time-independent particle $n(t,\vec{x})$ and spin $\vec{s}(t,\vec{x})$ densities, and, consequently, the Hamiltonian~(\ref{eq.def.H}) remains constant in the evolution. Furthermore, stationary states are eigenfunctions of the Hamilton operator. 

Next, we extend the study of the equilibrium configurations to the symmetry-enhanced sector, where $\lambda_s=0$ and in addition to the particle number $N$ the charges $\hat{Q}$ associated with the accidental symmetry are conserved. In analogy with the previous case, we define the energy functional
\begin{equation}\label{eq.functionalF.multi}
\mathcal{E}_{\hat{E}}[\vec{\psi}]:=\mathcal{E}[\vec{\psi}]+\frac{1}{2} \textrm{Tr}\left[ \hat{E}\left(\hat{Q}- \int (\vec{\psi}^*\otimes\vec{\psi}) dV \right) \right],
\end{equation}
with $\hat{E}$ a constant Hermitian transformation that plays the role of the Lagrange multiplier associated with $\hat{Q}$. In this case, the first variation of Eq.~(\ref{eq.functionalF.multi}) can be expressed in the form
\begin{equation}
\delta \mathcal{E}_{\hat{E}} = \textrm{Re}(\hat{\mathcal{H}}[\vec{\psi}]\vec{\psi}-\hat{E}\vec{\psi},\delta\vec{\psi}),
\end{equation}
and, therefore, imposing $\delta\mathcal{E}_{\hat{E}}=0$ for all $\delta\vec{\psi}$, yields
\begin{equation}
\label{eq.eigenvalue.multi}
\hat{E}\vec{\psi}=\hat{\mathcal{H}}[\vec{\psi}]\vec{\psi}.
\end{equation}
Now, if $\vec{\psi}(\vec{x})$ is a solution of Eq.~(\ref{eq.eigenvalue.multi}), then $\hat{U}\vec{\psi}(\vec{x})$ also satisfies this equation, where $\hat{U}=e^{i\hat{A}}$ is a constant unitary transformation with $\hat{A}$ Hermitian and commuting with $\hat{E}$. However, if we allow $\hat{A}$ to depend on time, and substitute this expression into the dynamical  Eq.~(\ref{integro-differential_equation}), we obtain
\begin{equation}\label{eq.multi.frequency}
 \vec{\psi}(t,\vec{x})=e^{-i \hat{E} t} \vec{\psi}(t=0,\vec{x}) ,
\end{equation}
where  again $\vec{\psi}(t=0,\vec{x})=\vec{\psi}(\vec{x})$. We will refer to these configurations as {\it multi-frequency states}, given that they involve more than one frequency of oscillation. These states maintain the particle number density $n(t,\vec{x})$ time-independent, although, in general, the spin density $s(t,\vec{x})$ depends on time. Nonetheless, this does not affect the Hamiltonian $\hat{\mathcal{H}}[\vec{\psi}]$, which is independent of the spin density when $\lambda_s=0$. Stationary states arise for the particular case in which $\hat{E}$ is proportional to the identity matrix; however, in the following, we will exclude this case when referring to multi-frequency states. Under this assumption, we can conclude that multi-frequency states are not eigenfunctions of the Hamilton operator. 

To summarize, the equilibrium configurations of the generic sector of the effective theory (where $\lambda_n$ is arbitrary and $\lambda_s\neq 0$), consist only of stationary states~(\ref{eq.single.frequency}). In the symmetry-enhanced sector (where $\lambda_n$ is arbitrary and $\lambda_s= 0$), in addition to the stationary (i.e. single-frequency) states, one must also consider the multi-frequency solutions~(\ref{eq.multi.frequency}).

In Secs.~\ref{equilibrium.stationary} and~\ref{equilibrium.multi}, we discuss in more detail the equilibrium configurations of the different sectors of the theory. Before doing so, however, in Sec.~\ref{sec.properties.E} we present some general properties of the energy functional and the equilibrium configurations that hold irrespectively of the values of the coupling constants $\lambda_n$ and $\lambda_s$.

\subsection{General properties of the energy functional and the equilibrium configurations}\label{sec.properties.E}

Following Ref.~\cite{Nambo:2024gvs}, we discuss some interesting properties of the energy functional and the equilibrium configurations that to a large extent can be deduced from a simple scaling argument.
To this purpose, it is convenient to express Eq.~(\ref{eq.energy}) in the form
\begin{equation}
   \mathcal{E}[\vec{\psi}\,] = T[\vec{\psi}\,] +\lambda_n F_n[n] + \lambda_s F_s[\vec{s}\,]- D[n, n],
   \label{DimensionlessFuncE} 
\end{equation}
where the functionals $T$, $F_n$, $F_s$ and $D$ are defined by 
\begin{subequations}\label{Eqs:relations}
\begin{align}
    T[\vec{\psi}\,] &:= \frac{1}{2m_0}\int |\nabla \vec{\psi}(\vec{x})|^2 dV,\\ 
    F_n[n] &:= \frac{1}{4m_0^2} \int n(\vec{x})^2 dV,\\
    F_s[\vec{s}\,] & := \frac{1}{4m_0^2} \int \vec{s}(\vec{x})^2 dV, \\
    D[n, n] &:= \frac{Gm_0^2}{2} \int \int \frac{n(\vec{x}) n(\vec{x}')}{|\vec{x} - \vec{x}'|} dV' dV,
\end{align}
\end{subequations}
and are positive definite. 

Next, let $\nu > 0$ be an arbitrary real and positive parameter, and $\vec{\psi}(\vec{x})$ a given wave function which does not vanish identically. Consider the rescaled function
\begin{equation}\label{eq.rescaling}
\vec{\psi}_\nu(\vec{x}) := \nu^{3/2}\vec{\psi}(\nu\vec{x}),
\end{equation}
which leaves the particle number and the global $U(3)$ charges invariant, i.e. $N[\vec{\psi}_\nu]=N[\vec{\psi}]$ and $\hat{Q}[\vec{\psi}_\nu]=\hat{Q}[\vec{\psi}]$ for all $\nu > 0$. Under the rescaling~(\ref{eq.rescaling}), the energy functional~(\ref{DimensionlessFuncE}) transforms according to
\begin{equation}
    \mathcal{E}[\vec{\psi}_\nu] = \nu^2 T[\vec{\psi}] + \nu^3 \lambda_n F_n[n] + \nu^3 \lambda_s F_s[\vec{s}] - \nu D[n,n].\label{Eq:Rescaling}
\end{equation}
Furthermore, the first and second variations of $\mathcal{E}[\vec{\psi}_\nu]$ at $\vec{\psi}_{\nu=1} = \vec{\psi}$ are given by
\begin{subequations}
\begin{align}
    \label{FirstDetivative}
    \left.\frac{d} {d\nu} \mathcal{E}[\vec{\psi}_\nu] \right|_{\nu = 1} &= 2T[\vec{\psi}] + 3\lambda_n F_n[n] + 3\lambda_s F_s[\vec{s}] - D[n,n], \\
    \label{SecondDerivative}
    \left.\frac{d^2} {d\nu^2} \mathcal{E}[\vec{\psi}_\nu] \right|_{\nu = 1}  &= 2T[\vec{\psi}] + 6\lambda_nF_n[n] + 6\lambda_sF_s[\vec{s}].
\end{align}
\end{subequations}
A number of interesting conclusions can be drawn from these results.

\subsubsection{Lower bound of the energy functional}\label{sec.lower.bound}

\begin{figure}
	\centering
\includegraphics[width=7.cm]{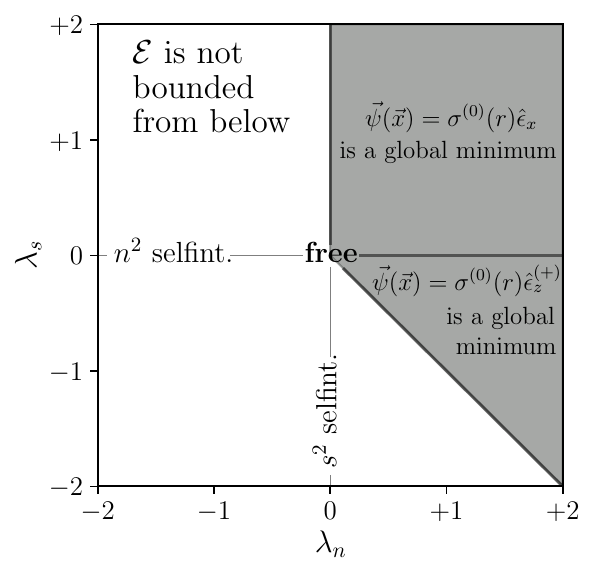}
\caption{{\bf Parameter space of the theory:} The shadow region ($\lambda_0\ge 0$) represents the parameter space of our effective theory for which, when $N$ is fixed, the energy functional is bounded from below. Furthermore, in this region there exists a global minimum of the energy functional, which is attained for a stationary and spherically symmetric state of constant polarization (linear if $\lambda_s >0$ and circular if $\lambda_s<0$) and negative energy. In general, we cannot guarantee the uniqueness of the ground state, unless $\lambda_n=\lambda_s=0$, in which case the ground state is unique up to translations and rigid unitary transformations.} \label{fig.lower.bound}
\end{figure}

First, we claim that, for fixed $N$, the energy functional~(\ref{DimensionlessFuncE}) is bounded from below if and only if $\lambda_0\geq 0$, where
\begin{equation}
\lambda_0 := \left\{ \begin{array}{ll}
 \lambda_n, & \textrm{if } \lambda_s\geq 0,\\
 \lambda_n - |\lambda_s|, & {\textrm{if }}\lambda_s < 0,
\end{array} \right.
\label{Eq:lambda0}
\end{equation}
see  Fig.~\ref{fig.lower.bound} for an illustration. Furthermore, the same holds true when $\lambda_s=0$ and $\hat{Q}$ is fixed.

To prove this, we first assume that $N= \textrm{Tr}(\hat{Q})$ is fixed but allow the trace-free part of $\hat{Q}$ to vary. Notice that if $\lambda_n\geq 0$ and $\lambda_s\geq 0$, then $\mathcal{E}[\psi]\geq T[\vec{\psi}] - D[n,n]$, which is known to be bounded from below when $N$ is fixed (see, for instance, Ref.~\cite{Lieb1977}). In contrast, when $\lambda_n <  0$ and $\lambda_s\geq 0$, we can choose $\vec{\psi}$ equal to a state of constant linear polarization, such that $F_s[\vec{s}] = 0$, and then it follows from Eq.~(\ref{Eq:Rescaling}) that $\mathcal{E}[\vec{\psi}_\nu]$ can be made arbitrarily negative by choosing $\nu$ large, showing that the energy functional is not bounded from below.\footnote{An energy functional that is unbounded from below is considered ill-defined, as it implies that an unlimited amount of energy can be extracted from the system,  rendering the theory non-physical. However, in the context of effective theories, this issue is expected to be resolved by the ultraviolet completion, where higher-order operators not included in Eq.~(\ref{eq.action.nonrel}) should ensure that the energy functional becomes bounded. This is what happens, for instance, with the well-known cosine potential of the QCD axion, where expanding in a Taylor series reveals that the quartic term has a negative coefficient (indicating an attractive self-interaction), while the potential remains positive definite.} Finally, if $\lambda_s < 0$, then the inequality $|\vec{s}|\leq n$ implies that $\lambda_n F_n[n] + \lambda_s F_s[\vec{s}]\geq (\lambda_n - |\lambda_s|) F_n[n]$, with the equality sign for states of constant circular polarization. Hence, in this case, the energy functional is bounded from below if and only if $\lambda_n - |\lambda_s|\geq 0$. 

When $\lambda_s = 0$ and $\hat{Q}$ is fixed it is easy to verify that one can apply the same arguments as above to show that the energy functional is bounded from below if and only if $\lambda_n\geq 0$. It should be noted that in this case one cannot always choose $\vec{\psi}$ to be a state of constant linear polarization since this requires that $\hat{Q}$ has rank $1$; however, this is not needed for the scaling argument since the term $\lambda_s F_s[\vec{s}]$ in Eq.~(\ref{Eq:Rescaling}) is automatically zero when $\lambda_s = 0$. Hence, one can use again the particular variation defined in Eq.~(\ref{eq.rescaling}) (which fixes $\hat{Q}$) to conclude that the energy functional is unbounded from below when $\lambda_n < 0$.

\subsubsection{Energy functional of equilibrium states}

Second, if $\vec{\psi}(t,\vec{x})$ is an equilibrium configuration, then the first variation of the energy functional vanishes, and Eq.~(\ref{FirstDetivative}) yields the relation
\begin{equation}\label{Eq:EnergFunctStat}
D[n,n] = 2T[\vec{\psi}] + 3\lambda_n F_n[n] + 3\lambda_s F_s[\vec{s}].
\end{equation}
This expression is valid for any equilibrium state and allows one to express the selfgravity term $D[n,n]$ as a function of the kinetic term  $T[\vec{\psi}]$ and the selfinteraction ones $F_n[n]$ and $F_s[\vec{s}]$, similar to the usual virial relation. Accordingly, the total energy of any equilibrium state can be expressed as
\begin{equation}\label{Eq:EnerRelat}
\mathcal{E}[\vec{\psi}] = -T[\vec{\psi}] - 2\lambda_nF_n[n]- 2\lambda_sF_s[\vec{s}],
\end{equation}
which does not require the computation of the selfgravity term $D[n,n]$. Since $\lambda_n F_n[n] + \lambda_s F_s[\vec{s}]\geq \lambda_0 F_n[n]$ and $T[\vec{\psi}]$ and $F_n[n]$ are positive definite, it follows that the energy of the equilibrium states is always negative when $\lambda_0\geq 0$.

\subsubsection{Discarding local minima of the energy functional}

Third, Eq.~(\ref{SecondDerivative}) implies that a critical point at $\nu=1$ corresponds to a local minimum of $\mathcal{E}[\vec{\psi}_\nu]$ if $T[\vec{\psi}] + 3\lambda_n F_n[n] + 3\lambda_s F_s[\vec{s}]$ is positive and to a local maximum if it is negative. In particular, an equilibrium state $\vec{\psi}(t,\vec{x})$ cannot be a (local) minimum of the energy functional with respect to arbitrary variations that fix $N$ (or $\hat{Q}$ if $\lambda_s=0$) when $\lambda_n F_n[n] + \lambda_s F_s[\vec{s}] < -T[\vec{\psi}]/3$.\\

Next, we come back to the discussion of the stationary and multi-frequency states in the different sectors of the effective theory.

\subsection{Generic sector: Stationary solutions}\label{equilibrium.stationary}

In the generic sector of the effective theory, equilibrium configurations are stationary states, which are characterized by the ansatz, c.f.~Eq.~(\ref{eq.single.frequency}):
\begin{equation}\label{eq.harmonic}
 \vec{\psi}(t,\vec{x})=e^{-i E t} \vec{\sigma}^{(0)}(\vec{x}) ,
\end{equation}
where $\vec{\sigma}^{(0)}$ is a complex vector-valued function of $\vec{x}$ and the energy eigenvalue $E$ is a real constant. As explained previously, these states have a time-independent Hamiltonian, and consequently, they give rise to a static gravitational potential $\mathcal{U}(t,\vec{x})=\mathcal{U}(\vec{x})$. Introducing Eq.~(\ref{eq.harmonic}) into the integro-differential equation~(\ref{eq.def.H}) yields the nonlinear eigenvalue problem:
\begin{equation}\label{Eq:NLsystem}
E\vec{\sigma}^{(0)} =\hat{\mathcal{H}}[ \vec{\sigma}^{(0)}]\vec{\sigma}^{(0)}.
\end{equation}
This equation determines the stationary solutions of the $s=1$ Gross-Pitaevskii-Poisson system, which will be solved numerically in Sec.~\ref{sec.numerical} under the assumption of spherical symmetry.

However, before doing so, we ask ourselves whether stationary states can arise as {\it global} minima of the energy functional. This question is particularly relevant since such a minimum is expected to represent a stationary state that is (orbitally) stable under small enough perturbations~\cite{Cazenave1982}.

In order to formulate the question in a more precise way, let
\begin{equation}
\mathcal{A}_N := \left\{ \vec{\psi}\in H^1(\Real^3,\Complex^3) : \int |\vec{\psi}(\vec{x})|^2 dV = N \right\},
\end{equation}
where here $H^1(\Real^3,\Complex^3)$ denotes the space of wave functions $\vec{\psi}: \Real^3\to \Complex^3$ such that $\vec{\psi}$ and its first partial derivatives are (Lebesgue-) square-integrable. It can be shown that $\mathcal{E}$ is well-defined on this space~\cite{Lieb1977}. Thus, we ask whether there exists a wave function $\vec{\psi}_*\in \mathcal{A}_N$ such that
\begin{equation}\label{Eq:Minimization}
\mathcal{E}[\vec{\psi}_*] = \inf\limits_{\vec{\psi}\in \mathcal{A}_N} \mathcal{E}[\vec{\psi}].
\end{equation}
The first observation is that a necessary condition for the existence of such a minimum is that $\mathcal{E}$ must be bounded from below on $\mathcal{A}_N$. As has been shown in Sec.~\ref{sec.lower.bound}, this is the case if and only if $\lambda_0\geq 0$, with $\lambda_0$ defined in Eq.~(\ref{Eq:lambda0}), so for the remainder of this section we shall assume that $\lambda_0\geq 0$.

Next, we claim that a global minimum is attained by a wave function of constant polarization and that, if multiple global minima exist, all share this form. To show this, we first decompose $\vec{\psi}(\vec{x})$ according to
\begin{equation} \label{eq.polar_decomp}
\vec{\psi}(\vec{x}) = f(\vec{x})\hat{\epsilon}(\vec{x}),
\end{equation}
where $f(\vec{x}) := |\vec{\psi}(\vec{x})|$ and $\hat{\epsilon}(\vec{x})$ has unit norm. Using the fact that $n = f^2$ and $|\nabla\vec{\psi}|^2 = |\nabla f|^2 + n|\nabla\hat{\epsilon}|^2$, and recalling the inequality $\lambda_n F_n[n] + \lambda_s F_s[\vec{s}]\geq \lambda_0 F_n[n]$, one obtains
\begin{equation}
\mathcal{E}[\vec{\psi}] \geq
\mathcal{E}_{\textrm{scalar}}[f] := T_{\textrm{scalar}}[f] + \lambda_0 F_n[n] - D[n,n],
\end{equation}
where $T_{\textrm{scalar}}[f] := \frac{1}{2m_0}\int |\nabla f(\vec{x})|^2 dV$. Furthermore, equality holds only if on the set of points where $f > 0$ the polarization vector 
$\hat{\epsilon}(\vec{x})$ is constant and satisfies $\lambda_n + \lambda_s|\hat{\epsilon}^*\times\hat{\epsilon}|^2 = \lambda_0$, which means that $\hat{\epsilon}(\vec{x})$ has linear (circular) polarization if $\lambda_s > 0$ ($\lambda_s < 0$); see App.~\ref{app.ConstantPol}.

One can further decrease the energy functional by replacing $f$ by its  ``symmetric decreasing rearrangement'' $f_*$ (which, by definition, is spherically symmetric, nonincreasing,  nonnegative and satisfies $\int f_*^p dV = \int f^p dV$ for all $p\geq 1$; see~\cite{Lieb1977} and references therein). It follows from~\cite{Lieb1977} that  $\mathcal{E}_{\textrm{scalar}}[f]\geq \mathcal{E}_{\textrm{scalar}}[f_*]$, with strict inequality unless $f(\vec{x}) = f_*(\vec{x}-\vec{x}_0)$ with a constant vector $\vec{x}_0$.

As a consequence, any minima $\vec{\psi}_*$ of the problem~(\ref{Eq:Minimization}) must lie in the subset of $\mathcal{A}_N$ consisting of constant polarization states for which the function $f(\vec{x}) := |\vec{\psi}(\vec{x})|$ is radially symmetric (up to translations), nonincreasing, and nonnegative and for which the polarization is linear for $\lambda_s > 0$ and circular for $\lambda_s < 0$. The function $f$ can then be found by minimizing the functional $\mathcal{E}_{\textrm{scalar}}$, which is known to have a minimum~\cite{Lieb1977, LIONS1984109, LIONS1984223} satisfying $f(\vec{x}) = f_*(\vec{x})$.

In conclusion, global minima of Eq.~(\ref{Eq:Minimization}) exist, and (up to translations) all of them are described by stationary and spherically symmetric states of constant polarization that have the form $\vec{\psi}(t,\vec{x})=e^{-iEt}\sigma^{(0)}(r)\hat{\epsilon}$, with $\sigma^{(0)}(r)$ monotonically decreasing and positive, the polarization $\hat{\epsilon}$ being linear when $\lambda_s > 0$ and circular when $\lambda_s < 0$. Whether or not the function $\sigma^{(0)}(r)$ characterizing the ground state is unique for $\lambda_0 > 0$ is an interesting open question that will not be addressed in this article.

Figure~\ref{fig.lower.bound} further illustrates the results of this section.

\subsection{Symmetry-enhanced sector: Stationary and multi-frequency solutions}\label{equilibrium.multi}

In absence of spin-spin selfinteractions ($\lambda_s=0$), stationary configurations still exist. Following similar arguments as in Sec.~\ref{equilibrium.stationary} we conclude that, for fixed $N$ and $\lambda_n\ge 0$, the energy is always minimized by a stationary and spherically symmetric state of constant polarization and no nodes. Furthermore, in the free theory ($\lambda_n=\lambda_s=0$), the ground state is unique, up to translations and rigid unitary transformations~\cite{Lieb1977}.

As explained in Sec.~\ref{sec.equilibrium}, in the symmetry-enhanced sector there also exists the possibility of equilibrium configurations that are realized as multi-frequency states. Expanding
\begin{equation}
\vec{\psi}(t=0,\vec{x}) = \sum\limits_{\lambda=1}^3 \sigma^{(0)}_\lambda(\vec{x})\hat{e}_\lambda
\end{equation}
in terms of an orthornormal basis $\hat{e}_\lambda$ of $\Complex^3$ which diagonalizes the transformation $\hat{E}$, Eq.~(\ref{eq.multi.frequency}) leads to the following expression for the multi-frequency states:
\begin{equation}\label{eq.harmonic.multi}
\vec{\psi}(t,\vec{x})=\sum_{\lambda=1}^3 e^{-iE_\lambda t}\sigma^{(0)}_\lambda(\vec{x})\hat{e}_\lambda,
\end{equation}
where $\sigma^{(0)}_\lambda$ are complex-valued functions depending on $\vec{x}$ and $E_\lambda$ denote the eigenvalues of $\hat{E}$ which are real and  correspond to the frequency of oscillation associated with $\sigma^{(0)}_\lambda$. As the stationary states, these solutions have a time-independent Hamiltonian and gravitational potential. Introducing Eq.~(\ref{eq.harmonic.multi}) into the integro-differential equation~(\ref{eq.def.H}) yields the nonlinear eigenvalue problem:\footnote{Recall that $\hat{\mathcal{H}}[\vec{\psi}]$ depends on $\vec{\psi}$ only through the combination $n=\sum_\lambda |\sigma_\lambda^{(0)}|^2$, which is independent of $t$.}
\begin{equation}\label{Eq:NLsystem.multi}
E_\lambda\sigma^{(0)}_\lambda =\hat{\mathcal{H}}[ \vec{\psi}]\sigma^{(0)}_\lambda.
\end{equation}
These equations determine the multi-frequency solutions of the symmetry-enhanced sector of the $s=1$ Gross-Pitaevskii-Poisson system, which will be solved numerically in Sec.~\ref{sec.sph.multi} under the assumption of spherical symmetry.

An open question is what states minimize the energy functional for arbitrary fixed values of $\hat{Q}$. Note that these states cannot have constant polarization unless $\hat{Q}$ has rank one. We do not address this question in this article.

An open question is whether constant polarization states, which minimize the energy functional for fixed $N$, also minimize it for arbitrary fixed values of $\hat{Q}$. We do not address this question in this article.

\section{Spherically symmetric equilibrium configurations}\label{Sec:.spherically}

Hereafter, we further specialize on equilibrium configurations which are spherically symmetric. As discussed in Sec.~\ref{sec.conservation}, our effective theory is invariant under the rotation group, hence we expect spherically symmetric configurations to play a relevant role in our discussion. Furthermore, as we have already argued in Sec.~\ref{sec.equilibrium}, for fixed $N$ and $\lambda_0\ge 0$, all ground state configurations are stationary and spherically symmetric (up to translations).

We define a spherically symmetric configuration as one that is invariant under rotational transformations.\footnote{ Clearly, a spherically symmetric configuration should be associated with a radially symmetric gravitational potential. Through Poisson's equation, this implies that the particle density $n$ should have the same symmetry. The question then is which wave functions $\vec{\psi}(t,\vec{x})$ give rise to such densities and, at the same time, are self-consistent stationary or multi-frequency solutions of the $s=1$ Gross-Pitaevskii-Poisson system. While in this article we do not provide a complete answer to this question (since we do not guarantee obtaining \emph{all possible} such wave functions), we nevertheless identify different families of such states based on symmetry arguments.} Recall that Eq.~(\ref{eq.action.nonrel}) is invariant with respect to any two of the following representations of the $SO(3)$ group: $\vec{\psi}(t,\vec{x})\mapsto R\vec{\psi}(t, R^{-1}\vec{x})$ and $\vec{\psi}(t,\vec{x})\mapsto \vec{\psi}(t, R^{-1}\vec{x})$. Therefore, a spherically symmetric configuration should lie in the trivial irreducible representation of any of these two representations of the rotation group.

In the first case, one obtains a state of the form
\begin{subequations}
\begin{equation}\label{eq.radial.generic}
\vec{\psi}(t,\vec{x}) = \psi_r(t,r)\hat{e}_r,
\end{equation}
with $\hat{e}_r$ the unit radial vector and $\psi_r(t,r)$ an arbitrary complex-valued function of $t$ and $r = |\vec{x}|$. In the second case, however, one gets
\begin{equation}\label{eq.constant.multi.generic}
\vec{\psi}(t,\vec{x}) = \sum\limits_{\lambda=1}^3 \psi_\lambda(t,r)\hat{e}_\lambda,
\end{equation}
\end{subequations}
where the three components of $\vec{\psi}(t,\vec{x})$ in an orthonormal constant basis $\hat{e}_\lambda$ of $\Complex^3$ are functions of $t$ and $r$ only. Note that in both cases the particle density $n$ is radially symmetric, which, according to Poisson's equation, guarantees that the gravitational potential $\mathcal{U}$ also respects this symmetry. The spin density $\vec{s}$ vanishes in the first case, however, in the second one it may be non-zero, although its three components are functions of $t$ and $r$ only.

Next, we identify the different types of spherically symmetric equilibrium configurations that can exist in our
effective theory. For doing this, we combine our definition of equilibrium configurations in Sec.~\ref{sec.equilibrium}, together with that of spherically symmetric configurations in this section.

\subsection{Stationary spherical solutions}\label{sec.sph.stationary}

\begin{table*}
\caption{{\bf The stationary and spherical $s=1$ Gross-Pitaevskii-Poisson system:} The stationary, spherically symmetric $s=1$ Gross-Pitaevskii-Poisson equations as compared to other systems studied in the framework of
multi-scalar field theories. The comparison depends on the polarization vector $\hat{\epsilon}$. A general constant polarization (different from the linear and circular ones) is prohibited in presence of spin-spin selfinteractions. Selfinteracting, radially polarized Proca stars are not related with previously known solutions. SP (Schr\"odinger-Poisson), GPP (Gross-Pitaevskii-Poisson).}
\begin{center}
\begin{tabular}{l l l l l l l l l}
 \hline
 \hline
   Polarization & $\quad$ & $\lambda_n=0$, $\lambda_s=0$   & $\quad$ & $\lambda_n\neq 0$, $\lambda_s=0$ &$\quad$& $\lambda_n=0$, $\lambda_s\neq 0$ &$\quad$& $\lambda_n\neq 0$, $\lambda_s\neq 0$\\
 \hline
  Constant: linear  &&  $\ell=0$ SP~\cite{Roque:2023sjl,Nambo:2024gvs}    && $\ell=0$ GPP~\cite{Nambo:2024gvs} &&  $\ell=0$ SP~\cite{Roque:2023sjl,Nambo:2024gvs} && $\ell=0$ GPP~\cite{Nambo:2024gvs}\\
  \hspace{1.53cm}circular && $\ell=0$ SP~\cite{Roque:2023sjl,Nambo:2024gvs}    && $\ell=0$ GPP~\cite{Nambo:2024gvs} && $\ell=0$ SP~\cite{Roque:2023sjl,Nambo:2024gvs} && $\ell=0$ GPP~\cite{Nambo:2024gvs} \\
  \hspace{1.53cm}arbitrary && $\ell=0$ SP~\cite{Roque:2023sjl,Nambo:2024gvs}    && $\ell=0$ GPP~\cite{Nambo:2024gvs} && non-existent && non-existent \\
  Radial  && $\ell=1$ SP~\cite{Roque:2023sjl}   && $\ell=1$ GPP (new) &&  $\ell=1$ GPP (new) && $\ell=1$ GPP (new) \\
 \hline
 \hline
\end{tabular}
\end{center}
\label{table.background}
\end{table*}

For the following, it will be convenient to express the field $\vec{\sigma}^{(0)}(\vec{x})$ of the stationary ansatz~(\ref{eq.harmonic}) in the form $\vec{\sigma}^{(0)}(\vec{x})=\sigma^{(0)}(r) \hat{\epsilon}(\vec{x})$. Here $\sigma^{(0)}$ is a real-valued function which, due to spherical symmetry, depends only on the radial coordinate $r$ and $\hat{\epsilon}$ is a complex polarization vector that, in general, depends on $\vec{x}$ and is normalized to have unit length. Then, combining Eqs.~(\ref{eq.harmonic}) and~(\ref{eq.radial.generic}) leads to $\hat{\epsilon}(\vec{x})=\hat{e}_r$, whereas combining  Eqs.~(\ref{eq.harmonic}) and~(\ref{eq.constant.multi.generic}) yields $\hat{\epsilon}(\vec{x})=\hat{\epsilon}(r)$.

In addition, the wave function $\vec{\psi}(t,\vec{x})$ needs to satisfy the $s = 1$ Gross-Pitaevskii equation~(\ref{eq.dimensionless.GPP.1.GP}), and this puts a further condition on the possible form of the polarization vector. Although we have not analyzed the full implications of this condition,\footnote{However, when $\lambda_s=0$ one can prove that $\hat{\epsilon}(r)$ must be constant since in this case the three components of $\vec{\sigma}^{(0)}(\vec{x})$ satisfy the same time-independent Schr\"odinger equation with the same energy level $E$, and thus they must be proportional to each other according to the Sturm oscillation theorem~\cite{Simon2005}. Further, in App.~\ref{app.ConstantPol}, we prove that under the assumption that $\hat{\epsilon}(r)$ is constant, only linear or circular polarizations are possible when $\lambda_s\neq 0$, although in this case we have not been able to demonstrate that the polarization vector is necessarily constant.} we identified three different types of polarization vectors that are compatible with the structure of the field equations and lead to stationary and spherically symmetric  configurations:
\begin{enumerate}
\item[i)] A linear polarization vector, for which $\hat{\epsilon}(\vec{x})=\hat{e}_x$; see Eq.~(\ref{Eq:LinPolarization}).
\item[ii)] A circular polarization vector, for which $\hat{\epsilon}(\vec{x})=\hat{\epsilon}_z^{(+)}$; see Eq.~(\ref{Eq:CircularPolarization}).
\item[iii)] A radial polarization vector, for which $\hat{\epsilon}(\vec{x}) = \hat{e}_r$, with $\hat{e}_r$ the unit radial vector.
\end{enumerate}
The first two cases were already introduced in Sec.~\ref{sec.constantpolariztion} and represent (up to a global symmetry transformation) the most general stationary states with constant polarization. For $\lambda_s=0$, they degenerate; however, when $\lambda_s\neq 0$ they lead to inequivalent states (see App.~\ref{app.ConstantPol}). Furthermore, as discussed in the previous section, when $\lambda_0\ge 0$ there exists 
a spherical and constant polarization state that minimizes the energy functional. These solutions were previously explored in Refs.~\cite{Jain:2021pnk, Zhang:2021xxa}. The radially polarized states can be obtained by substituting $\gamma_1=\gamma_2=0$ and $\theta=\vartheta$ and $\phi=\varphi$ into Eq.~(\ref{eq.general.pol}), where $\vartheta$ and $\varphi$ represent the polar and azimuthal angles in three-dimensional space, respectively. These solutions constitute the nonrelativistic limit of the spherically symmetric Proca stars originally reported in~\cite{Brito:2015pxa} (see also Ref.~\cite{Wang:2023tly}, which explores radial polarization under the name of hedgehog field configurations).

Introducing the ans\"atze i), ii), and iii) into the $s=1$ Gross-Pitaevskii-Poisson system~(\ref{eq.dimensionless.GPP.1}), 
we obtain that the stationary and spherically symmetric configurations of linear, circular and radial polarization
must satisfy the nonlinear eigenvalue problem
\begin{widetext}
\begin{subequations}\label{s=1GPP.stationary}
\begin{eqnarray}
E\sigma^{(0)}&=&\left[-\frac{1}{2m_0}\left(\Delta_s-\frac{2\gamma}{r^2}\right)+\frac{\lambda_n+\alpha\lambda_s}{2m_0^2}\sigma^{(0)2} + m_0\mathcal{U}\right]\sigma^{(0)},\label{s=1GPP.stationary.1}\\
\Delta_s\mathcal{U} &=&4\pi Gm_0\sigma^{(0)2}.\label{s=1GPP.stationary.2}
\end{eqnarray}
\end{subequations}
\end{widetext}
Here, $\Delta_s:=\frac{1}{r}\frac{d^2}{dr^2}r$ denotes the radial part of the Laplace operator, and the parameters $\gamma$ and $\alpha$ depend on the polarization vector $\hat{\epsilon}$ and take the following values:
\begin{enumerate}
    \item[i)] $\gamma=0$, $\alpha=0$ for linearly polarized Proca stars, 
    \item[ii)] $\gamma=0$, $\alpha= 1$ for circularly polarized Proca stars,
    \item[iii)]  $\gamma=1$, $\alpha=0$ for radially polarized Proca stars.
\end{enumerate}

It is interesting to note that linearly and circularly polarized Proca stars are described by exactly the same equations as nonrelativistic boson stars, c.f. Eqs.~(32) in Ref.~\cite{Nambo:2024gvs} (see also~\cite{Chavanis:2011zi,Chavanis:2011zm} for previous studies of the equilibrium configurations of the $s=0$ Gross-Pitaevskii-Poisson system). As we clarified earlier, in absence of spin-spin selfinteraction, all constant polarization states are equivalent to each other; therefore, when $\lambda_s=0$, elliptically polarized Proca stars beyond the linear and circular cases also exist and are described by Eqs.~(\ref{s=1GPP.stationary}) with $\gamma=0$. Non-selfinteracting ($\lambda_n=\lambda_s=0$) radially polarized Proca stars, on the other hand, are described by the same equations as nonrelativistic, $\ell=1$ boson stars, c.f. Eqs.~(41) in Ref.~\cite{Roque:2023sjl}. We explain the reason for this coincidence in App.~\ref{app.radiallyVSl=1}. Finally, selfinteracting, radially polarized Proca stars satisfy the same system of equations as selfinteracting, $\ell=1$ boson stars; however, up to our knowledge, such solutions have not been reported in the literature. We review the equivalence between these different systems in Table~\ref{table.background}.

Equations~(\ref{s=1GPP.stationary}) must be complemented with appropriate boundary conditions that guarantee that the solutions remain regular at the origin and possess finite total energy. Near $r=0$, we can expand the solutions in power series of the form $\sigma^{(0)}(r)= \sigma_0r^{\alpha}+\sigma_1r^{\alpha+1}+\sigma_2r^{\alpha+2}+\ldots$, $\mathcal{U}(r)= \mathcal{U}_0r^{\beta}+\mathcal{U}_1 r^{\beta+1}+\mathcal{U}_2r^{\beta+2}+\ldots$, with $\sigma_0 \neq 0$, $\mathcal{U}_0\neq 0$, $\alpha\ge 0$ and $\beta\ge 0$ taking constant values.\footnote{There is no loss of generality in assuming that $\sigma_0$ and $\mathcal{U}_0$ are nonzero; if e.g. $\sigma_0$ were to vanish, we could simply redefine $\alpha\to\alpha+1$ to account for it. On the other hand, we need to impose that $\alpha$ and $\beta$ are non-negative to ensure that the solutions remain regular at the origin.} Introducing this ansatz into Eqs.~(\ref{s=1GPP.stationary}) yields four solutions, but only one of them meets the criteria $\alpha\ge 0$ and $\beta\ge 0$. When $\gamma=0$, the regular solution has  $(\alpha,\beta)=(0,0)$ and $(\sigma_1,\mathcal{U}_1)=(0,0)$, whereas for $\gamma=1$ it has $(\alpha,\beta) = (1,0)$ and $(\sigma_1,\mathcal{U}_1) = (0,0)$. This implies the following regular boundary conditions at the center: 
\begin{subequations}\label{Eq.BounCondr0}
\begin{align}
   \sigma^{(0)}(r=0)&=(1-\gamma)\sigma_{0},\quad \sigma^{(0)\prime}(r=0)=\gamma\sigma_0,\\
   \mathcal{U}(r=0)&=\mathcal{U}_0,\hspace{1.73cm} \mathcal{U}'(r=0)=0,
\end{align}
\end{subequations}
with $\sigma_0$ and $\mathcal{U}_0$ constants and where the primes refer to derivation with respect to $r$. Notice that, at the origin, linearly and circularly polarized states have a nonzero field value and a vanishing first derivative. In contrast, for radially polarized Proca stars, regularity implies that the wave function vanishes at $r=0$, whereas the first derivative at the origin does not vanish. Nonetheless, we will sometimes refer to $\sigma_0$ as the central ``amplitude'' of the configuration, although rigorously this is only true if the polarization is constant. Note also that Eqs.~(\ref{s=1GPP.stationary}) are invariant under simultaneous shifts of $E$ and $m_0\mathcal{U}(r)$, and we can use this symmetry to fix arbitrarily the value of $\mathcal{U}_0$.

At infinity, we impose $\lim\limits_{r\to\infty}\sigma^{(0)}(r)=0$, which is required for the solutions to have a finite total energy. This defines a nonlinear eigenvalue problem for the frequency $E$, where, for each central amplitude of the configuration, $\sigma_0$, there exists a discrete set of frequencies $E_n(\sigma_0)$, $n = 0, 1, 2, \ldots$, for which the boundary conditions are satisfied. We discuss this problem in more detail in Sec.~\ref{sec.numerical}, where we present our numerical results.

\subsection{Multi-frequency spherical solutions}\label{sec.sph.multi}

\begin{table*}
\caption{{\bf The multi-frequency and spherical $s=1$ Gross-Pitaevskii-Poisson system:} The multi-frequency, spherically symmetric $s=1$ Gross-Pitaevskii-Poisson equations as compared to other systems studied in the framework of
multi-scalar field theories.  Selfinteracting, multi-frequency Proca stars are not related with previously known solutions. SP (Schr\"odinger-Poisson), GPP (Gross-Pitaevskii-Poisson).}
\begin{center}
\begin{tabular}{ l l l l l l l}
 \hline
 \hline
 $\lambda_n=0$, $\lambda_s=0$   & $\quad$ & $\lambda_n\neq 0$, $\lambda_s=0$ &$\quad$& $\lambda_n=0$, $\lambda_s\neq 0$ &$\quad$& $\lambda_n\neq 0$, $\lambda_s\neq 0$\\
 \hline
 $\ell=0$ multi-state SP~\cite{Matos:2007zza}    && $\ell=0$ multi-state GPP (new) &&  $\ell=0$ multi-state GPP (new) && $\ell=0$ multi-state GPP (new)\\
 \hline
 \hline
\end{tabular}
\end{center}
\label{table.background2}
\end{table*}

Next, we explore multi-frequency states, which are only possible in absence of spin-spin selfinteractions ($\lambda_s=0$). 

Combining Eqs.~(\ref{eq.harmonic.multi}) and~(\ref{eq.radial.generic}), we find that the only possible states are the stationary radially polarized solutions that were already discussed in the previous section. On the other hand, combining Eqs.~(\ref{eq.harmonic.multi}) and~(\ref{eq.constant.multi.generic}), we find
\begin{equation}\label{eq.ansatz.multi.cartesian}
\vec{\psi}(t,\vec{x})= \sum_{\lambda=1}^3 e^{-iE_\lambda t}\sigma_\lambda^{(0)}(r)\hat{e}_\lambda,
\end{equation}
where $\sigma_\lambda^{(0)}$ are complex-valued functions depending only on $r$. As far as we know, these solutions have not been previously reported in the literature of Proca stars.

Introducing Eq.~(\ref{eq.ansatz.multi.cartesian}) into the $s=1$, $\lambda_s=0$ Gross-Pitaevskii-Poisson system~(\ref{eq.dimensionless.GPP.1}), we obtain: 
\begin{subequations}\label{eqs.multi.frequency}
\begin{eqnarray}
E_i\sigma_i^{(0)}&=&\left[-\frac{1}{2m_0}\Delta_s+\frac{\lambda_n}{2m_0^2}\sum_{j}|\sigma_{j}^{(0)}|^2 + m_0\mathcal{U}\right]\sigma^{(0)}_i,\qquad \label{SEc2.2.2.2}\\
\Delta_s\mathcal{U} &=&4\pi Gm_0\sum_{j}|\sigma^{(0)}_{j}|^2,\label{s=1GPP.stationary.3}
\end{eqnarray}
\end{subequations}
where, as in Eqs.~(\ref{s=1GPP.stationary.1}), $\Delta_s$ denotes the radial Laplace operator, and without loss of generality we have chosen the Cartesian basis $\hat{e}_\lambda =  \hat{e}_i$ with $i=x,y,z$.\footnote{ By applying a unitary transformation one can always map an arbitrary orthonormal basis $\hat{e}_\lambda$ to the standard Cartesian one $\hat{e}_i$, which changes the original multi-frequency state $\vec{\psi}(t,\vec{x})$ of Eq.~(\ref{eq.harmonic.multi}) to an equivalent one.} Furthermore, it is sufficient to consider real-valued functions $\sigma_i^{(0)}(r)$,\footnote{Indeed, one can take the real and imaginary parts of Eq.~(\ref{SEc2.2.2.2}) and conclude that $\mbox{Re}(\sigma_i^{(0)})$ and $\mbox{Im}(\sigma_i^{(0)})$ satisfy the same one-dimensional Schr\"odinger equation. As a consequence of the nodal theorem, the imaginary part of the wave function must be proportional to the real part, which means that the functions $\sigma_i^{(0)}$ are real, up to a global phase. After applying a constant unitary transformation one achieves that all $\sigma_i^{(0)}$'s are real.} and from now on we will stick to this assumption.

It is worth noting that non-selfinteracting ($\lambda_n=0$), multi-frequency Proca stars are described by exactly the same equations as nonrelativistic, multi-state boson stars with two or three occupied energy levels in which the angular momentum number $\ell$ vanishes, c.f. Eqs.~(9) in Ref.~\cite{Matos:2007zza}. To the best of our knowledge, the generalization of this system for $\lambda_n\neq 0$ has not been reported in the literature, with the exception of the relativistic scenario discussed in Ref.~\cite{Li:2020ffy}. We review the different possible scenarios in Table~\ref{table.background2}.

To proceed, Eqs.~(\ref{eqs.multi.frequency}) must be accompanied by some appropriate boundary conditions. Near $r=0$, we demand that the solutions remain regular. For this, we again expand the functions $\sigma_i^{(0)}(r)$ and $\mathcal{U}(r)$ in power series $\sigma^{(0)}_i(r)= \sigma_{i0}r^{\alpha_i}+\sigma_{i1}r^{\alpha_i+1}+\sigma_{i2}r^{\alpha_i+2}+\ldots$, $\mathcal{U}(r)= \mathcal{U}_{0}r^{\beta_i}+\mathcal{U}_{1} r^{\beta_i}+\mathcal{U}_{2}r^{\beta_i+2}+\ldots$, and impose $\sigma_{i0} \neq 0$, $\mathcal{U}_{0}\neq 0$, $\alpha_i\ge 0$ and $\beta_i\ge 0$. Introducing this expansion into Eqs.~(\ref{eqs.multi.frequency}) results in eight solutions, although only the one for which $(\alpha_i,\beta)=(0,0)$ and $(\sigma_{i1},\mathcal{U}_{1})=(0,0)$  is regular. This leads to the following boundary conditions at $r=0$:
\begin{subequations}\label{Eq.BounCond.multy}
\begin{align}
\sigma_i^{(0)}(r=0)&=\sigma_{i0},\quad \sigma_i^{(0)\prime}(r=0)=0,\\
\mathcal{U}(r=0)&=\mathcal{U}_{0},\hspace{0.75cm} \mathcal{U}'(r=0)=0.
\end{align}
\end{subequations}
As for the stationary states, the system~(\ref{eqs.multi.frequency}) is invariant under common constant shifts in $E_i$ and $m_0\mathcal{U}(r)$, and we can choose arbitrarily the value of $\mathcal{U}_0$. 

Finally, to guarantee finite total energy, we impose $\lim\limits_{r\to\infty}\sigma^{(0)}_i(r)=0$. This defines a nonlinear multi-eigenvalue problem for the frequencies $E_i$, where, for each combination of central amplitudes $(\sigma_{x0},\sigma_{y0}, \sigma_{z0})$, there exists a discrete set of frequencies $E_{x n_x}(\sigma_{x0},\sigma_{y0},\sigma_{z0})$, $E_{y n_y}(\sigma_{x0},\sigma_{y0}, \sigma_{z0})$, and $E_{z n_z}(\sigma_{x0},\sigma_{y0}, \sigma_{z0})$, $n_x,n_y,n_z=0,1,2,\ldots$, satisfying the boundary conditions. 

For the interpretation of our results, the following observation will be important. According to the nodal (or Sturm oscillation) theorem (see~\cite{Simon2005} for a pedagogical review), the eigenfunction $\psi_n(r)$ corresponding to the $n$th energy level $E_n$ of a one-dimensional Schr\"odinger operator has exactly $n$ nodes. Once the nonlinear system~(\ref{eqs.multi.frequency}) is solved, Eq.~(\ref{SEc2.2.2.2}) can be interpreted as a Schr\"odinger equation for the wave functions $\sigma^{(0)}_i(r)$ with a fixed effective potential 
\begin{equation}
\frac{\lambda_n}{2m_0^2}\sum_{j}|\sigma_{j}^{(0)}|^2
+m_0\mathcal{U}.
\end{equation}
Therefore, the frequencies $E_{i n_i}$ can be ordered according to the node number of the functions $\sigma^{(0)}_i(r)$. Consequently, two functions $\sigma^{(0)}_i(r)$ and $\sigma^{(0)}_j(r)$ with $i\neq j$ are proportional to each other if they coincide in their node numbers, whereas they are mutually orthogonal if they have different numbers of nodes. In particular, this implies that a solution whose wave functions $\sigma^{(0)}_i(r)$, $i=x,y,z$, have equal number of nodes are proportional to each other and satisfy $E_x = E_y = E_z$; therefore it yields a stationary solution. Thus, for fixed central amplitudes $(\sigma_{x0},\sigma_{y0}, \sigma_{z0})$, multi-frequency solutions can be labeled by their node numbers $(n_x,n_y,n_z)$ with $n_x\leq n_y\leq n_z$ and not all $n_i$'s equal to each other.

In the next section, we numerically solve the nonlinear eigenvalue problems belonging to spherically symmetric stationary and multi-frequency states.

\begin{figure*}
\centering
\includegraphics[width=18.cm]{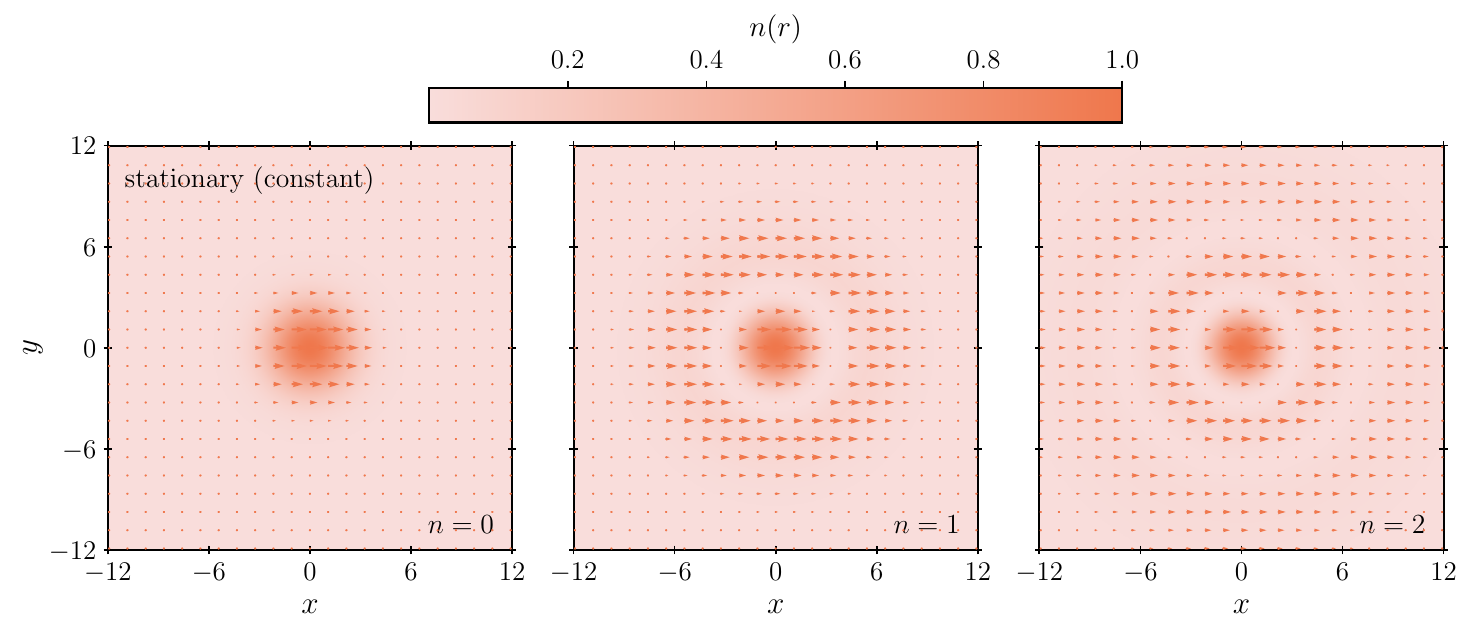}
\caption{{\bf Stationary Proca stars of constant polarization:} The normalized real part of the vector field, $\vec{\psi}_R(t,\vec{x})$, and the normalized particle number density, $n(t,\vec{x})$, for three Proca stars of constant polarization, unit central ``amplitude'', $\sigma_0=1$, and repulsive selfinteraction, $\lambda^{phys}_n+\alpha\lambda^{phys}_s>0$, at time $t=0$. {\it Left panel:} No nodes, $n=0$. {\it Center panel:} One node, $n=1$. {\it Right panel:} Two nodes, $n=2$. Note the appearance of an additional ``layer'' in the configuration for each increment of the variable $n$. Code variables use the scale $\lambda_*=|\lambda_n^{phys}+\alpha\lambda_s^{phys}|$.} \label{Fig.PolLinCirc}
\end{figure*}

\begin{figure*}
\centering
\includegraphics[width=18.cm]{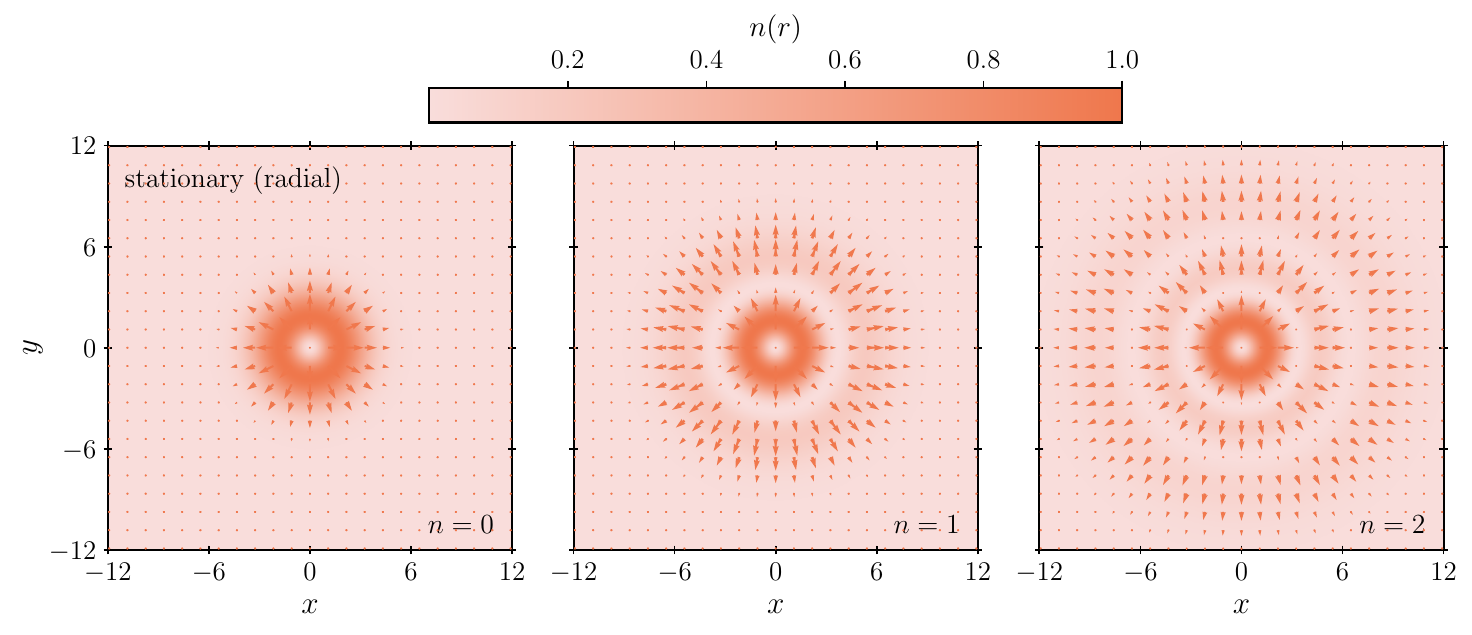}
\caption{{\bf Stationary Proca stars of radial polarization:} Same as in Fig.~\ref{Fig.PolLinCirc} but for the case of radially polarized Proca stars. The main difference with respect to the linear and circular cases, apart from the fact that the vector points radially and does not pick a preferred direction, is the presence of a ``hole'' in the center of the configuration. This is a consequence of the regularity conditions at the origin.} \label{Fig.PolRadial}
\end{figure*}

\begin{figure*}
\centering
\includegraphics[width=18.cm]{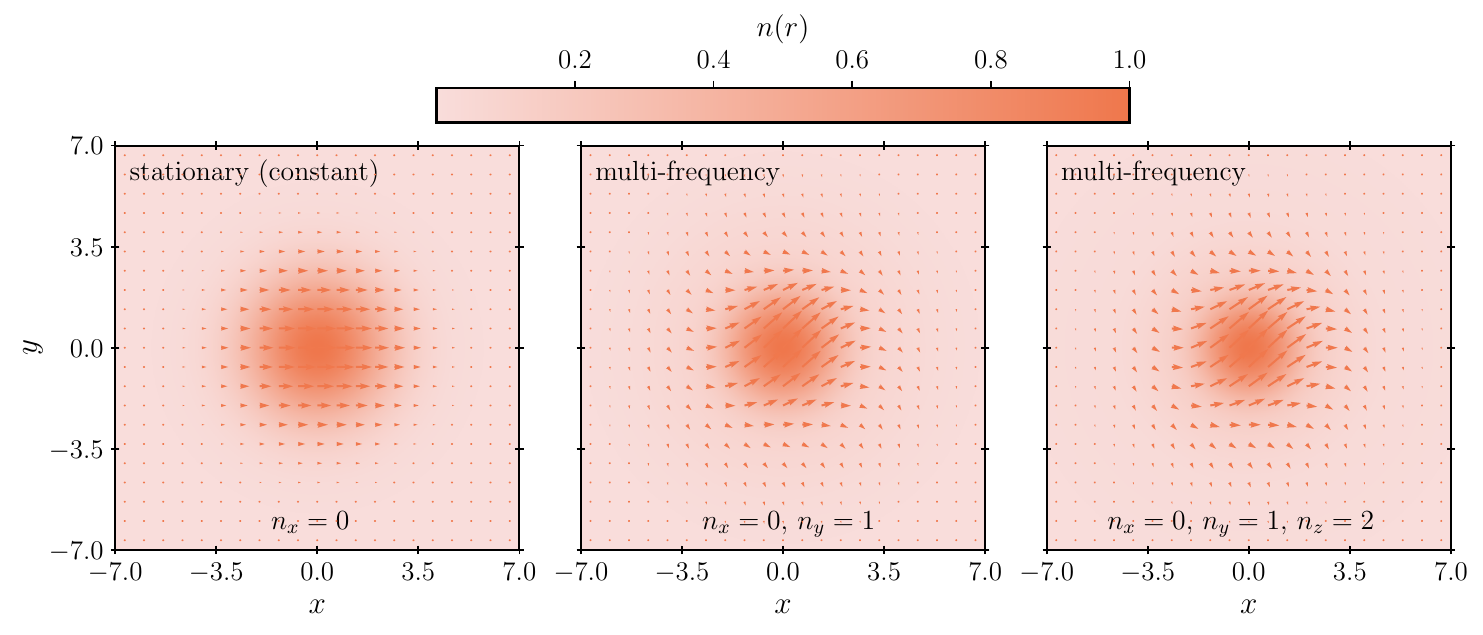}
\caption{{\bf Stationary and multi-frequency Proca stars:} Similar as in Figs.~\ref{Fig.PolLinCirc} and~\ref{Fig.PolRadial}, but comparing a stationary Proca star with multi-frequency ones. {\it Left panel:} $(\sigma_{x0},\sigma_{y0},\sigma_{z0})=(1,0,0)$ and $(n_x,n_y,n_z)=(0,0,0)$. {\it Center panel:} $(\sigma_{x0},\sigma_{y0},\sigma_{z0})=(1,1,0)$ and $(n_x,n_y,n_z)=(0,1,0)$. {\it Right panel:} $(\sigma_{x0},\sigma_{y0},\sigma_{z0})=(1,1,1)$ and $(n_x,n_y,n_z)=(0,1,2)$. Although the stationary constantly polarized and multi-frequency solutions are spherically symmetric according to the representation $\vec{\psi}(t,\vec{x})\mapsto \vec{\psi}(t, R^{-1}\vec{x})$ of the $SO(3)$ group, this symmetry is not manifest given the pattern formed by the vector field.} \label{fig.multiFreqProf}
\end{figure*}

\begin{figure*}
	\centering 
\includegraphics[width=18.cm]{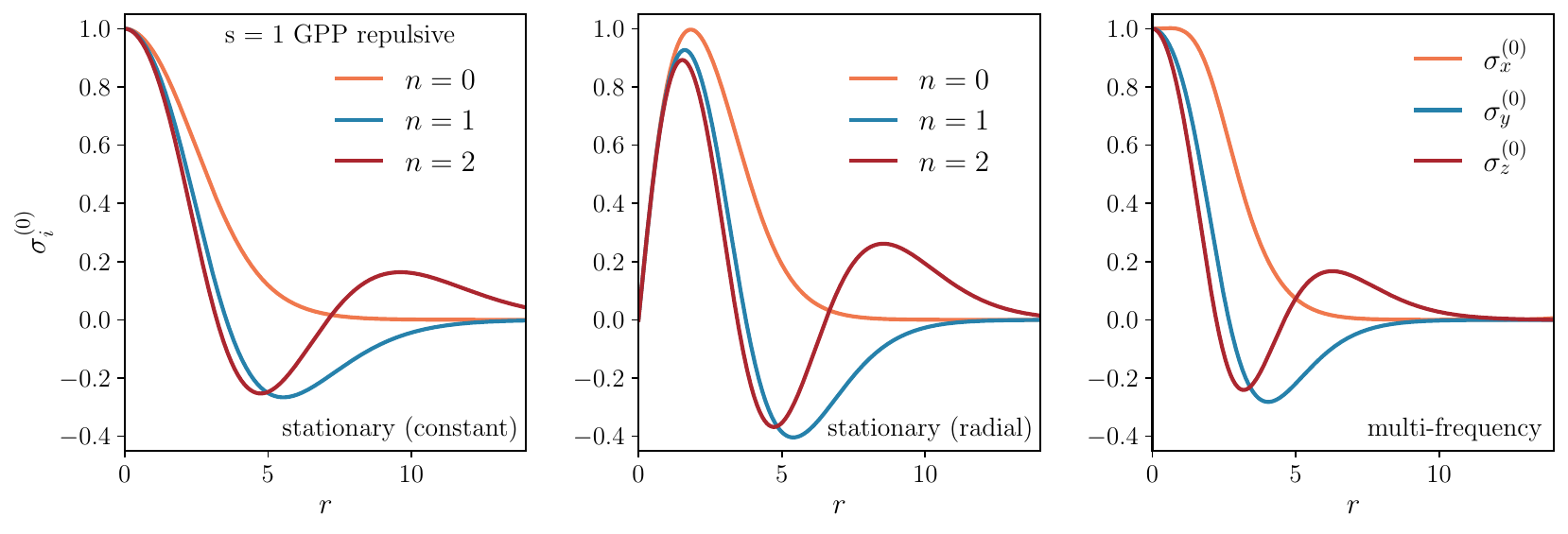}
\caption{{\bf Radial profiles of some representative stationary and multi-frequency configurations:} Radial profiles $\sigma_i^{(0)}(r)$ of some configurations reported in Figs.~\ref{Fig.PolLinCirc},~\ref{Fig.PolRadial} and~\ref{fig.multiFreqProf}. {\it Left panel:} The profiles $\sigma^{(0)}(r)$ of the three linear/circular configurations of Fig.~\ref{Fig.PolLinCirc}. {\it Center panel:} The profiles $\sigma^{(0)}(r)$ of the three radial configurations of Fig.~\ref{Fig.PolRadial}. {\it Right panel:} The profile $\sigma_{x}^{(0)}(r)$, $\sigma_{y}^{(0)}(r)$, and $\sigma_{z}^{(0)}(r)$ of the multi-frequency configuration in the right panel of Fig.~\ref{fig.multiFreqProf}.} \label{fig.radial.prof}
\end{figure*}

\section{Numerical results}\label{sec.numerical}

In this section, we provide numerical solutions of the $s=1$ Gross-Pitaevskii-Poisson system and discuss their properties. To proceed, we introduce the dimensionless quantities:\footnote{ In this section, $t$, $\vec{x}$, $\mathcal{U}$, $\dots$ denote dimensionless variables. Whenever needed, we will label dimensionfull quantities by the superscript {\it phys}.}

\begin{subequations}\label{eq.code.numbers1}
\begin{eqnarray}
{\displaystyle t:=\frac{4\pi G m_{0}^3 }{\lambda_* }t^{phys},} &\quad {\displaystyle  \vec{x} := \frac{\sqrt{8\pi G} m_{0}^2}{\lambda_*^{1/2}} \vec{x}^{phys},} \\
{\displaystyle \mathcal{U} := \frac{\lambda_* }{4\pi G m_0^2}\mathcal{U}^{phys},} &\quad {\displaystyle \vec{\psi}:= \frac{\lambda_* }{\sqrt{8\pi G}m_0^{5/2}}\vec{\psi}^{phys},} \\
{\displaystyle \lambda_n := \frac{\lambda_n^{phys}}{\lambda_*},} 
&\quad {\displaystyle  \lambda_s := \frac{\lambda_s^{phys}}{\lambda_*}},
\end{eqnarray}
\end{subequations}
where $\lambda_*>0$ is a characteristic selfinteraction scale that we can choose at our convenience. Note that the dimensions of $E$ are the inverse of those of $t$.

In terms of the dimensionless variables, Eqs.~(\ref{s=1GPP.stationary}) and~(\ref{eqs.multi.frequency}) can be conveniently combined into a single system of the form
\begin{subequations}\label{eqs.numerical}
    \begin{align}
\Delta_s \sigma_i^{(0)}&= \left(\frac{2\gamma}{r^2}\pm \sum_{j}\sigma_j^{(0)2}-u_i^{(0)}\right) \sigma_i^{(0)},\label{SEc2.2.2.1}\\
\Delta_s u_i^{(0)}&= - \sum_{j}\sigma_j^{(0)2},
    \end{align}
\end{subequations}
where Latin indices and summations range from $1$ to $3$ for multi-frequency states and are omitted for stationary states. Here, $\gamma=0$ for multi-frequency states and stationary states with constant polarization, whereas $\gamma=1$ for radially polarized stationary states. To simplify the numerical implementation, we have also introduced $u_i^{(0)}(r):=E_i - \mathcal{U}(r)$ as the difference between the frequencies $E_i$ and the gravitational potential $\mathcal{U}(r)$. (Although the $u_i^{(0)}$ differ only by a constant number, for multi-frequency states we still find it convenient to solve the three Poisson equations for the shooting algorithm described below.) In addition, we have fixed the characteristic selfinteraction scale of Eqs.~(\ref{eq.code.numbers1}) to $\lambda_*=|\lambda_n^{phys}+\alpha\lambda_s^{phys}|$,\footnote{In absence of selfinteractions, when the polarization vector is linear or radial and $\lambda_n^{phys}=0$, or when it is circular and $\lambda_n^{phys}=-\lambda_s^{phys}$, the second term on the right-hand side of Eqs.~(\ref{s=1GPP.stationary.1}) and~(\ref{SEc2.2.2.2}) vanishes. In these cases, the second term on the right-hand-side of Eq.~(\ref{SEc2.2.2.1}) should be discarded and the scale $\lambda_*$ is arbitrary.} where $\alpha$ was defined in Sec.~\ref{sec.sph.stationary}. The $\pm$ signs in Eq.~(\ref{SEc2.2.2.1}) make reference to the ``repulsive'', $\lambda^{phys}_n+\alpha\lambda^{phys}_s>0$, and the ``attractive'', $\lambda^{phys}_n+\alpha\lambda^{phys}_s<0$, cases, respectively. These equations must be complemented with the following boundary conditions at $r=0$; cf.~Eqs.~(\ref{Eq.BounCondr0}) and~(\ref{Eq.BounCond.multy}): 
\begin{subequations}\label{Eq.BounCondr0.2}
\begin{align}
   \sigma^{(0)}_i(r=0)&=(1-\gamma)\sigma_{i0},\quad \sigma^{(0)\prime}(r=0)=\gamma\sigma_{i0},\\
   u_i^{(0)}(r=0)&=u_{i0},\hspace{1.43cm} u_i^{(0)\prime}(r=0)=0.
\end{align}
\end{subequations}

To find the appropriate values of $\sigma_{i0}$ and $u_{i0}$, we use a methodology similar to the one described in Ref.~\cite{Roque:2023sjl}, where, given $\sigma_{i0}$, the possible values for $u_{i0}$ are fine-tuned using a numerical shooting method based on the conditions $\lim\limits_{r\to\infty}\sigma_i^{(0)}(r)=0$, which are required for the solutions to have finite total energy. This results in a discrete family of solutions $\sigma_i^{(0)}(\sigma_{i0},n_i;r)$, where $n_i=0,1,2,\ldots$ label the number of nodes of the functions $\sigma_i^{(0)}(r)$ in the interval $0<r<\infty$. For the numerical integration, we use an adaptive explicit 5(4)-order Runge-Kutta routine~\cite{2020SciPy-NMeth, DORMAND198019, Lawrence1986SomePR}, which requires rewriting the equations as a first-order system for the fields $(\sigma_i^{(0)}, u_i^{(0)})$, and for the shooting method we employ a technique based on bisection. Some additional information about the numerical implementation is presented in App.~\ref{app:shooting}.

In Figs.~\ref{Fig.PolLinCirc} and~\ref{Fig.PolRadial} we plot some representative solutions of the stationary and spherically symmetric $s=1$ Gross-Pitaevskii-Poisson system for $\sigma_0=1$, $n=0$, $1$ and $2$, and different polarizations $\hat{\epsilon}$ at time $t=0$. In addition, Fig.~\ref{fig.multiFreqProf} presents a comparison between the constantly polarized stationary configuration defined by the parameters $\sigma_0 = 1$ and $n=0$, and two prototypical multi-frequency states, where for concreteness we have chosen $(\sigma_{x0},\sigma_{y0},\sigma_{z0})=(1,1,0)$, $(n_x,n_y,n_z)=(0,1,0)$, and $(\sigma_{x0},\sigma_{y0},\sigma_{z0})=(1,1,1)$, $(n_x,n_y,n_z)=(0,1,2)$, all of them evaluated at $t=0$. For completeness, in Fig.~\ref{fig.radial.prof}, we include the radial profiles $\sigma_i^{(0)}(r)$ associated with some of these configurations. It is important to stress that in all these figures we have focused on the repulsive case.

In the remainder of this section we center our attention on key physical quantities that characterize Proca stars, such as their mass and radius in Sec.~\ref{sec.mass.radius}, and their orbital and spin angular momentum in Sec.~\ref{sec.angular.momentum}. In Sec.~\ref{sec.global.charges}, we construct the global charges associated with the accidental symmetry that appears when spin-spin selfinteractions are absent. Finally, in Sec.~\ref{sec.energy.functional}, we focus on the behavior of the energy functional, which is expected to play a relevant role for the stability of the configurations.

\subsection{Mass and radius}\label{sec.mass.radius}

\begin{figure*}
\centering
\includegraphics[width=18.cm]{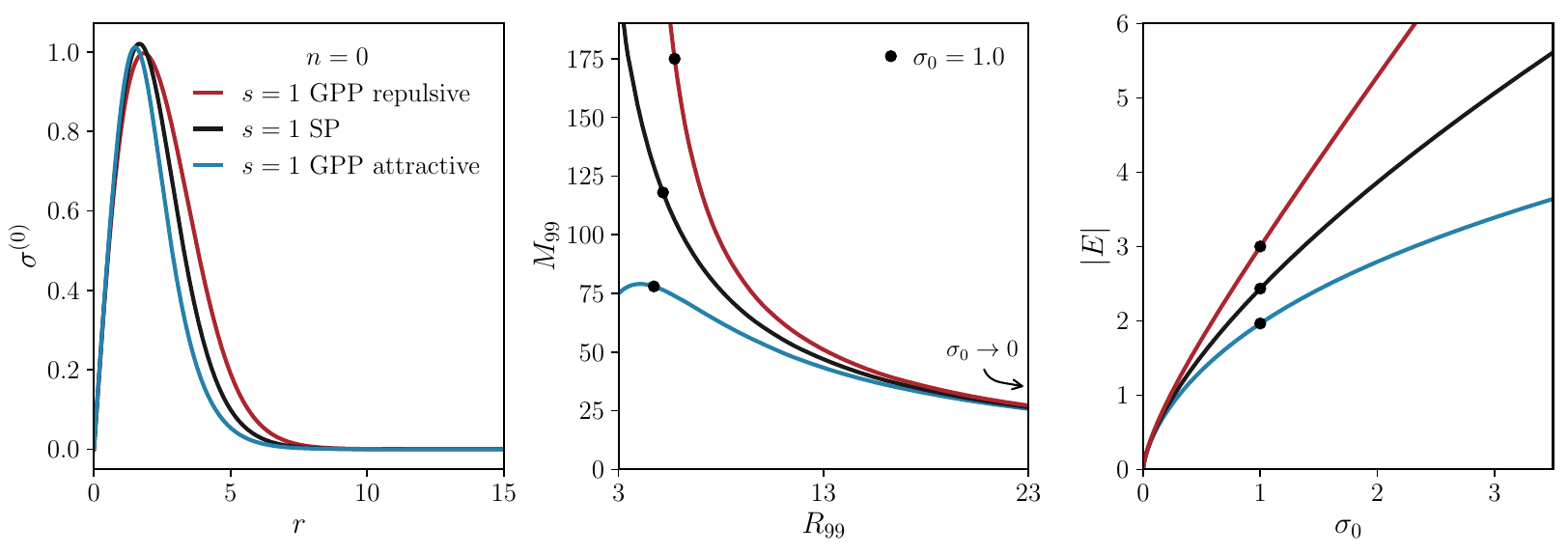}
\caption{{\bf Radially polarized Proca stars with no nodes:} Stationary and spherically symmetric solutions of the $s=1$ Gross-Pitaevskii-Poisson system with no nodes ($n=0$) and radial polarization. Red (blue) lines correspond to the repulsive (attractive) case, and we have included the solutions to the $s=1$ Schr\"odinger-Poisson system  (black lines) for reference. {\it Left panel:} The profile of $\sigma^{(0)}(r)$ for $\sigma_0=1$. {\it Center panel:} The effective mass of the configurations $M_{99}$ as a function of the effective radius $R_{99}$. {\it Right panel:} The magnitude of the energy eigenvalue $|E|$ as a function of the central amplitude $\sigma_0$. The dots in the last two panels correspond to the configurations of unit amplitude. For $\sigma_0\to 0$ the effects of the selfinteractions become negligible and we recover non-selfinteracting radially polarized Proca star configurations, which are equivalent to $\ell=1$ boson star.} \label{FigFondn0}
\end{figure*}

\begin{figure*}
\centering
\includegraphics[width=18.cm]{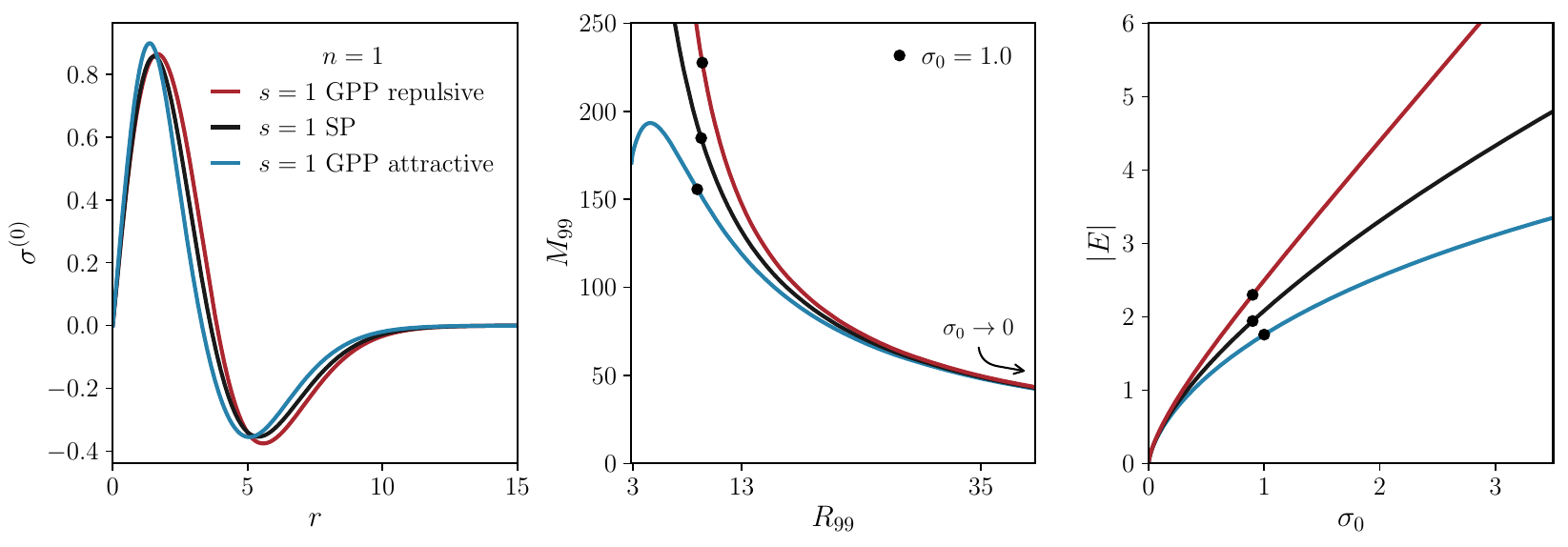}
\caption{{\bf Radially polarized Proca stars with one node:} Same as in Fig.~\ref{FigFondn0} but for the stationary and spherically symmetric solutions of the $s=1$ Gross-Pitaevskii-Poisson system with one node ($n=1$).}\label{FigFondn1}
\end{figure*}

\begin{figure*}
	\centering
\includegraphics[width=18cm]{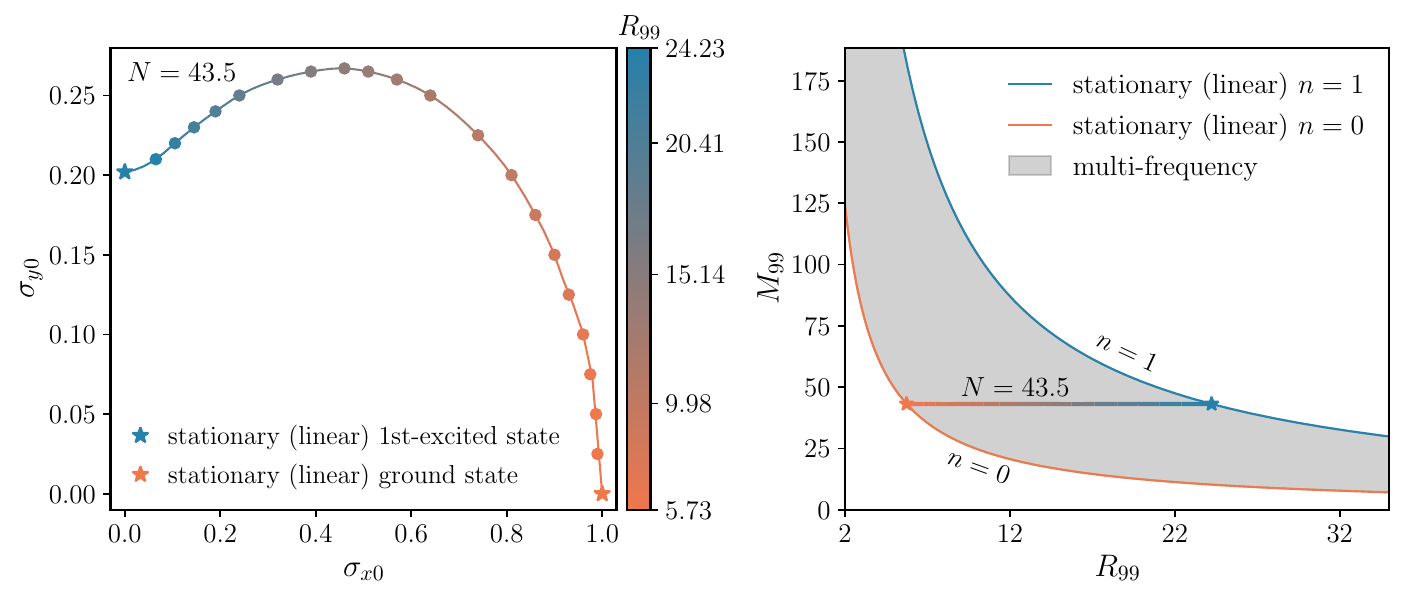}
\caption{{\bf Phase diagrams for multi-frequency states:} Multi-frequency Proca stars with amplitude $(\sigma_{x0},\sigma_{y0},0)$ and node numbers $(n_x,n_y,n_z)=(0,1,0)$ in the free theory ($\lambda_n=\lambda_s=0$). {\it Left panel:} The  central amplitudes $\sigma_{x0}$ and $\sigma_{y0}$ that are consistent with $N=43.5$ number of particles. {\it Right panel:} The $M_{99}$ vs. $R_{99}$ plot. Contrary to stationary states (described by the border lines), multi-frequency states correspond to a region instead of a curve.} \label{Fig.MassRelation}
\end{figure*}

The mass of a Proca star can be computed as the product of $m_0$ with the particle number defined in Eq.~(\ref{eq.number.particles}), which yields $M^{phys}=m_0N^{phys}$. Here $N^{phys}=[1/(\sqrt{8\pi G\lambda_*} m_0)]N$, where $N$ represents the number of particles in the dimensionless variables of  Eq.~(\ref{eq.code.numbers1}), and is given by
\begin{equation}\label{eq.N.code}
N=  4\pi \sum_i \int_0^\infty \sigma^{(0)2}_i r^2 dr.
\end{equation}
To simplify the presentation, we have chosen the same conventions as those defined just below Eqs.~(\ref{eqs.numerical}).

Formally, the size of a Proca star extends to infinity, and for that reason it is usual to define an effective radius $R_{99}$ as the one containing $99\%$ of the total mass of the configuration, $M_{99}$, which in physical units is given by $R_{99}^{phys}=[\sqrt{\lambda_*} /(\sqrt{8\pi G}m_0^2)]R_{99}$.

First, we concentrate on stationary solutions. In Figs.~\ref{FigFondn0} and~\ref{FigFondn1}, we show the profile $\sigma^{(0)}(r)$ for $\sigma_0=1.0$, the $M_{99}$ vs. $R_{99}$ plot, and the behavior of the energy eigenvalue $E$ as a function of the central amplitude $\sigma_0$, for radially polarized Proca stars of zero and one nodes, respectively, in the attractive, repulsive, and free theories. The corresponding figures for linearly and circularly polarized Proca stars, which coincides with those of nonrelativistic boson stars, can be found in Figs.~1 and~2 of Ref.~\cite{Nambo:2024gvs} (see Sec.~IVA in that paper for a discussion). In the limit $\sigma_0\to 0$, the attractive and repulsive branches of the radially polarized Proca stars converge to those of the free theory (see the central and right panels of Figs.~\ref{FigFondn0} and~\ref{FigFondn1}). This occurs because, at low densities, short range selfinteractions (which are cubic in the fields) become negligible and the Gross-Pitaevskii equation approaches the Schr\"odinger one. The same property has been observed for self-interacting boson stars in Ref.~\cite{Nambo:2024gvs} and thus it also holds for linearly and circularly polarized Proca stars, and for all constant polarization states when $\lambda_s=0$. In all cases, the numerical data suggest that the mass of a stationary Proca star increases without bound as the radius of the object decreases, except when an attractive selfinteraction is present. In this situation, the objects reach a state whose mass cannot be exceeded.

For multi-frequency states, the situation becomes more involved, as the configurations are labeled by the three independent amplitudes $(\sigma_{x0}, \sigma_{y0}, \sigma_{z0})$, in addition to the corresponding number of nodes $(n_x, n_y, n_z)$. As a result, we anticipate that each curve in the $M_{99}$ vs. $R_{99}$ plot belonging to stationary states of fixed $n$  transforms into a region when considering multi-frequency states of fixed $(n_x,n_y,n_z)$.

To analyze this, we focus on the free theory ($\lambda_n=\lambda_s = 0$) for simplicity. In this case, the $s=1$ Gross-Pitaevskii-Poisson system is invariant under the scaling transformation
\begin{equation}\label{eq.scaling.free}
t\mapsto\lambda_*^{-1}t, \quad 
\vec{x}\mapsto\lambda_*^{-1/2}\vec{x}, \quad 
\mathcal{U}\mapsto\lambda_* \mathcal{U},\quad \vec{\psi} \mapsto \lambda_*\vec{\psi},
\end{equation}
where $\lambda_*$ is an arbitrary nonvanishing constant. This invariance is associated with the arbitrary choice of the scale $\lambda_*$ in Eqs.~(\ref{eq.code.numbers1}), which does not affect the eigenvalue problem~(\ref{eqs.numerical}) when both $\lambda_n$ and $\lambda_s$ vanish. This simplifies the analysis of the free theory, since, for any node numbers $(n_x,n_y,n_z)$,  configurations $(\sigma_{x0},\sigma_{y0},\sigma_{z0})$ that maintain the same ratio between the different $\sigma_{i0}$ are related to each other by a rescaling transformation.

In particular, we concentrate on the case where $(n_x,n_y,n_z)=(0,1,0)$, which describes multi-frequency states with two components, $(\sigma_{x0}, \sigma_{y0}, 0)$. In the left panel of Fig.~\ref{Fig.MassRelation}, we show the family of states corresponding to a fixed particle number $N=43.5$, i.e. $M_{99}=43.1$, leading to configurations with sizes between $5.7 \le R_{99}\leq 24.2$. Here, the number $N=43.5$ is the one obtained from the configuration with parameters $\sigma_{x0}=1$ and $\sigma_{y0}=0$, which represents a stationary state of linear polarization. As we just mentioned, this family can be rescaled to any value of $N$ using the transformation~(\ref{eq.scaling.free}), where every element in the family transforms according to  $M_{99}\sim 1/R_{99}$. We show the $M_{99}$ vs. $R_{99}$ plot for multi-frequency states of $(n_x,n_y,n_z)=(0,1,0)$ in the right panel of Fig.~\ref{Fig.MassRelation}. Note that the left border line of this diagram corresponds to the $M_{99}$ vs. $R_{99}$ plot of the $n=0$ stationary states of linear polarization, whereas the right border to the their first excited states $n=1$ (these border lines correspond to the black curves in the central panels of Figs.~1 and~2 in Ref.~\cite{Nambo:2024gvs}). As we anticipated, multi-frequency states fill whole regions in the $M_{99}$ vs. $R_{99}$ diagram.

\subsection{Angular momentum}\label{sec.angular.momentum}

All states constructed in this paper are spherically symmetric, which inherently results in configurations of vanishing angular momentum, $\vec{L}^{phys} = 0$. This can be directly verified by substituting Eqs.~(\ref{eq.radial.generic}) and~(\ref{eq.constant.multi.generic}) into the general expression for the orbital angular momentum~(\ref{eq.orbitalL}). On the other hand,  whereas the spin angular momentum vanishes for multi-frequency states, for stationary states Eq.~(\ref{eq.spin.density}) yields $\vec{S}^{phys}=[1/(\sqrt{8\pi G \lambda_*}m_0)]\vec{S}$, where\footnote{ In the absence of spin-spin selfinteraction, elliptical polarization states~(\ref{eq.pol.elliptical}) have $\vec{S}=N\sin(2\phi)\sin\gamma_1\hat{e}_z$.}
\begin{equation}
\vec{S} = \alpha N \hat{e}_z,
\end{equation}
which is non zero for states of circular polarization. From a quantum perspective, the macroscopic spin angular momentum of circularly polarized Proca stars originates from the intrinsic  microscopic spin of the individual particles that conform the configuration. 

\subsection{Global charges}\label{sec.global.charges}

\begin{figure*}
\centering
\includegraphics[width=18.cm]{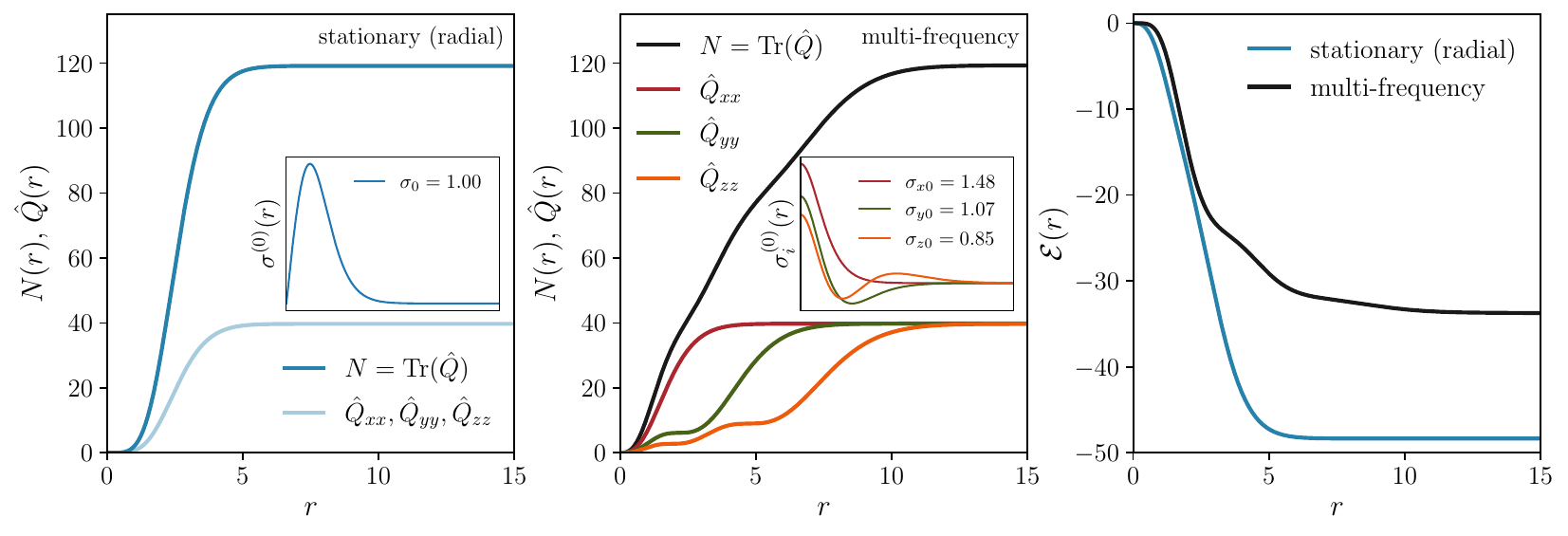}
\caption{{\bf Radial profiles of $\hat{Q}$, $N$, and $\mathcal{E}$:} {\it Left panel:} The radial profile of $\hat{Q}_{xx}$, $\hat{Q}_{yy}$, $\hat{Q}_{zz}$, and $N=\text{Tr}\, \hat{Q}$ for a stationary radially polarized Proca star of $\hat{Q}_{xx}=\hat{Q}_{yy}=\hat{Q}_{zz}=40$ and $N=120$. {\it Center panel:} Same as in the left panel but for a multi-frequency state with the same values of $\hat{Q}_{xx}$, $\hat{Q}_{yy}$, $\hat{Q}_{zz}$ and $N$. In both panels, the insets correspond to the radial profiles of the components of $\vec{\sigma}^{(0)}(\vec{x})$ (remember that for a Proca star of radial polarization $\sigma_x^{(0)}=\sigma^{(0)}(r)\sin\vartheta\cos\varphi$, $\sigma_y^{(0)}=\sigma^{(0)}(r)\sin\vartheta\sin\varphi$, and $\sigma_z^{(0)}=\sigma^{(0)}(r)\cos\vartheta$). {\it Right panel:} The radial profile of $\mathcal{E}$ for both configurations. Note that the radial Proca star has a lower energy than the multi-frequency configuration. In all cases we have assumed $\lambda_n=\lambda_s=0$. Given the rescaling symmetry of the free theory we can extend these conclusions to arbitrary values of $N$.} \label{Fig.EnergyProf}
\end{figure*}

If the spin-spin selfinteraction term vanishes, one can construct the charges associated with the accidental symmetry, which are also conserved in the time evolution. In physical units, Eq.~(\ref{eq.globalQ}) can be expressed in the form $\hat{Q}^{phys}=[1/(\sqrt{8\pi G\lambda_*} m_0)]\hat{Q}$, with
\begin{equation}
\hat{Q}_{ij} = 4\pi \int_0^\infty \sigma_i^{(0)*}\sigma_j^{(0)}r^2 dr,
\end{equation}
where $\sigma_i^{(0)}$ are the Cartesian components of $\vec{\sigma}^{(0)}(\vec{x})$.

Given that we do not distinguish between unitarily equivalent configurations, we can limit our study to states for which $\hat{Q}$ is diagonal, $\hat{Q}=\textrm{diag}(\hat{Q}_{xx},\hat{Q}_{yy},\hat{Q}_{zz})$. Specifically, stationary, linearly polarized Proca stars have $\hat{Q}=N\textrm{diag}(1,0,0)$, while stationary, radially polarized Proca stars have $\hat{Q}=\frac{N}{3}\textrm{diag}(1,1,1)$ (remember that in the symmetry-enhanced sector of the effective theory linearly and circularly polarized states are degenerated). In contrast, multi-frequency configurations allow the diagonal components of $\hat{Q}$ to be arbitrary, with the particle number given by $\textrm{Tr}(\hat{Q})=N$.

As an illustration, we compute the global charges of the configurations that we have constructed in Figs.~\ref{Fig.PolLinCirc},~\ref{Fig.PolRadial}, and~\ref{fig.multiFreqProf}. To do that, we assume $\lambda_s=0$. In particular, for the constant polarization states of Fig.~\ref{Fig.PolLinCirc} we obtain $\hat{Q}=\textrm{diag}(85, 0, 0)$, $\hat{Q}=\textrm{diag}(134, 0, 0)$, and $\hat{Q}=\textrm{diag}(190, 0, 0)$, where we have rounded the numbers to the closest integer. The radially polarized states of Fig.~\ref{Fig.PolRadial} yield $\hat{Q}=59\,\textrm{diag}(1, 1, 1)$, $\hat{Q}=81\,\textrm{diag}(1,1,1)$, and $\hat{Q}=104\,\textrm{diag}(1,1,1)$, whereas the stationary and multi-frequency states of Fig.~\ref{fig.multiFreqProf} have $\hat{Q}=\textrm{diag}(80, 0, 0)$, $\hat{Q}=\textrm{diag}(79, 18, 0)$, and $\hat{Q}=\textrm{diag}(79, 18, 7)$.

In the first two panels of Fig.~\ref{Fig.EnergyProf} we present, for the free theory ($\lambda_n=\lambda_s=0$), the profiles $\hat{Q}_{ij}(r)=4\pi\int_0^r \sigma_i^{(0)*}\sigma_j^{(0)}r'^2dr'$ for two configurations of charge $\hat{Q}=\textrm{diag}(40,40,40)$, i.e. $N=120$, although using the rescaling of Eq.~(\ref{eq.scaling.free}) we can extend these results to an arbitrary $N$: a stationary radially polarized Proca star with $\sigma_{0}=1.0$ and $n=0$, and a multi-frequency star with $(\sigma_{x0},\sigma_{y0},\sigma_{z0})=(1.48,1.07,0.85)$ and $(n_x,n_y,n_z)=(0,1,2)$. Even though the charge is distributed differently in these objects (the multi-frequency Proca star being more extended than the radially polarized one), their total charges coincide. According to Eqs.~(\ref{Eq:EnergFunctStat}) and~(\ref{Eq:EnerRelat}), when $\lambda_n=\lambda_s=0$, the total energy is $\mathcal{E}[\vec{\psi}]=-T[\vec{\psi}]=-\frac{1}{2}D[n,n]$, which shows that for fixed $N$, the more extended objects have smaller values of $D[n,n]$, resulting in higher energies. This suggests that, for fixed $\hat{Q}$ proportional to the identity matrix, $(n_x,n_y,n_z)=(0,1,2)$ multi-frequency Proca stars are more energetic than $n=0$ radially polarized configurations. The plots in the right panel of Fig.~\ref{Fig.EnergyProf} and the findings in Sec.~\ref{sec.energy.functional} confirm this expectation.

\subsection{Energy functional} \label{sec.energy.functional}

\begin{figure*}
	\centering
    \includegraphics[width=18.cm]{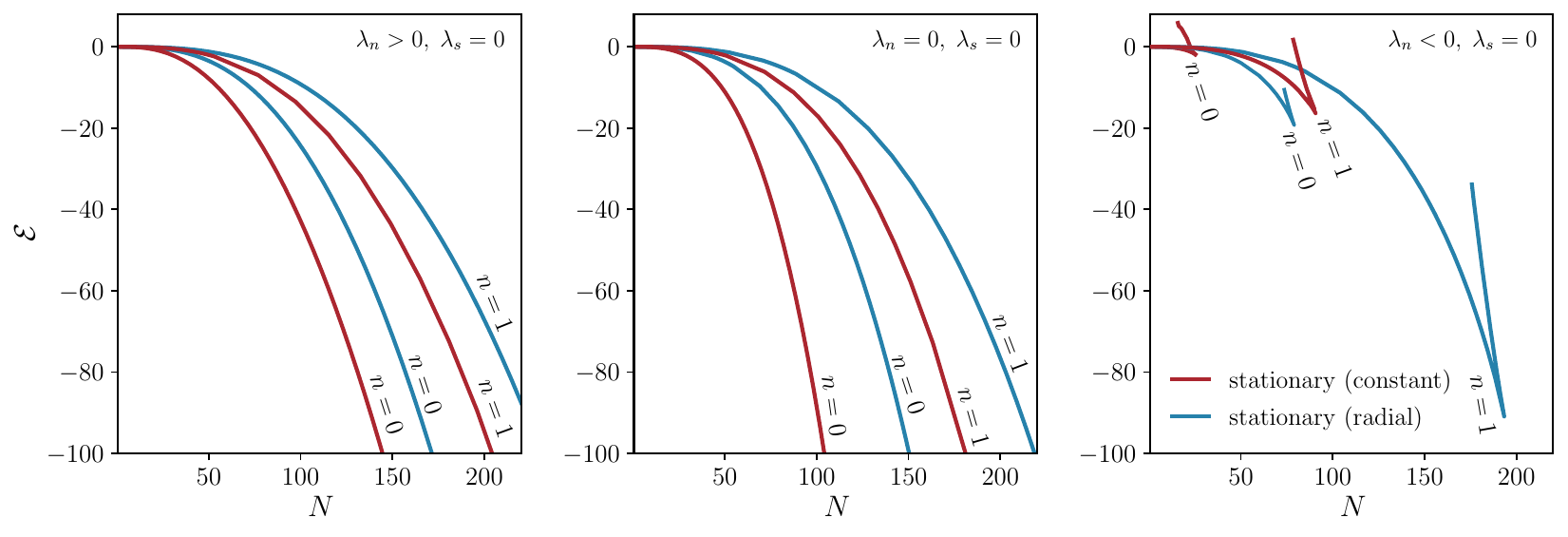}
\caption{{\bf Energy functional of stationary states (symmetry-enhanced sector):} The energy $\mathcal{E}$ of stationary Proca stars with $\lambda_s=0$ as function of their particle number $N$ and polarization vector $\hat{\epsilon}$. Constant polarization states are indicated by red lines and are degenerated when $\lambda_s=0$, whereas states with radial polarization are indicated by blue lines. When the selfinteraction is repulsive ($\lambda_n>0$) or is absent ($\lambda_n=0$) the ground state configuration is provided by nodeless Proca stars of constant polarization and negative energy, $\mathcal{E}<0$, in agreement with the analytical results of Sec.~\ref{sec.equilibrium}. When the selfinteraction is attractive ($\lambda_n<0$) it is not possible to define a ground state configuration and the polarization of the lowest energy stationary state changes with $N$.}\label{FigEnerg}
\end{figure*}

\begin{figure*}
	\centering
    \includegraphics[width=18.cm]{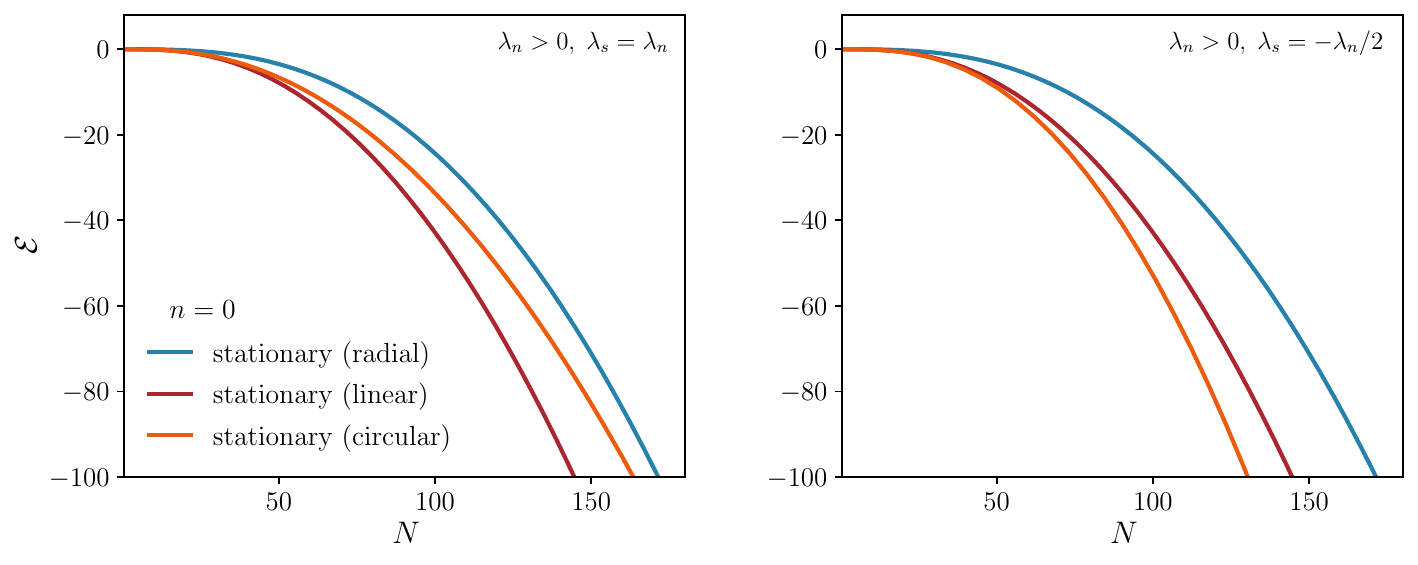}
\caption{{\bf Energy functional of stationary states (generic sector):} Similar as in Fig.~\ref{FigEnerg} but for the case in which $\lambda_s\neq 0$. The spin-spin selfinteraction breaks the degeneracy of constant polarization states, which are only possible for linear (red lines) or circular (orange lines) cases. Radial polarization states are indicated in blue lines and represent excited configurations. The left panel belongs to the shaded region in the first quadrant of Fig.~\ref{fig.lower.bound}, where $\lambda_0> 0$, $\lambda_s> 0$, and a spherically symmetric ground state of linear polarization exists, while the right panel belongs to the shaded triangle in the fourth quadrant of the same figure, where $\lambda_0> 0$, $\lambda_s<0$, and the polarization of the ground state is circular. For simplicity, we have only considered nodeless ($n=0$) configurations. In this figure $\mathcal{E}^{phys}=[\sqrt{8\pi G}m_0^2|\lambda_n^{phys}|^{-3/2}]\mathcal{E}$ and $N^{phys}=[1/(\sqrt{8\pi G}m_0|\lambda_n^{phys}|^{1/2} )]N$.
} \label{FigEnerg.2}
\end{figure*}

\begin{figure*}
	\centering
    \includegraphics[width=17.cm]{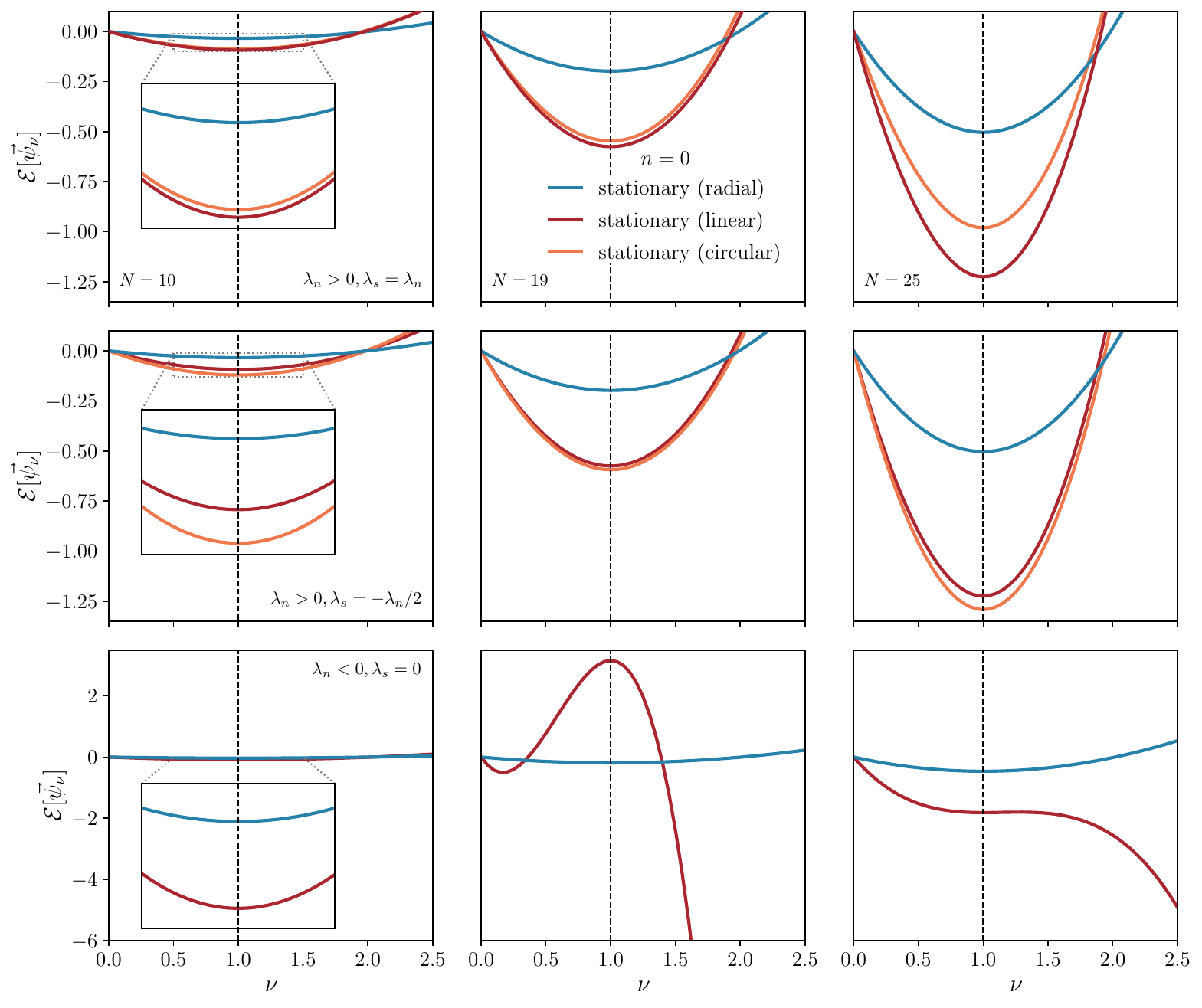}
\caption{{\bf Stationary states as critical points of the energy functional:} The energy functional $\mathcal{E}[\vec{\psi}_\nu]$ of the rescaled states $\vec{\psi}_{\nu}(\vec{x})$ associated to stationary configurations $\vec{\psi}_{\nu=1}(\vec{x})$ of particle number $N=10$, $19$ and $25$ (see Eq.~(\ref{eq.rescaling})). {\it First row:} For $\lambda_0> 0$, $\lambda_s>0$, there exists a global minimum of the energy functional provided by a stationary state of linear polarization and no nodes, where $\mathcal{E}<0$. {\it Second row:} For $\lambda_0> 0$, $\lambda_s<0$, there is also a global minimum of the energy functional, in this case provided by a stationary state of circular polarization and no nodes, where $\mathcal{E}<0$. {\it Third row:} For $\lambda_0<0$, $\lambda_s=0$, stationary states of linear and circular polarization are degenerated and the energy functional is not bounded from below (this applies to the radial case as well, although the maximum of the energy functional appears for larger values of $\nu$). In all cases, the energy functional has a critical point at $\nu=1$, which is a global minimum if $\lambda_0\ge 0$, and a local minimum, a saddle point or a maximum if $\lambda_0<0$, depending on the value of $N$. Here, $\mathcal{E}^{phys}=[\sqrt{8\pi G}m_0^2|\lambda_n^{phys}|^{-3/2}]\mathcal{E}$ and $N^{phys}=[1/(\sqrt{8\pi G}m_0|\lambda_n^{phys}|^{1/2} )]N$.} \label{Fig.extrema}
\end{figure*}

\begin{figure*}
\centering
\includegraphics[width=18cm]{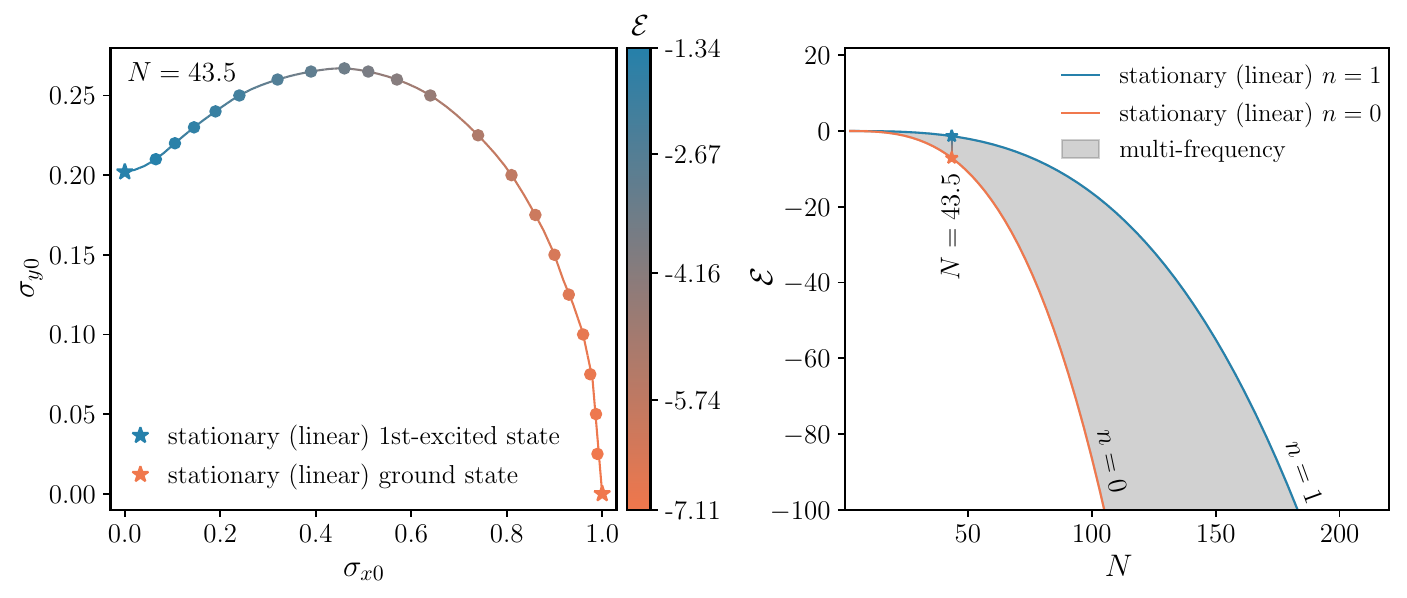}
\caption{{\bf The energy spectrum of multi-frequency states:} The energy of the same multi-frequency Proca stars as in Fig.~\ref{Fig.MassRelation}. The lowest energy state is obtained for $\sigma_{x0}=1$ and $\sigma_{y0}=0$ (orange star), while the highest energy state is for $\sigma_{x0}=0$ and $\sigma_{y0}\approx 0.2$ (blue star). Both configurations correspond to a stationary, linearly polarized state of constant polarization. Multi-frequency states are represented by circle markers, with their energy values indicated by color. In addition to the discrete set of stationary states shown in the center panel of Fig.~\ref{FigEnerg}, there exists a continuum of multi-frequency solutions that connect the ground state with the first excited state.
}\label{Fig.EnergyNfixes}
\end{figure*}

The energy functional plays a central role to determine the equilibrium configurations and deserves an independent discussion. Using Eq.~(\ref{Eq:EnerRelat}), the total energy of a Proca star is given by $\mathcal{E}^{phys}=[\sqrt{8\pi G}m_0^2/\sqrt{\lambda_*^{3}}]\mathcal{E}$, with
\begin{equation}\label{eq.E.code}
\mathcal{E} = -4\pi \sum_i\int_0^\infty \left[\frac{1}{2}\left(\sigma_i^{(0)\prime 2}+\frac{2\gamma\sigma_i^{(0)2}}{r^2}\right)\pm\frac{1}{2}\sigma_i^{(0)4}\right]r^2 dr,
\end{equation}
where we are using the same conventions as in Eqs.~(\ref{eqs.numerical}).

As before, we first focus on stationary states. For $\lambda_s=0$, Fig.~\ref{FigEnerg} shows the energy $\mathcal{E}$ of Proca stars as function of their particle number $N$ and polarization vector $\hat{\epsilon}$. In absence of spin-spin selfinteractions, constant polarization states are degenerated, and this is the reason why for fixed $N$ all constantly polarized Proca stars possess the same total energy. Furthermore, if $\lambda_n \ge 0$, the ground state configuration (i.e. the lowest possible energy state that exists for a given particle number) is given by nodeless ($n=0$), spherically symmetric constant polarization states, irrespectively of the value of $N$, as anticipated at the beginning of Sec.~\ref{equilibrium.multi}. Similarly, when $N$ and $n$ are fixed, radially polarized configurations have more energy than constant polarization states, signaling that radially polarized Proca stars represent excited states of the $s=1$ Gross-Pitaevskii-Poisson system with $\lambda_n\ge 0$ and $\lambda_s=0$. On the other hand, when the selfinteraction is attractive ($\lambda_n<0$), the energy is unbounded from below (see Sec.~\ref{sec.properties.E}) and one cannot define a ground state configuration. Nevertheless, it is interesting to note from the right panel of Fig.~\ref{FigEnerg} that for some values of $N$, radial polarization states possess less energy than constant polarization states whose energy can even become positive. Note that these curves exhibit a spike-like behavior, signaling an extremal point in the energy functional $\mathcal{E}$ as well as the particle number $N$. This feature signals the appearance of a zero mode, which is indicative of a transition in the system's stability. A more detailed discussion of this will be presented in Ref.~\cite{Nambo:preparation}. Finally, as we argued in Sec.~\ref{sec.mass.radius}, when $N\to 0$ (i.e. $\sigma_0\to 0$), the effect of the selfinteraction is negligible and we recover the same results as in the free theory, no matter the value of $\lambda_n$ (see the behavior of the four curves in each panel of Fig.~\ref{FigEnerg} close to the origin).

When $\lambda_s\neq 0$, the situation is more involved given that the characteristic selfinteraction scale $\lambda_*=\lambda_n^{phys}+\alpha \lambda_s^{phys}$ that we have used to normalize physical quantities depends on the state of the system ($\alpha=0$ if the polarization is linear or radial, and $\alpha= 1$ if it is circular). To proceed, we will focus on two cases: $\lambda_n>0$, $\lambda_s=\lambda_n$, and  $\lambda_n>0$, $\lambda_s=-\frac{1}{2}\lambda_n$, both of which lie within the region shown in Fig.~\ref{fig.lower.bound} where the energy functional is bounded from below and a ground state exists. As is evident from Fig.~\ref{FigEnerg.2}, the spin-spin selfinteraction term breaks the degeneration between the constant polarization states that is present when $\lambda_s=0$. This becomes more pronounced as $N$ increases and the effects of the selfinteraction grow in significance. In particular, as we anticipated in Sec.~\ref{sec.properties.E} (see Eq.~(\ref{Eq:lambda0})), when $\lambda_0> 0$, $\lambda_s> 0$, the ground state configuration is given by a stationary state of linear polarization, whereas when $\lambda_0> 0$, $\lambda_s< 0$, the polarization of the lowest energy state is circular. Moreover, in both cases radially polarized Proca stars represent excited configurations. 

We further illustrate this in Fig.~\ref{Fig.extrema}, where we study the behavior of the energy functional under variations of the vector $\vec{\psi}$ which are consistent with the rescaling of Eq.~(\ref{eq.rescaling}). As we demonstrated in Sec.~\ref{sec.properties.E}, when $\lambda_0\ge 0$, the energy functional is bounded from below. Furthermore, if $\lambda_s> 0$, there exists a global minimum of the energy functional that is provided by a stationary and spherically symmetric state of linear polarization (first row of Fig.~\ref{Fig.extrema}), whereas if $\lambda_s<0$ the polarization of the state that minimizes the energy is circular (second row). This suggests the existence of Proca stars that are stable under small perturbations. On the contrary, if $\lambda_0<0$, the energy functional is not bounded from below, as can be appreciated in the third row of Fig.~\ref{Fig.extrema}. Furthermore, for large $N$, the critical points turn into maxima, signaling the onset of an instability.

To study the multi-frequency solutions, we again concentrate on the free theory ($\lambda_n=\lambda_s = 0$). In Fig.~\ref{Fig.EnergyNfixes}, we show the same family of multi-frequency states $N=43.5$ that we introduced in Fig.~\ref{Fig.MassRelation}. Interestingly, this family connects the linearly polarized stationary ground state (in the $x$ direction) with the first excited linearly polarized stationary state (in the $y$ direction). Also shown in Fig.~\ref{Fig.EnergyNfixes} through the color bar is the energy of each state in this family. As can be appreciated, the configuration with $\sigma_{x0}=1$ and $\sigma_{y0}=0$ has the lowest energy, as expected, whereas the energy is growing monotonously when moving along the family towards the state with $\sigma_{x0}=0$ and $\sigma_{y0} \approx 0.2$.  Finally, the right panel of this figure has been constructed from the family $N=43.5$ using the rescaling~(\ref{eq.scaling.free}), which implies that, when $\lambda_n=\lambda_s=0$, the energy $\mathcal{E}$ and the particle number $N$ are related through $\mathcal{E}\sim N^3$. As we previously identified in Fig.~\ref{Fig.MassRelation} for the $M_{99}$ vs. $R_{99}$ plot, multi-frequency states with $(n_x,n_y,n_z)=(0,1,0)$ fill a region in the $\mathcal{E}$ vs. $N$ diagram, which are delimited by the curves associated with the  $n=0$ and $n=1$ stationary states of linear polarization (see the center panel of Fig.~\ref{FigEnerg}).

In the right panel of Fig.~\ref{Fig.EnergyProf}, we compare the energy profile of a stationary nodeless state of radial polarization and charge $\hat{Q}=\textrm{diag}(40,40,40)$, with the one of the multi-frequency solution of the same charge and $(n_x,n_y,n_z)=(0,1,2)$, when $\lambda_n=\lambda_s=0$. Again, this plot can be rescaled to any value of $N$. As was anticipated in Sec.~\ref{sec.global.charges}, for fixed $\hat{Q}$, multi-frequency solutions are more energetic than $n=0$ radially polarized states. 

\section{Conclusions}\label{Sec:Conclussions}

In this paper, we have analyzed the effective theory that describes the nonrelativistic limit of a selfgravitating and selfinteracting massive vector field. This theory consists of a  Gross-Pitaevskii type equation coupled to the Newtonian potential, which, according to Poisson's equation, is sourced by the mass density of the vector field. By expanding the effective theory up to operators of mass dimension 6 in the vector field, we have identified two possible selfinteraction terms in the Gross-Pitaevskii equation: one that depends on the number density squared and is proportional to the coupling constant $\lambda_n$, and another one that depends on the spin density squared and is proportional to $\lambda_s$. These equations define the $s=1$ Gross-Pitaevskii-Poisson system, which reduce to the $s=1$ Schr\"odinger-Poisson system when $\lambda_n=\lambda_s=0$.

We have focused on equilibrium configurations, defined as the critical points of the total energy functional under suitable constraints. Depending on the specific sector of the effective theory that we are exploring, we have identified the existence of stationary states (when $\lambda_s$ is arbitrary), for which the wave function evolves harmonically in time, and multi-frequency states (when $\lambda_s=0$), for which the wave function oscillates with two or three distinct frequencies. Furthermore, we have shown that for $\lambda_0\ge 0$ a ground state configuration which minimizes the total energy for fixed particle number exists (see Eq.~(\ref{Eq:lambda0}) for the definition of $\lambda_0$). 

In spherical symmetry, stationary equilibrium configurations can have linear, circular or radial polarization, where the first two cases are degenerated when $\lambda_s = 0$.  It is interesting to stress that although all these configurations have vanishing orbital angular momentum, circularly polarized states have nonzero spin angular momentum. The angular momentum of these states arises from a macroscopic amplification of the spin of the individual particles that conform the condensate, suggesting that in the relativistic regime the corresponding stars are non-spherical. 
With respect to their gravitational field,  constant polarization states are indistinguishable from  standard nonrelativistic $\ell=0$ boson stars with selfinteraction parameter $\lambda=\lambda_0$.
On the other hand, 
in the free theory ($\lambda_n=\lambda_s=0$), stationary radially polarized Proca stars are indistinguishable from nonrelativistic $\ell=1$ boson stars, which have been previously explored in the context of multi-scalar field theories. In the symmetry-enhanced sector ($\lambda_s=0$), the existence of spherically symmetric multi-frequency solutions allow for a continuous transition in the solution space between different stationary states of constant polarization. In the framework of multi-scalar field theories, these configurations correspond to nonrelativistic $\ell=0$ multi-state solutions.

When $\lambda_0\ge 0$, there is a spherically symmetric stationary state of constant polarization that minimizes the total energy for fixed particle number, which in the free theory is unique. Although we have not been able to prove the uniqueness of the ground state for $\lambda_0>0$, we have demonstrated that if additional states minimize the total energy, they must also be spherical, stationary, and possess constant polarization which is linear if $\lambda_s>0$ and circular if $\lambda_s<0$.  Furthermore, as our numerical solutions  confirm, for the same particle number, the spherically symmetric stationary and radially polarized configurations have more energy than the ground state, and thus they correspond to excited states. In the symmetry-enhanced sector, our numerical results suggest that for fixed $N$ multi-frequency states are more energetic than the ground state and also represent excited configurations. The linear stability of (ground and excited) spherical stationary and multi-frequency states will be analyzed in future work. 

\begin{acknowledgments}

This work was partially supported by CONAHCyT Projects No. 376127 ``Sombras, lentes y ondas gravitatorias generadas por objetos compactos astrofísicos'' and No. 286897 ``Materia oscura: Implicaciones de sus propiedades fundamentales en las observaciones astrofísicas y cosmológicas'', and by CONAHCyT-SNII. E.C.N. and E.P.G. were supported by a CONAHCyT doctoral scholarship. A.D.T. acknowledges support from DAIP Project No. CIIC 2024 198/2024. A.A.R. also acknowledges funding from a postdoctoral fellowship ``Estancias Posdoctorales por México para la Formación y Consolidación de las y los Investigadores por México''. O.S. was partially supported by a CIC grant to Universidad Michoacana de San Nicolás de Hidalgo. We also acknowledge the use of the computing server COUGHS from the UGDataLab at the Physics Department of Guanajuato University.
\end{acknowledgments}

\appendix

\section{Nonrelativistic effective field theory}\label{App:EFT} 

In this appendix, we motivate the origin of our effective theory described by Eq.~(\ref{eq.action.nonrel}). Our starting point is Newtonian gravity, which is expressed in terms of the action
\begin{equation}\label{eq.Newton.action}
S = \int dt\int dV \left[\frac{1}{8\pi G}\mathcal{U}\Delta\mathcal{U}-m_0\mathcal{U} n + \mathcal{L}_m\right].
\end{equation}
The first two terms of this equation correspond to the ``kinetic'' term of the gravitational field $\mathcal{U}(t,\vec{x})$ and its coupling to the mass density $m_0 n(t,\vec{x})$, respectively.

Conversely, the last term of Eq.~(\ref{eq.Newton.action}) comprises the matter sector, that for the purposes of this paper consists of a vector field $\vec{\psi}(t,\vec{x})$ of mass dimension 3/2, i.e. $[\vec{\psi}]=E^{3/2}$, where $E$ denotes dimensions of energy. Following an effective theory approach, we write down the most general expression for $\mathcal{L}_m$ that is compatible with the allowed symmetries of the theory, that we assume to consist of locality (i.e. all fields are evaluated at the same spacetime point) and Galilean transformations~\cite{Merzbacher}:
\begin{equation}\label{eq.trans.psi}
\vec{\psi}(t,\vec{x}) \mapsto 
 e^{-i\left(m_0\vec{v}\cdot\vec{x}+\frac{1}{2}m_0\vec{v}^2t\right)} R\vec{\psi}(t-t_0,R^{-1}\vec{x}+\vec{v}t-\vec{x}_0).
\end{equation}
In this equation $t_0$, $\vec{x}_0$, $\vec{v}$, and $R$ are constant, with $R$ an element of the orthogonal group. With these assumptions, there is an infinite number of terms that can contribute to the effective action, with higher-dimensional operators being suppressed at low energies. For concreteness, in this paper, we will restrict ourselves to operators of mass dimension $6$ or lower.

The vector field transforms non-trivially under the Galilei group, Eq.~(\ref{eq.trans.psi}), which implies that derivative terms must appear in combinations of the form:
\begin{equation}
\int dt\int dV \vec{\psi}^*\cdot m_0^{1-n}\left(i\frac{\partial}{\partial t} +\frac{1}{2m_0}\Delta \right)^n\vec{\psi}.
\end{equation}
When $n=1$, this leads to the standard Schr\"odinger operator appearing in Eq.~(\ref{eq.action.nonrel}). At this point one might be tempted to include terms with $n=2, 3, 4, \ldots$ However, they involve operators of mass dimension 7 or higher, and this is why we have excluded them from our effective theory.

Moreover, non-derivative terms must appear in combinations of the form: $\delta^{ij}\psi_i^*\psi_j$, $\delta^{i\ell}\delta^{jk}\psi_i^*\psi_j\psi_k^*\psi_\ell$, $\delta^{ik}\delta^{j\ell}\psi_i^*\psi_j\psi_k^*\psi_\ell$, $\delta^{i\ell}\delta^{jm}\delta^{kn}\psi_i^*\psi_j\psi_k^*\psi_\ell\psi_m^*\psi_n$, $\ldots$, although only the first three of this series are of mass dimension $6$ or lower. The first term in this list is equal to the particle number density, $\delta^{ij}\psi_i^*\psi_j=n$, and is already included in the first line of Eq.~(\ref{eq.action.nonrel}).\footnote{An operator of the form $\lambda_0 m_0 n$ (where $\lambda_0$ is a dimensionless coupling constant) can be absorbed into the second term of Eq.~(\ref{eq.action.nonrel}) by redefining the gravitational potential as $\mathcal{U}'=\mathcal{U}+\lambda_0$.} The second term of the list gives rise to a selfinteraction operator that depends on the number density squared, $\delta^{i\ell}\delta^{jk}\psi_i^*\psi_j\psi_k^*\psi_\ell=(\psi_i^*\psi^i)(\psi_j^*\psi^j)=n^2$, and it is also present in our effective theory. Apparently, the third term is absent from  Eq.~(\ref{eq.action.nonrel}); however, using the identity $\varepsilon^{mij}\varepsilon_{m}{}^{k\ell}=\delta^{ik}\delta^{j\ell}-\delta^{i\ell}\delta^{jk}$, we can express this operator as $\delta^{ik}\delta^{j\ell}\psi_i^*\psi_j\psi_k^*\psi_\ell= (\varepsilon^{mij}\varepsilon_{m}{}^{k\ell}+\delta^{i\ell}\delta^{jk})\psi_i^*\psi_j\psi_k^*\psi_\ell=-s_i s^i+n^2$, where $s^m=-i\varepsilon^{mk\ell}\psi_k^*\psi_\ell$ represents the spin density. Thus, in general, our theory includes two selfinteraction terms of mass dimension 6: one depending on the square of the number density, $n^2$, and the other on the square of the spin density, $\vec{s}{\,}^2=\vec{s}\cdot\vec{s}$.

\section{Nonrelativistic limit of a selfinteracting Einstein-Proca theory}\label{app.nonrelativistic.Proca}

In this appendix, we compute the nonrelativistic limit of a massive vector field theory.

Our starting point is the Einstein-Proca theory for a complex-valued\footnote{For the nonrelativistic limit of a real-valued vector field theory, see e.g. Ref.~\cite{Zhang:2021xxa}.} vector field $A^{\mu}(t, \vec{x})$ of mass $m_0$ and quartic selfinteraction $\lambda_1(A_\mu^* A^\mu)^2 + \lambda_2 (A_\mu A^\mu)(A^*_\nu A^{\nu*})$.\footnote{Recent works~\cite{Clough:2022ygm, Mou:2022hqb, Coates:2022qia} have pointed out to a fundamental problem with relativistic selfinteracting vector fields due to the appearance of unstable modes that could render these theories unphysical. However, the authors of Refs.~\cite{Barausse:2022rvg, Aoki:2022woy, Rubio:2024ryv} have argued that these instabilities are not indicative of ghosts and/or tachyons, but rather of the breakdown of the well-posedness of the Cauchy problem and the regime of validity of the effective theory.} In natural units, this theory is described by the action
\begin{subequations}\label{eq.action}
\begin{align}\label{eq.action1}
S[g_{\mu\nu},A^\mu]=& \int d^4x\sqrt{-g}\left(\frac{1}{16\pi G}R+\mathcal{L}_M\right),
\end{align}
which consists of the Einstein-Hilbert term 
with matter sector
\begin{eqnarray}\label{eq.action2}
\mathcal{L}_M &=& -\frac{1}{2}F^*_{\mu\nu}F^{\mu\nu}-m_{0}^{2}A_\mu^* A^\mu \nonumber \\
&&-\lambda_1(A_\mu^* A^\mu)^2-\lambda_2 (A_{\mu}A^{\mu})(A^*_\nu A^{\nu *}),
\end{eqnarray}
\end{subequations}
where $F_{\mu\nu} \equiv \nabla_\mu A_\nu -\nabla_\nu A_\mu$ is the ``electromagnetic'' tensor. 

To proceed, we will explore the nonrelativistic limit of the theory~(\ref{eq.action}) at increasing levels of complexity. First, we will focus on the free theory, which excludes the effects of selfinteractions and gravity. Subsequently, we will incorporate the selfinteraction terms, and finally, we will consider the influence of gravity. At the end of the presentation, we will arrive at an expression that coincides with the effective action that we introduced in Eq.~(\ref{eq.action.nonrel}). In this appendix we employ the $(-,+,+,+)$ signature convention for the spacetime metric, and for convenience we occasionally represent Newton's constant $G$ in terms of the Planck mass, denoted as $M_{\textrm{Pl}} \equiv 1/\sqrt{G}$.

\subsection{Free theory}

In absence of gravity, a non-selfinteracting complex-valued vector field $A^{\mu}(t,\vec{x})$ of mass $m_0$ is described in terms of the action
\begin{equation}\label{Ap.eq.action.Proca}
S=\int d^4x \left[-\frac{1}{2}F^*_{\mu\nu}F^{\mu\nu}-m_{0}^{2}A_\mu^* A^\mu\right].
\end{equation}
If we perform a $1+3$ decomposition of the vector field $A^\mu = (A^0, A^i)$ we can write this expression in the form:
\begin{align}\label{Ap.eq.action.Proca.2}
S&=\int d^4x\bigg(\dot{A}_i^*\dot{A}^i+\partial_iA_0^*\partial^iA_0-\partial_iA_j^*\partial^iA^j-\dot{A}_i^*\partial^iA_0 \nonumber\\
&-\partial_{i}A_0^*\dot{A}^i+\partial_{i} A_j^*\partial^{j}A^i+m_0^2A_0^*A_0-m_0^2A_i^*A^i\bigg),
\end{align}
where the overdot indicates time derivative and indices are raised and lowered with the flat spacetime metric.

Now, we express the time $A_0(t,\vec{x})$ and spatial $A_i(t,\vec{x})$ components of $A_\mu(t,\vec{x})$ in the form
\begin{subequations}\label{Ap.eq.nonrel.decomposition}
\begin{align}
A_0(t,\vec{x}) &= \frac{1}{\sqrt{2m_0}}e^{-im_0t}a_0(t,\vec{x}), \\
A_i(t,\vec{x}) &= \frac{1}{\sqrt{2m_0}}e^{-im_0t}\psi_i(t,\vec{x}).
\end{align}
\end{subequations}
This allows us to write Eq.~(\ref{Ap.eq.action.Proca.2}) as
\begin{align}\label{Ap.eq.action.Proca.3}
S&=\int d^4x\bigg(\frac{i}{2}\psi_{i}^{*}\dot{\psi}^{i}-\frac{i}{2}\dot{\psi}_{i}^{*}\psi^{i}+\frac{1}{2m_{0}}\dot{\psi}_{i}^{*}\dot{\psi}^{i}+\frac{1}{2m_{0}}\partial_{i}a_{0}^{*}\partial^{i}a_{0} \nonumber\\
&-\frac{1}{2m_{0}}\partial_{i}\psi_{j}^{*}\partial^{i}\psi^{j}-\frac{i}{2}\psi_{i}^{*}\partial^{i}a_{0}-\frac{1}{2m_{0}}\dot{\psi}_{i}^{*}\partial^{i} a_{0}+\frac{i}{2}\partial_{i} a_{0}^{*}\psi^{i}\nonumber\\
&-\frac{1}{2m_{0}}\partial_{i}a_{0}^{*}\dot{\psi}^{i}+\frac{1}{2m_{0}}\partial_{i}\psi_{j}^{*}\partial^{j}\psi^{i}+\frac{m_0}{2}a_{0}^{*}a_{0}\bigg).
\end{align}
In the nonrelativistic limit, the different quantities scale as $\partial_t\sim\epsilon m_0$, $\partial_i \sim \epsilon^{1/2}m_0$, and $a_0\sim \epsilon^{1/2}|\psi_i|$, with $\epsilon$ a small positive number, so to leading order in $\epsilon$ we can approximate:
\begin{align}\label{Ap.eq.action.Proca.3.nonrel}
S&=\int d^4x\bigg(\frac{i}{2}\psi_{i}^{*}\dot{\psi}^{i}-\frac{i}{2}\dot{\psi}_{i}^{*}\psi^{i}-\frac{1}{2m_{0}}\partial_{i}\psi_{j}^{*}\partial^{i}\psi^{j}-\frac{i}{2}\psi_{i}^{*}\partial^{i}a_{0}\nonumber\\
&+\frac{i}{2}\partial_{i}a_{0}^{*}\psi^{i}+\frac{1}{2m_{0}}\partial_{i}\psi_{j}^{*}\partial^{j}\psi^{i}+\frac{m_{0}}{2}a_{0}^{*}a_{0}\bigg).
\end{align}

Note that there are no time derivatives of $a_0$ in this expression, indicating that this component is not dynamical and can be integrated out from the action. In order to do that, we vary Eq.~(\ref{Ap.eq.action.Proca.3.nonrel}) with respect to $a_0$, and obtain 
\begin{equation}\label{Ap.eq.constraint}
a_0 = \frac{i}{m_0}\partial_j\psi^j,
\end{equation}
which is a condition that must be satisfied by $a_0$. 
Introducing the constraint~(\ref{Ap.eq.constraint}) back into Eq.~(\ref{Ap.eq.action.Proca.3.nonrel}), integrating by parts, and neglecting surface terms, yields
\begin{equation}\label{Ap.eq.action.Proca.4}
S= \int d^4x \left[\psi_i^*\left(i\frac{\partial}{\partial t}+\frac{1}{2m_0}\Delta\right)\psi^i\right].
\end{equation}
This is just the Schr\"odinger action for a vector wave function $\psi_i(t,\vec{x})$ that describes free particles of spin $s=1$.

\subsection{Selfinteractions}\label{app.selfinteractions}

In presence of selfinteractions, Eq.~(\ref{Ap.eq.action.Proca.3}) requires the addition of the new terms:
\begin{eqnarray}
&&\int d^4 x \left\{\frac{\lambda_1}{4m_0^2}\left[
-(a_0^*a_0)^2+2a_0^*a_0\psi_i^*\psi^i-(\psi_i^*\psi^i)^2
\right]\right.\\
&&\left. +\frac{\lambda_2}{4m_0^2}\left[ -(a_0^*a_0)^2 +a_0^2\psi_i^*\psi^{i*}  +a^{*2}_0\psi_i\psi^{i}-\psi_i\psi^i\psi_j^*\psi^{j*}\right]
\right\}\nonumber.
\end{eqnarray}
In the nonrelativistic limit, the third and seventh terms dominate over the other five, and hence the constraint in Eq.~(\ref{Ap.eq.constraint}) is unaffected. If again we integrate out the field $a_0$, we obtain
\begin{equation}\label{eq.action.Proca.self2}
S= \int d^4x \bigg[\psi_j^*\left(i\frac{\partial}{\partial t}+\frac{1}{2m_0}\Delta\right)\psi^j - \frac{\lambda_n}{4m_0^2}n^2-\frac{\lambda_s}{4m_0^2}s_js^j\bigg],
\end{equation}
where $n=\psi_i^*\psi^i$ is the number density, $s^m=-i\varepsilon^{mij}\psi_i^*\psi_j$ is the spin density and we have defined $\lambda_n:= \lambda_1 + \lambda_2$ and $\lambda_s:= -\lambda_2$. In order to obtain Eq.~(\ref{eq.action.Proca.self2}), we have used some properties of the Levi-Civita symbol that were previously introduced in App.~\ref{App:EFT}. Note that the only difference with respect to the free theory, Eq.~(\ref{Ap.eq.action.Proca.4}), is the appearance of two selfinteraction terms.

\subsection{Gravity}

Finally, we include the effects of gravity, which are codified in the spacetime metric $g_{\mu\nu}(t,\vec{x})$. For that purpose, it is convenient to decompose the spacetime line element in the form~\cite{Nambo:2024gvs}
\begin{equation}\label{eq.metric.approx}
 ds^2=-\left[1+2\Phi(t, \vec{x})\right]dt^2+\left[1-2\Psi(t, \vec{x})\right]\delta_{jk}dx^{j}dx^{k},
\end{equation}
which has been expressed in the Newtonian gauge and codifies only the scalar degrees of freedom of the gravitational field (vector and tensor modes do not couple to nonrelativistic matter and we have omitted them here). The functions $\Phi(t,\vec{x})$ and $\Psi(t,\vec{x})$ transform as scalars under spatial rotations and constitute the gravitational potentials.

This introduces the additional terms
\begin{equation}\label{eq.action.terms.gravity}
\int d^4x \left[\frac{1}{8\pi G}\Psi\Delta\left(2\Phi-\Psi\right)-m_0\Phi n\right]
\end{equation}
into the action~(\ref{eq.action.Proca.self2}), where we have taken into account that, in the nonrelativistic limit, $\Phi\sim\Psi\sim \epsilon$ and $|\psi_i|\sim \sqrt{M_{\textrm{Pl}}^2m_0}\epsilon$. Here, the first term of Eq.~(\ref{eq.action.terms.gravity}) originates from the Einstein-Hilbert action, whereas the second one from the kinetic term of the vector field. Note that the field $a_0$ is absent from Eq.~(\ref{eq.action.terms.gravity}), and hence the constraint (\ref{Ap.eq.constraint}) is not affected.

Combining Eqs.~(\ref{eq.action.Proca.self2}) and~(\ref{eq.action.terms.gravity}), we obtain 
\begin{eqnarray}\label{eq.action.nonrel.}
&& S[\Phi, \Psi, \psi_j]=\int dt\int d^3x \bigg[\frac{1}{8\pi G}\Psi\Delta\left(2\Phi-\Psi\right)-m_0\Phi n\nonumber\\ 
&&+\psi_{j}^*\left(i\frac{\partial}{\partial t}+\frac{1}{2m_{0}}\Delta \right)\psi^{j}-\frac{\lambda_n}{4m_{0}^{2}}n^{2}-\frac{\lambda_{s}}{4m_{0}^{2}}s_{j}s^{j}\bigg].
\end{eqnarray}
Varying this expression with respect to  $\Psi$ yields $\Delta(\Phi - \Psi) = 0$, and assuming that $\Phi$ and $\Psi$ vanish at infinity this implies that $\Phi=\Psi$.  Introducing this result back into Eq.~(\ref{eq.action.nonrel.}), we arrive at Eq.~(\ref{eq.action.nonrel}), where we have defined $\Phi=\Psi:=\mathcal{U}$ as the Newtonian gravitational potential.

\section{Constant polarization states in the presence of spin-spin selfinteraction} \label{app.ConstantPol}

In this appendix, we prove that when $\lambda_s\neq 0$, constant polarization states satisfying the field equations are necessarily linear or circular. To do so, we substitute the definition~(\ref{Eq.Const.Polariz}) into Eq.~(\ref{integro-differential_equation}) and obtain 
\begin{align}
    i\frac{\partial f}{\partial t} \hat{\epsilon} &= -(\Delta f) \hat{\epsilon} + \lambda_n |f|^2 f \hat{\epsilon}\nonumber \\
    &+ \lambda_s |f|^2 f (\hat{\epsilon}^* \times \hat{\epsilon}) \times \hat{\epsilon} + \Delta^{-1}(|f|^2)f \hat{\epsilon}.
\end{align}
Therefore, the following conditions must be satisfied:
\begin{equation}
\label{Eq:ConditionEpsilon}
    (\hat{\epsilon}^* \times \hat{\epsilon}) \times \hat{\epsilon} = \hat{\epsilon} - (\hat{\epsilon} \cdot \hat{\epsilon}) \hat{\epsilon}^* = C\hat{\epsilon},
\end{equation}
with $C$ a complex constant. Taking the dot product with $\hat{\epsilon}$ on both sides of Eq.~(\ref{Eq:ConditionEpsilon}) yields
\begin{equation}
\label{Eq:EpsilonDotEpsilon}
    C(\hat{\epsilon} \cdot \hat{\epsilon}) = 0.
\end{equation}
There are two solution to this equation: $C = 0$ and $\hat{\epsilon} \cdot \hat{\epsilon} = 0$. In the first case, Eq.~(\ref{Eq:ConditionEpsilon}) leads to
\begin{equation}
\label{Eq:EpsilonReal}
    \hat{\epsilon} = (\hat{\epsilon} \cdot \hat{\epsilon}) \hat{\epsilon}^*,
\end{equation}
which, together with the condition $\hat{\epsilon}^* \cdot \hat{\epsilon} = 1$, implies that $\hat{\epsilon}$ is real-valued up to a global phase factor. This is the condition for linear polarization. In the second case, if one writes $\hat{\epsilon} = \hat{\epsilon}_R + i \hat{\epsilon}_I$ (with $\hat{\epsilon}_R$ and $\hat{\epsilon}_I$ denoting the real and imaginary parts of $\hat{\epsilon}$, respectively) then the equation $\hat{\epsilon} \cdot \hat{\epsilon} = 0$ together with $\hat{\epsilon}^* \cdot \hat{\epsilon} = 1$ implies that
\begin{subequations}
\begin{align}
    |\hat{\epsilon}_R| = |\hat{\epsilon}_I| &= \frac{1}{\sqrt{2}}, \\
    \hat{\epsilon}_R \cdot \hat{\epsilon}_I &= 0,
\end{align}
\end{subequations}
which is the condition for circular polarization in the direction $\hat{\epsilon}_R\times\hat{\epsilon}_I$.

\section{Radially polarized Proca stars as $\ell=1$ boson stars} \label{app.radiallyVSl=1}

Radially polarized Proca stars,
\begin{equation}\label{eq.app.radially}
\vec{\psi}(t,\vec{x})= e^{-iEt}\sigma^{(0)}(r)\hat{e}_r,
\end{equation}
are described in terms of Eqs.~(\ref{s=1GPP.stationary}) with $\gamma=1$ and $\alpha=0$, which, in absence of selfinteractions, $\lambda_n=\lambda_s=0$, coincide with those of a $\ell=1$ boson star consisting on three independent scalar fields of the form
\begin{equation}\label{eq.components.ell=1}
\phi_m(t,\vec{x}) = e^{-iEt}\phi(r)Y^{1,m}(\vartheta,\varphi), \quad m=1,2,3,
\end{equation}
cf. Eqs.~(41) of Ref.~\cite{Roque:2023sjl}.

To clarify this, we can use the identity (see e.g. Sec.~IIIC of Ref.~\cite{Nambo:2023yut})
\begin{equation}
\hat{e}_r = \sqrt{\frac{4\pi}{3}}\hat{U}\left[Y^{1,-1}(\vartheta,\varphi)\hat{e}_x + Y^{1,0}(\vartheta,\varphi)\hat{e}_y + Y^{1,1}(\vartheta,\varphi)\hat{e}_z\right],
\end{equation}
where $\hat{U}$ is the unitary constant operator with matrix elements
\begin{equation}
\hat{U} = \frac{1}{\sqrt{2}}\left(\begin{matrix}
1 & 0 & -1\\
i & 0 & i\\
0 & \sqrt{2} & 0
\end{matrix}
\right),
\end{equation}
to express the ansatz~(\ref{eq.app.radially}) in the form
\begin{eqnarray}\label{eq.ansatz.proca2}
\vec{\psi}(t,\vec{x}) &=& \sqrt{\frac{4\pi}{3}} e^{-iEt}\sigma^{(0)}(r)\\
&& \left[Y^{1,-1}(\vartheta,\varphi)\hat{e}_x+Y^{1,0}(\vartheta,\varphi)\hat{e}_y+Y^{1,1}(\vartheta,\varphi)\hat{e}_z\right], \nonumber
\end{eqnarray}
where we do not distinguish between unitary equivalent configurations (in absence of selfinteractions the action~(\ref{eq.action.nonrel}) is invariant under unitary transformations). Now, it is evident that the Cartesian components of the vector field $\vec{\psi}(t,\vec{x})$ in Eq.~(\ref{eq.ansatz.proca2}) coincide with the three independent scalar fields $\phi_m(t,\vec{x})$ in Eq.~(\ref{eq.components.ell=1}) that constitute a $\ell=1$ boson star configuration.

\section{Some comments about the shooting method}\label{app:shooting}

Due to the numerical precision (about $16$ decimal digits in our code), the shooting method described in Sec.~\ref{sec.numerical} only allows us to reach a finite radius. Beyond this radius, we utilize the asymptotic solutions
\begin{align}\label{Eq.AsymForm}
    \sigma_{i}^{(0)}(r) \approx\frac{C_i}{r^{1+\gamma}} e^{-\sqrt{\abs{E_i}} r}, \quad u_i^{(0)}(r) \approx E_i+\frac{N}{4\pi r}
\end{align}
of Eqs.~(\ref{eqs.numerical}), with $C_i, E_i$, and $N$ constants, where $E_i$ and $N$ represent the dimensionless frequencies and trace of the Hermitian operator $\hat{Q}$, respectively, and $C_i$ is an amplitude scale. The value of $E_i$ and $N$ are computed according to\footnote{Alternatively, one can use the asymptotic form described in Eq.~(\ref{Eq.AsymForm}) to obtain $E_i$. This alternative form to compute the energy eigenvalue was used to check the validity of the results obtained from Eq.~(\ref{EqIntEnerg}).}
\begin{subequations}
\begin{align}
E_i &= u_{i0}-\sum_{i}\int_{0}^{\infty} \sigma_i^{(0)2}(r) r^{2\gamma + 1} dr,\label{EqIntEnerg}\\
N &= 4\pi \sum_{i}\int_{0}^{\infty} \sigma_i^{(0)2}(r) r^{2(\gamma+1)} dr,
\end{align}
\end{subequations}
whereas the coefficients $C_i$ are obtained using a linear fitting methodology. The value of  $\gamma$ is one for radial polarization and zero for all other cases.
For more details, see App.~C in Ref.~\cite{Roque:2023sjl}.

\bibliography{ref.bib} 

\begin{thebibliography}{64}%
\makeatletter
\providecommand \@ifxundefined [1]{%
 \@ifx{#1\undefined}
}%
\providecommand \@ifnum [1]{%
 \ifnum #1\expandafter \@firstoftwo
 \else \expandafter \@secondoftwo
 \fi
}%
\providecommand \@ifx [1]{%
 \ifx #1\expandafter \@firstoftwo
 \else \expandafter \@secondoftwo
 \fi
}%
\providecommand \natexlab [1]{#1}%
\providecommand \enquote  [1]{``#1''}%
\providecommand \bibnamefont  [1]{#1}%
\providecommand \bibfnamefont [1]{#1}%
\providecommand \citenamefont [1]{#1}%
\providecommand \href@noop [0]{\@secondoftwo}%
\providecommand \href [0]{\begingroup \@sanitize@url \@href}%
\providecommand \@href[1]{\@@startlink{#1}\@@href}%
\providecommand \@@href[1]{\endgroup#1\@@endlink}%
\providecommand \@sanitize@url [0]{\catcode `\\12\catcode `\$12\catcode
  `\&12\catcode `\#12\catcode `\^12\catcode `\_12\catcode `\%12\relax}%
\providecommand \@@startlink[1]{}%
\providecommand \@@endlink[0]{}%
\providecommand \url  [0]{\begingroup\@sanitize@url \@url }%
\providecommand \@url [1]{\endgroup\@href {#1}{\urlprefix }}%
\providecommand \urlprefix  [0]{URL }%
\providecommand \Eprint [0]{\href }%
\providecommand \doibase [0]{https://doi.org/}%
\providecommand \selectlanguage [0]{\@gobble}%
\providecommand \bibinfo  [0]{\@secondoftwo}%
\providecommand \bibfield  [0]{\@secondoftwo}%
\providecommand \translation [1]{[#1]}%
\providecommand \BibitemOpen [0]{}%
\providecommand \bibitemStop [0]{}%
\providecommand \bibitemNoStop [0]{.\EOS\space}%
\providecommand \EOS [0]{\spacefactor3000\relax}%
\providecommand \BibitemShut  [1]{\csname bibitem#1\endcsname}%
\let\auto@bib@innerbib\@empty
\bibitem [{\citenamefont {Kaup}(1968)}]{Kaup:1968zz}%
  \BibitemOpen
  \bibfield  {author} {\bibinfo {author} {\bibfnamefont {D.~J.}\ \bibnamefont
  {Kaup}},\ }\bibfield  {title} {\bibinfo {title} {{Klein-Gordon Geon}},\
  }\href {https://doi.org/10.1103/PhysRev.172.1331} {\bibfield  {journal}
  {\bibinfo  {journal} {Phys. Rev.}\ }\textbf {\bibinfo {volume} {172}},\
  \bibinfo {pages} {1331} (\bibinfo {year} {1968})}\BibitemShut {NoStop}%
\bibitem [{\citenamefont {Ruffini}\ and\ \citenamefont
  {Bonazzola}(1969)}]{Ruffini:1969qy}%
  \BibitemOpen
  \bibfield  {author} {\bibinfo {author} {\bibfnamefont {R.}~\bibnamefont
  {Ruffini}}\ and\ \bibinfo {author} {\bibfnamefont {S.}~\bibnamefont
  {Bonazzola}},\ }\bibfield  {title} {\bibinfo {title} {{Systems of
  selfgravitating particles in general relativity and the concept of an
  equation of state}},\ }\href {https://doi.org/10.1103/PhysRev.187.1767}
  {\bibfield  {journal} {\bibinfo  {journal} {Phys. Rev.}\ }\textbf {\bibinfo
  {volume} {187}},\ \bibinfo {pages} {1767} (\bibinfo {year}
  {1969})}\BibitemShut {NoStop}%
\bibitem [{\citenamefont {Jetzer}(1992)}]{Jetzer:1991jr}%
  \BibitemOpen
  \bibfield  {author} {\bibinfo {author} {\bibfnamefont {P.}~\bibnamefont
  {Jetzer}},\ }\bibfield  {title} {\bibinfo {title} {{Boson stars}},\ }\href
  {https://doi.org/10.1016/0370-1573(92)90123-H} {\bibfield  {journal}
  {\bibinfo  {journal} {Phys. Rept.}\ }\textbf {\bibinfo {volume} {220}},\
  \bibinfo {pages} {163} (\bibinfo {year} {1992})}\BibitemShut {NoStop}%
\bibitem [{\citenamefont {Liddle}\ and\ \citenamefont
  {Madsen}(1992)}]{Liddle:1992fmk}%
  \BibitemOpen
  \bibfield  {author} {\bibinfo {author} {\bibfnamefont {A.~R.}\ \bibnamefont
  {Liddle}}\ and\ \bibinfo {author} {\bibfnamefont {M.~S.}\ \bibnamefont
  {Madsen}},\ }\bibfield  {title} {\bibinfo {title} {{The Structure and
  formation of boson stars}},\ }\href
  {https://doi.org/10.1142/S0218271892000057} {\bibfield  {journal} {\bibinfo
  {journal} {Int. J. Mod. Phys. D}\ }\textbf {\bibinfo {volume} {1}},\ \bibinfo
  {pages} {101} (\bibinfo {year} {1992})}\BibitemShut {NoStop}%
\bibitem [{\citenamefont {Schunck}\ and\ \citenamefont
  {Mielke}(2003)}]{Schunck:2003kk}%
  \BibitemOpen
  \bibfield  {author} {\bibinfo {author} {\bibfnamefont {F.~E.}\ \bibnamefont
  {Schunck}}\ and\ \bibinfo {author} {\bibfnamefont {E.~W.}\ \bibnamefont
  {Mielke}},\ }\bibfield  {title} {\bibinfo {title} {{General relativistic
  boson stars}},\ }\href {https://doi.org/10.1088/0264-9381/20/20/201}
  {\bibfield  {journal} {\bibinfo  {journal} {Class. Quant. Grav.}\ }\textbf
  {\bibinfo {volume} {20}},\ \bibinfo {pages} {R301} (\bibinfo {year}
  {2003})},\ \Eprint {https://arxiv.org/abs/0801.0307} {arXiv:0801.0307
  [astro-ph]} \BibitemShut {NoStop}%
\bibitem [{\citenamefont {Liebling}\ and\ \citenamefont
  {Palenzuela}(2023)}]{Liebling:2012fv}%
  \BibitemOpen
  \bibfield  {author} {\bibinfo {author} {\bibfnamefont {S.~L.}\ \bibnamefont
  {Liebling}}\ and\ \bibinfo {author} {\bibfnamefont {C.}~\bibnamefont
  {Palenzuela}},\ }\bibfield  {title} {\bibinfo {title} {{Dynamical boson
  stars}},\ }\href {https://doi.org/10.1007/s41114-023-00043-4} {\bibfield
  {journal} {\bibinfo  {journal} {Living Rev. Rel.}\ }\textbf {\bibinfo
  {volume} {26}},\ \bibinfo {pages} {1} (\bibinfo {year} {2023})},\ \Eprint
  {https://arxiv.org/abs/1202.5809} {arXiv:1202.5809 [gr-qc]} \BibitemShut
  {NoStop}%
\bibitem [{\citenamefont {Zhang}(2019)}]{Zhang:2018slz}%
  \BibitemOpen
  \bibfield  {author} {\bibinfo {author} {\bibfnamefont {H.}~\bibnamefont
  {Zhang}},\ }\bibfield  {title} {\bibinfo {title} {{Axion Stars}},\ }\href
  {https://doi.org/10.3390/sym12010025} {\bibfield  {journal} {\bibinfo
  {journal} {Symmetry}\ }\textbf {\bibinfo {volume} {12}},\ \bibinfo {pages}
  {25} (\bibinfo {year} {2019})},\ \Eprint {https://arxiv.org/abs/1810.11473}
  {arXiv:1810.11473 [hep-ph]} \BibitemShut {NoStop}%
\bibitem [{\citenamefont {Visinelli}(2021)}]{Visinelli:2021uve}%
  \BibitemOpen
  \bibfield  {author} {\bibinfo {author} {\bibfnamefont {L.}~\bibnamefont
  {Visinelli}},\ }\bibfield  {title} {\bibinfo {title} {{Boson stars and
  oscillatons: A review}},\ }\href {https://doi.org/10.1142/S0218271821300068}
  {\bibfield  {journal} {\bibinfo  {journal} {Int. J. Mod. Phys. D}\ }\textbf
  {\bibinfo {volume} {30}},\ \bibinfo {pages} {2130006} (\bibinfo {year}
  {2021})},\ \Eprint {https://arxiv.org/abs/2109.05481} {arXiv:2109.05481
  [gr-qc]} \BibitemShut {NoStop}%
\bibitem [{\citenamefont {Ch\'avez~Nambo}\ \emph {et~al.}(tion)\citenamefont
  {Ch\'avez~Nambo}, \citenamefont {Diez-Tejedor}, \citenamefont
  {Preciado-Govea}, \citenamefont {Roque},\ and\ \citenamefont
  {Sarbach}}]{Nambo:preparation}%
  \BibitemOpen
  \bibfield  {author} {\bibinfo {author} {\bibfnamefont {E.}~\bibnamefont
  {Ch\'avez~Nambo}}, \bibinfo {author} {\bibfnamefont {A.}~\bibnamefont
  {Diez-Tejedor}}, \bibinfo {author} {\bibfnamefont {E.}~\bibnamefont
  {Preciado-Govea}}, \bibinfo {author} {\bibfnamefont {A.~A.}\ \bibnamefont
  {Roque}},\ and\ \bibinfo {author} {\bibfnamefont {O.}~\bibnamefont
  {Sarbach}},\ }\href@noop {} {\  (\bibinfo {year} {{in
  preparation}})}\BibitemShut {NoStop}%
\bibitem [{\citenamefont {Brito}\ \emph {et~al.}(2016)\citenamefont {Brito},
  \citenamefont {Cardoso}, \citenamefont {Herdeiro},\ and\ \citenamefont
  {Radu}}]{Brito:2015pxa}%
  \BibitemOpen
  \bibfield  {author} {\bibinfo {author} {\bibfnamefont {R.}~\bibnamefont
  {Brito}}, \bibinfo {author} {\bibfnamefont {V.}~\bibnamefont {Cardoso}},
  \bibinfo {author} {\bibfnamefont {C.~A.~R.}\ \bibnamefont {Herdeiro}},\ and\
  \bibinfo {author} {\bibfnamefont {E.}~\bibnamefont {Radu}},\ }\bibfield
  {title} {\bibinfo {title} {{Proca stars: Gravitating
  Bose\textendash{}Einstein condensates of massive spin 1 particles}},\ }\href
  {https://doi.org/10.1016/j.physletb.2015.11.051} {\bibfield  {journal}
  {\bibinfo  {journal} {Phys. Lett. B}\ }\textbf {\bibinfo {volume} {752}},\
  \bibinfo {pages} {291} (\bibinfo {year} {2016})},\ \Eprint
  {https://arxiv.org/abs/1508.05395} {arXiv:1508.05395 [gr-qc]} \BibitemShut
  {NoStop}%
\bibitem [{\citenamefont {Salazar~Landea}\ and\ \citenamefont
  {Garc\'\i{}a}(2016)}]{SalazarLandea:2016bys}%
  \BibitemOpen
  \bibfield  {author} {\bibinfo {author} {\bibfnamefont {I.}~\bibnamefont
  {Salazar~Landea}}\ and\ \bibinfo {author} {\bibfnamefont {F.}~\bibnamefont
  {Garc\'\i{}a}},\ }\bibfield  {title} {\bibinfo {title} {{Charged Proca
  Stars}},\ }\href {https://doi.org/10.1103/PhysRevD.94.104006} {\bibfield
  {journal} {\bibinfo  {journal} {Phys. Rev. D}\ }\textbf {\bibinfo {volume}
  {94}},\ \bibinfo {pages} {104006} (\bibinfo {year} {2016})},\ \Eprint
  {https://arxiv.org/abs/1608.00011} {arXiv:1608.00011 [hep-th]} \BibitemShut
  {NoStop}%
\bibitem [{\citenamefont {Brihaye}\ \emph {et~al.}(2017)\citenamefont
  {Brihaye}, \citenamefont {Delplace},\ and\ \citenamefont
  {Verbin}}]{Brihaye:2017inn}%
  \BibitemOpen
  \bibfield  {author} {\bibinfo {author} {\bibfnamefont {Y.}~\bibnamefont
  {Brihaye}}, \bibinfo {author} {\bibfnamefont {T.}~\bibnamefont {Delplace}},\
  and\ \bibinfo {author} {\bibfnamefont {Y.}~\bibnamefont {Verbin}},\
  }\bibfield  {title} {\bibinfo {title} {{Proca Q Balls and their Coupling to
  Gravity}},\ }\href {https://doi.org/10.1103/PhysRevD.96.024057} {\bibfield
  {journal} {\bibinfo  {journal} {Phys. Rev. D}\ }\textbf {\bibinfo {volume}
  {96}},\ \bibinfo {pages} {024057} (\bibinfo {year} {2017})},\ \Eprint
  {https://arxiv.org/abs/1704.01648} {arXiv:1704.01648 [gr-qc]} \BibitemShut
  {NoStop}%
\bibitem [{\citenamefont {Minamitsuji}(2018)}]{Minamitsuji:2018kof}%
  \BibitemOpen
  \bibfield  {author} {\bibinfo {author} {\bibfnamefont {M.}~\bibnamefont
  {Minamitsuji}},\ }\bibfield  {title} {\bibinfo {title} {{Vector boson star
  solutions with a quartic order self-interaction}},\ }\href
  {https://doi.org/10.1103/PhysRevD.97.104023} {\bibfield  {journal} {\bibinfo
  {journal} {Phys. Rev. D}\ }\textbf {\bibinfo {volume} {97}},\ \bibinfo
  {pages} {104023} (\bibinfo {year} {2018})},\ \Eprint
  {https://arxiv.org/abs/1805.09867} {arXiv:1805.09867 [gr-qc]} \BibitemShut
  {NoStop}%
\bibitem [{\citenamefont {Herdeiro}\ \emph {et~al.}(2020)\citenamefont
  {Herdeiro}, \citenamefont {Panotopoulos},\ and\ \citenamefont
  {Radu}}]{Herdeiro:2020kba}%
  \BibitemOpen
  \bibfield  {author} {\bibinfo {author} {\bibfnamefont {C.~A.~R.}\
  \bibnamefont {Herdeiro}}, \bibinfo {author} {\bibfnamefont {G.}~\bibnamefont
  {Panotopoulos}},\ and\ \bibinfo {author} {\bibfnamefont {E.}~\bibnamefont
  {Radu}},\ }\bibfield  {title} {\bibinfo {title} {{Tidal Love numbers of Proca
  stars}},\ }\href {https://doi.org/10.1088/1475-7516/2020/08/029} {\bibfield
  {journal} {\bibinfo  {journal} {JCAP}\ }\textbf {\bibinfo {volume} {08}},\
  \bibinfo {pages} {029}},\ \Eprint {https://arxiv.org/abs/2006.11083}
  {arXiv:2006.11083 [gr-qc]} \BibitemShut {NoStop}%
\bibitem [{\citenamefont {Herdeiro}\ and\ \citenamefont
  {Radu}(2020)}]{Herdeiro:2020jzx}%
  \BibitemOpen
  \bibfield  {author} {\bibinfo {author} {\bibfnamefont {C.~A.~R.}\
  \bibnamefont {Herdeiro}}\ and\ \bibinfo {author} {\bibfnamefont
  {E.}~\bibnamefont {Radu}},\ }\bibfield  {title} {\bibinfo {title}
  {{Asymptotically flat, spherical, self-interacting scalar, Dirac and Proca
  stars}},\ }\href {https://doi.org/10.3390/sym12122032} {\bibfield  {journal}
  {\bibinfo  {journal} {Symmetry}\ }\textbf {\bibinfo {volume} {12}},\ \bibinfo
  {pages} {2032} (\bibinfo {year} {2020})},\ \Eprint
  {https://arxiv.org/abs/2012.03595} {arXiv:2012.03595 [gr-qc]} \BibitemShut
  {NoStop}%
\bibitem [{\citenamefont {Herdeiro}\ \emph {et~al.}(2024)\citenamefont
  {Herdeiro}, \citenamefont {Radu}, \citenamefont {Sanchis-Gual}, \citenamefont
  {Santos},\ and\ \citenamefont {dos Santos Costa~Filho}}]{Herdeiro:2023wqf}%
  \BibitemOpen
  \bibfield  {author} {\bibinfo {author} {\bibfnamefont {C.~A.~R.}\
  \bibnamefont {Herdeiro}}, \bibinfo {author} {\bibfnamefont {E.}~\bibnamefont
  {Radu}}, \bibinfo {author} {\bibfnamefont {N.}~\bibnamefont {Sanchis-Gual}},
  \bibinfo {author} {\bibfnamefont {N.~M.}\ \bibnamefont {Santos}},\ and\
  \bibinfo {author} {\bibfnamefont {E.}~\bibnamefont {dos Santos
  Costa~Filho}},\ }\bibfield  {title} {\bibinfo {title} {{The non-spherical
  ground state of Proca stars}},\ }\href
  {https://doi.org/10.1016/j.physletb.2024.138595} {\bibfield  {journal}
  {\bibinfo  {journal} {Phys. Lett. B}\ }\textbf {\bibinfo {volume} {852}},\
  \bibinfo {pages} {138595} (\bibinfo {year} {2024})},\ \Eprint
  {https://arxiv.org/abs/2311.14800} {arXiv:2311.14800 [gr-qc]} \BibitemShut
  {NoStop}%
\bibitem [{\citenamefont {Joaquin}\ and\ \citenamefont
  {Alcubierre}(2025)}]{Joaquin:2024quo}%
  \BibitemOpen
  \bibfield  {author} {\bibinfo {author} {\bibfnamefont {C.}~\bibnamefont
  {Joaquin}}\ and\ \bibinfo {author} {\bibfnamefont {M.}~\bibnamefont
  {Alcubierre}},\ }\bibfield  {title} {\bibinfo {title} {{Proca stars in
  excited states}},\ }\href {https://doi.org/10.1007/s10714-025-03375-4}
  {\bibfield  {journal} {\bibinfo  {journal} {Gen. Rel. Grav.}\ }\textbf
  {\bibinfo {volume} {57}},\ \bibinfo {pages} {45} (\bibinfo {year} {2025})},\
  \Eprint {https://arxiv.org/abs/2411.09032} {arXiv:2411.09032 [gr-qc]}
  \BibitemShut {NoStop}%
\bibitem [{\citenamefont {Sanchis-Gual}\ \emph {et~al.}(2017)\citenamefont
  {Sanchis-Gual}, \citenamefont {Herdeiro}, \citenamefont {Radu}, \citenamefont
  {Degollado},\ and\ \citenamefont {Font}}]{Sanchis-Gual:2017bhw}%
  \BibitemOpen
  \bibfield  {author} {\bibinfo {author} {\bibfnamefont {N.}~\bibnamefont
  {Sanchis-Gual}}, \bibinfo {author} {\bibfnamefont {C.}~\bibnamefont
  {Herdeiro}}, \bibinfo {author} {\bibfnamefont {E.}~\bibnamefont {Radu}},
  \bibinfo {author} {\bibfnamefont {J.~C.}\ \bibnamefont {Degollado}},\ and\
  \bibinfo {author} {\bibfnamefont {J.~A.}\ \bibnamefont {Font}},\ }\bibfield
  {title} {\bibinfo {title} {{Numerical evolutions of spherical Proca stars}},\
  }\href {https://doi.org/10.1103/PhysRevD.95.104028} {\bibfield  {journal}
  {\bibinfo  {journal} {Phys. Rev. D}\ }\textbf {\bibinfo {volume} {95}},\
  \bibinfo {pages} {104028} (\bibinfo {year} {2017})},\ \Eprint
  {https://arxiv.org/abs/1702.04532} {arXiv:1702.04532 [gr-qc]} \BibitemShut
  {NoStop}%
\bibitem [{\citenamefont {Sanchis-Gual}\ \emph {et~al.}(2019)\citenamefont
  {Sanchis-Gual}, \citenamefont {Herdeiro}, \citenamefont {Font}, \citenamefont
  {Radu},\ and\ \citenamefont {Di~Giovanni}}]{Sanchis-Gual:2018oui}%
  \BibitemOpen
  \bibfield  {author} {\bibinfo {author} {\bibfnamefont {N.}~\bibnamefont
  {Sanchis-Gual}}, \bibinfo {author} {\bibfnamefont {C.}~\bibnamefont
  {Herdeiro}}, \bibinfo {author} {\bibfnamefont {J.~A.}\ \bibnamefont {Font}},
  \bibinfo {author} {\bibfnamefont {E.}~\bibnamefont {Radu}},\ and\ \bibinfo
  {author} {\bibfnamefont {F.}~\bibnamefont {Di~Giovanni}},\ }\bibfield
  {title} {\bibinfo {title} {{Head-on collisions and orbital mergers of Proca
  stars}},\ }\href {https://doi.org/10.1103/PhysRevD.99.024017} {\bibfield
  {journal} {\bibinfo  {journal} {Phys. Rev. D}\ }\textbf {\bibinfo {volume}
  {99}},\ \bibinfo {pages} {024017} (\bibinfo {year} {2019})},\ \Eprint
  {https://arxiv.org/abs/1806.07779} {arXiv:1806.07779 [gr-qc]} \BibitemShut
  {NoStop}%
\bibitem [{\citenamefont {Wang}\ \emph {et~al.}(2024)\citenamefont {Wang},
  \citenamefont {Helfer},\ and\ \citenamefont {Amin}}]{Wang:2023tly}%
  \BibitemOpen
  \bibfield  {author} {\bibinfo {author} {\bibfnamefont {Z.}~\bibnamefont
  {Wang}}, \bibinfo {author} {\bibfnamefont {T.}~\bibnamefont {Helfer}},\ and\
  \bibinfo {author} {\bibfnamefont {M.~A.}\ \bibnamefont {Amin}},\ }\bibfield
  {title} {\bibinfo {title} {{General relativistic polarized Proca stars}},\
  }\href {https://doi.org/10.1103/PhysRevD.109.024019} {\bibfield  {journal}
  {\bibinfo  {journal} {Phys. Rev. D}\ }\textbf {\bibinfo {volume} {109}},\
  \bibinfo {pages} {024019} (\bibinfo {year} {2024})},\ \Eprint
  {https://arxiv.org/abs/2309.04345} {arXiv:2309.04345 [gr-qc]} \BibitemShut
  {NoStop}%
\bibitem [{\citenamefont {Calder\'on~Bustillo}\ \emph
  {et~al.}(2021)\citenamefont {Calder\'on~Bustillo}, \citenamefont
  {Sanchis-Gual}, \citenamefont {Torres-Forn\'e}, \citenamefont {Font},
  \citenamefont {Vajpeyi}, \citenamefont {Smith}, \citenamefont {Herdeiro},
  \citenamefont {Radu},\ and\ \citenamefont
  {Leong}}]{CalderonBustillo:2020fyi}%
  \BibitemOpen
  \bibfield  {author} {\bibinfo {author} {\bibfnamefont {J.}~\bibnamefont
  {Calder\'on~Bustillo}}, \bibinfo {author} {\bibfnamefont {N.}~\bibnamefont
  {Sanchis-Gual}}, \bibinfo {author} {\bibfnamefont {A.}~\bibnamefont
  {Torres-Forn\'e}}, \bibinfo {author} {\bibfnamefont {J.~A.}\ \bibnamefont
  {Font}}, \bibinfo {author} {\bibfnamefont {A.}~\bibnamefont {Vajpeyi}},
  \bibinfo {author} {\bibfnamefont {R.}~\bibnamefont {Smith}}, \bibinfo
  {author} {\bibfnamefont {C.}~\bibnamefont {Herdeiro}}, \bibinfo {author}
  {\bibfnamefont {E.}~\bibnamefont {Radu}},\ and\ \bibinfo {author}
  {\bibfnamefont {S.~H.~W.}\ \bibnamefont {Leong}},\ }\bibfield  {title}
  {\bibinfo {title} {{GW190521 as a Merger of Proca Stars: A Potential New
  Vector Boson of $8.7\times 10^{-13}$ eV}},\ }\href
  {https://doi.org/10.1103/PhysRevLett.126.081101} {\bibfield  {journal}
  {\bibinfo  {journal} {Phys. Rev. Lett.}\ }\textbf {\bibinfo {volume} {126}},\
  \bibinfo {pages} {081101} (\bibinfo {year} {2021})},\ \Eprint
  {https://arxiv.org/abs/2009.05376} {arXiv:2009.05376 [gr-qc]} \BibitemShut
  {NoStop}%
\bibitem [{\citenamefont {Herdeiro}\ \emph {et~al.}(2021)\citenamefont
  {Herdeiro}, \citenamefont {Pombo}, \citenamefont {Radu}, \citenamefont
  {Cunha},\ and\ \citenamefont {Sanchis-Gual}}]{Herdeiro:2021lwl}%
  \BibitemOpen
  \bibfield  {author} {\bibinfo {author} {\bibfnamefont {C.~A.~R.}\
  \bibnamefont {Herdeiro}}, \bibinfo {author} {\bibfnamefont {A.~M.}\
  \bibnamefont {Pombo}}, \bibinfo {author} {\bibfnamefont {E.}~\bibnamefont
  {Radu}}, \bibinfo {author} {\bibfnamefont {P.~V.~P.}\ \bibnamefont {Cunha}},\
  and\ \bibinfo {author} {\bibfnamefont {N.}~\bibnamefont {Sanchis-Gual}},\
  }\bibfield  {title} {\bibinfo {title} {{The imitation game: Proca stars that
  can mimic the Schwarzschild shadow}},\ }\href
  {https://doi.org/10.1088/1475-7516/2021/04/051} {\bibfield  {journal}
  {\bibinfo  {journal} {JCAP}\ }\textbf {\bibinfo {volume} {04}},\ \bibinfo
  {pages} {051}},\ \Eprint {https://arxiv.org/abs/2102.01703} {arXiv:2102.01703
  [gr-qc]} \BibitemShut {NoStop}%
\bibitem [{\citenamefont {Sanchis-Gual}\ \emph {et~al.}(2022)\citenamefont
  {Sanchis-Gual}, \citenamefont {Calder\'on~Bustillo}, \citenamefont
  {Herdeiro}, \citenamefont {Radu}, \citenamefont {Font}, \citenamefont
  {Leong},\ and\ \citenamefont {Torres-Forn\'e}}]{Sanchis-Gual:2022mkk}%
  \BibitemOpen
  \bibfield  {author} {\bibinfo {author} {\bibfnamefont {N.}~\bibnamefont
  {Sanchis-Gual}}, \bibinfo {author} {\bibfnamefont {J.}~\bibnamefont
  {Calder\'on~Bustillo}}, \bibinfo {author} {\bibfnamefont {C.}~\bibnamefont
  {Herdeiro}}, \bibinfo {author} {\bibfnamefont {E.}~\bibnamefont {Radu}},
  \bibinfo {author} {\bibfnamefont {J.~A.}\ \bibnamefont {Font}}, \bibinfo
  {author} {\bibfnamefont {S.~H.~W.}\ \bibnamefont {Leong}},\ and\ \bibinfo
  {author} {\bibfnamefont {A.}~\bibnamefont {Torres-Forn\'e}},\ }\bibfield
  {title} {\bibinfo {title} {{Impact of the wavelike nature of Proca stars on
  their gravitational-wave emission}},\ }\href
  {https://doi.org/10.1103/PhysRevD.106.124011} {\bibfield  {journal} {\bibinfo
   {journal} {Phys. Rev. D}\ }\textbf {\bibinfo {volume} {106}},\ \bibinfo
  {pages} {124011} (\bibinfo {year} {2022})},\ \Eprint
  {https://arxiv.org/abs/2208.11717} {arXiv:2208.11717 [gr-qc]} \BibitemShut
  {NoStop}%
\bibitem [{\citenamefont {Rosa}\ and\ \citenamefont
  {Rubiera-Garcia}(2022)}]{Rosa:2022tfv}%
  \BibitemOpen
  \bibfield  {author} {\bibinfo {author} {\bibfnamefont {J.~a.~L.}\
  \bibnamefont {Rosa}}\ and\ \bibinfo {author} {\bibfnamefont {D.}~\bibnamefont
  {Rubiera-Garcia}},\ }\bibfield  {title} {\bibinfo {title} {{Shadows of boson
  and Proca stars with thin accretion disks}},\ }\href
  {https://doi.org/10.1103/PhysRevD.106.084004} {\bibfield  {journal} {\bibinfo
   {journal} {Phys. Rev. D}\ }\textbf {\bibinfo {volume} {106}},\ \bibinfo
  {pages} {084004} (\bibinfo {year} {2022})},\ \Eprint
  {https://arxiv.org/abs/2204.12949} {arXiv:2204.12949 [gr-qc]} \BibitemShut
  {NoStop}%
\bibitem [{\citenamefont {Sengo}\ \emph {et~al.}(2024)\citenamefont {Sengo},
  \citenamefont {Cunha}, \citenamefont {Herdeiro},\ and\ \citenamefont
  {Radu}}]{Sengo:2024pwk}%
  \BibitemOpen
  \bibfield  {author} {\bibinfo {author} {\bibfnamefont {I.}~\bibnamefont
  {Sengo}}, \bibinfo {author} {\bibfnamefont {P.~V.~P.}\ \bibnamefont {Cunha}},
  \bibinfo {author} {\bibfnamefont {C.~A.~R.}\ \bibnamefont {Herdeiro}},\ and\
  \bibinfo {author} {\bibfnamefont {E.}~\bibnamefont {Radu}},\ }\bibfield
  {title} {\bibinfo {title} {{The imitation game reloaded: effective shadows of
  dynamically robust spinning Proca stars}},\ }\href
  {https://doi.org/10.1088/1475-7516/2024/05/054} {\bibfield  {journal}
  {\bibinfo  {journal} {JCAP}\ }\textbf {\bibinfo {volume} {05}},\ \bibinfo
  {pages} {054}},\ \Eprint {https://arxiv.org/abs/2402.14919} {arXiv:2402.14919
  [gr-qc]} \BibitemShut {NoStop}%
\bibitem [{\citenamefont {Su\'arez}\ \emph {et~al.}(2014)\citenamefont
  {Su\'arez}, \citenamefont {Robles},\ and\ \citenamefont
  {Matos}}]{Suarez:2013iw}%
  \BibitemOpen
  \bibfield  {author} {\bibinfo {author} {\bibfnamefont {A.}~\bibnamefont
  {Su\'arez}}, \bibinfo {author} {\bibfnamefont {V.~H.}\ \bibnamefont
  {Robles}},\ and\ \bibinfo {author} {\bibfnamefont {T.}~\bibnamefont
  {Matos}},\ }\bibfield  {title} {\bibinfo {title} {{A Review on the Scalar
  Field/Bose-Einstein Condensate Dark Matter Model}},\ }\href
  {https://doi.org/10.1007/978-3-319-02063-1_9} {\bibfield  {journal} {\bibinfo
   {journal} {Astrophys. Space Sci. Proc.}\ }\textbf {\bibinfo {volume} {38}},\
  \bibinfo {pages} {107} (\bibinfo {year} {2014})},\ \Eprint
  {https://arxiv.org/abs/1302.0903} {arXiv:1302.0903 [astro-ph.CO]}
  \BibitemShut {NoStop}%
\bibitem [{\citenamefont {Marsh}(2016)}]{Marsh:2015xka}%
  \BibitemOpen
  \bibfield  {author} {\bibinfo {author} {\bibfnamefont {D.~J.~E.}\
  \bibnamefont {Marsh}},\ }\bibfield  {title} {\bibinfo {title} {{Axion
  Cosmology}},\ }\href {https://doi.org/10.1016/j.physrep.2016.06.005}
  {\bibfield  {journal} {\bibinfo  {journal} {Phys. Rept.}\ }\textbf {\bibinfo
  {volume} {643}},\ \bibinfo {pages} {1} (\bibinfo {year} {2016})},\ \Eprint
  {https://arxiv.org/abs/1510.07633} {arXiv:1510.07633 [astro-ph.CO]}
  \BibitemShut {NoStop}%
\bibitem [{\citenamefont {Ure\~na L\'opez}(2019)}]{Urena-Lopez:2019kud}%
  \BibitemOpen
  \bibfield  {author} {\bibinfo {author} {\bibfnamefont {L.~A.}\ \bibnamefont
  {Ure\~na L\'opez}},\ }\bibfield  {title} {\bibinfo {title} {{Brief Review on
  Scalar Field Dark Matter Models}},\ }\href
  {https://doi.org/10.3389/fspas.2019.00047} {\bibfield  {journal} {\bibinfo
  {journal} {Front. Astron. Space Sci.}\ }\textbf {\bibinfo {volume} {6}},\
  \bibinfo {pages} {47} (\bibinfo {year} {2019})}\BibitemShut {NoStop}%
\bibitem [{\citenamefont {Ferreira}(2021)}]{Ferreira:2020fam}%
  \BibitemOpen
  \bibfield  {author} {\bibinfo {author} {\bibfnamefont {E.~G.~M.}\
  \bibnamefont {Ferreira}},\ }\bibfield  {title} {\bibinfo {title}
  {{Ultra-light dark matter}},\ }\href
  {https://doi.org/10.1007/s00159-021-00135-6} {\bibfield  {journal} {\bibinfo
  {journal} {Astron. Astrophys. Rev.}\ }\textbf {\bibinfo {volume} {29}},\
  \bibinfo {pages} {7} (\bibinfo {year} {2021})},\ \Eprint
  {https://arxiv.org/abs/2005.03254} {arXiv:2005.03254 [astro-ph.CO]}
  \BibitemShut {NoStop}%
\bibitem [{\citenamefont {Antypas}\ \emph {et~al.}(2022)\citenamefont {Antypas}
  \emph {et~al.}}]{Antypas:2022asj}%
  \BibitemOpen
  \bibfield  {author} {\bibinfo {author} {\bibfnamefont {D.}~\bibnamefont
  {Antypas}} \emph {et~al.},\ }\bibfield  {title} {\bibinfo {title} {{New
  Horizons: Scalar and Vector Ultralight Dark Matter}},\ }\href@noop {} {\
  (\bibinfo {year} {2022})},\ \Eprint {https://arxiv.org/abs/2203.14915}
  {arXiv:2203.14915 [hep-ex]} \BibitemShut {NoStop}%
\bibitem [{\citenamefont {Jain}\ and\ \citenamefont
  {Amin}(2022)}]{Jain:2021pnk}%
  \BibitemOpen
  \bibfield  {author} {\bibinfo {author} {\bibfnamefont {M.}~\bibnamefont
  {Jain}}\ and\ \bibinfo {author} {\bibfnamefont {M.~A.}\ \bibnamefont
  {Amin}},\ }\bibfield  {title} {\bibinfo {title} {{Polarized solitons in
  higher-spin wave dark matter}},\ }\href
  {https://doi.org/10.1103/PhysRevD.105.056019} {\bibfield  {journal} {\bibinfo
   {journal} {Phys. Rev. D}\ }\textbf {\bibinfo {volume} {105}},\ \bibinfo
  {pages} {056019} (\bibinfo {year} {2022})},\ \Eprint
  {https://arxiv.org/abs/2109.04892} {arXiv:2109.04892 [hep-th]} \BibitemShut
  {NoStop}%
\bibitem [{\citenamefont {Zhang}\ \emph {et~al.}(2022)\citenamefont {Zhang},
  \citenamefont {Jain},\ and\ \citenamefont {Amin}}]{Zhang:2021xxa}%
  \BibitemOpen
  \bibfield  {author} {\bibinfo {author} {\bibfnamefont {H.-Y.}\ \bibnamefont
  {Zhang}}, \bibinfo {author} {\bibfnamefont {M.}~\bibnamefont {Jain}},\ and\
  \bibinfo {author} {\bibfnamefont {M.~A.}\ \bibnamefont {Amin}},\ }\bibfield
  {title} {\bibinfo {title} {{Polarized vector oscillons}},\ }\href
  {https://doi.org/10.1103/PhysRevD.105.096037} {\bibfield  {journal} {\bibinfo
   {journal} {Phys. Rev. D}\ }\textbf {\bibinfo {volume} {105}},\ \bibinfo
  {pages} {096037} (\bibinfo {year} {2022})},\ \Eprint
  {https://arxiv.org/abs/2111.08700} {arXiv:2111.08700 [astro-ph.CO]}
  \BibitemShut {NoStop}%
\bibitem [{\citenamefont {Amin}\ \emph {et~al.}(2022)\citenamefont {Amin},
  \citenamefont {Jain}, \citenamefont {Karur},\ and\ \citenamefont
  {Mocz}}]{Amin:2022pzv}%
  \BibitemOpen
  \bibfield  {author} {\bibinfo {author} {\bibfnamefont {M.~A.}\ \bibnamefont
  {Amin}}, \bibinfo {author} {\bibfnamefont {M.}~\bibnamefont {Jain}}, \bibinfo
  {author} {\bibfnamefont {R.}~\bibnamefont {Karur}},\ and\ \bibinfo {author}
  {\bibfnamefont {P.}~\bibnamefont {Mocz}},\ }\bibfield  {title} {\bibinfo
  {title} {{Small-scale structure in vector dark matter}},\ }\href
  {https://doi.org/10.1088/1475-7516/2022/08/014} {\bibfield  {journal}
  {\bibinfo  {journal} {JCAP}\ }\textbf {\bibinfo {volume} {08}}\bibfield
  {number} {\bibinfo  {number} { (08)},\ \bibinfo {pages} {014}},\ }\Eprint
  {https://arxiv.org/abs/2203.11935} {arXiv:2203.11935 [astro-ph.CO]}
  \BibitemShut {NoStop}%
\bibitem [{\citenamefont {Jain}\ and\ \citenamefont
  {Amin}(2023)}]{Jain:2022agt}%
  \BibitemOpen
  \bibfield  {author} {\bibinfo {author} {\bibfnamefont {M.}~\bibnamefont
  {Jain}}\ and\ \bibinfo {author} {\bibfnamefont {M.~A.}\ \bibnamefont
  {Amin}},\ }\bibfield  {title} {\bibinfo {title} {{i-SPin: an integrator for
  multicomponent Schr\"odinger-Poisson systems with self-interactions}},\
  }\href {https://doi.org/10.1088/1475-7516/2023/04/053} {\bibfield  {journal}
  {\bibinfo  {journal} {JCAP}\ }\textbf {\bibinfo {volume} {04}},\ \bibinfo
  {pages} {053}},\ \Eprint {https://arxiv.org/abs/2211.08433} {arXiv:2211.08433
  [astro-ph.CO]} \BibitemShut {NoStop}%
\bibitem [{\citenamefont {Jain}\ \emph {et~al.}(2024)\citenamefont {Jain},
  \citenamefont {Wanichwecharungruang},\ and\ \citenamefont
  {Thomas}}]{Jain:2023tsr}%
  \BibitemOpen
  \bibfield  {author} {\bibinfo {author} {\bibfnamefont {M.}~\bibnamefont
  {Jain}}, \bibinfo {author} {\bibfnamefont {W.}~\bibnamefont
  {Wanichwecharungruang}},\ and\ \bibinfo {author} {\bibfnamefont
  {J.}~\bibnamefont {Thomas}},\ }\bibfield  {title} {\bibinfo {title} {{Kinetic
  relaxation and nucleation of Bose stars in self-interacting wave dark
  matter}},\ }\href {https://doi.org/10.1103/PhysRevD.109.016002} {\bibfield
  {journal} {\bibinfo  {journal} {Phys. Rev. D}\ }\textbf {\bibinfo {volume}
  {109}},\ \bibinfo {pages} {016002} (\bibinfo {year} {2024})},\ \Eprint
  {https://arxiv.org/abs/2310.00058} {arXiv:2310.00058 [astro-ph.CO]}
  \BibitemShut {NoStop}%
\bibitem [{\citenamefont {Adshead}\ and\ \citenamefont
  {Lozanov}(2021)}]{Adshead:2021kvl}%
  \BibitemOpen
  \bibfield  {author} {\bibinfo {author} {\bibfnamefont {P.}~\bibnamefont
  {Adshead}}\ and\ \bibinfo {author} {\bibfnamefont {K.~D.}\ \bibnamefont
  {Lozanov}},\ }\bibfield  {title} {\bibinfo {title} {{Self-gravitating Vector
  Dark Matter}},\ }\href {https://doi.org/10.1103/PhysRevD.103.103501}
  {\bibfield  {journal} {\bibinfo  {journal} {Phys. Rev. D}\ }\textbf {\bibinfo
  {volume} {103}},\ \bibinfo {pages} {103501} (\bibinfo {year} {2021})},\
  \Eprint {https://arxiv.org/abs/2101.07265} {arXiv:2101.07265 [gr-qc]}
  \BibitemShut {NoStop}%
\bibitem [{\citenamefont {Gorghetto}\ \emph {et~al.}(2022)\citenamefont
  {Gorghetto}, \citenamefont {Hardy}, \citenamefont {March-Russell},
  \citenamefont {Song},\ and\ \citenamefont {West}}]{Gorghetto:2022sue}%
  \BibitemOpen
  \bibfield  {author} {\bibinfo {author} {\bibfnamefont {M.}~\bibnamefont
  {Gorghetto}}, \bibinfo {author} {\bibfnamefont {E.}~\bibnamefont {Hardy}},
  \bibinfo {author} {\bibfnamefont {J.}~\bibnamefont {March-Russell}}, \bibinfo
  {author} {\bibfnamefont {N.}~\bibnamefont {Song}},\ and\ \bibinfo {author}
  {\bibfnamefont {S.~M.}\ \bibnamefont {West}},\ }\bibfield  {title} {\bibinfo
  {title} {{Dark photon stars: formation and role as dark matter
  substructure}},\ }\href {https://doi.org/10.1088/1475-7516/2022/08/018}
  {\bibfield  {journal} {\bibinfo  {journal} {JCAP}\ }\textbf {\bibinfo
  {volume} {08}}\bibfield  {number} {\bibinfo  {number} { (08)},\ \bibinfo
  {pages} {018}},\ }\Eprint {https://arxiv.org/abs/2203.10100}
  {arXiv:2203.10100 [hep-ph]} \BibitemShut {NoStop}%
\bibitem [{\citenamefont {Chen}\ \emph {et~al.}(2024)\citenamefont {Chen},
  \citenamefont {Nguyen},\ and\ \citenamefont {Marsh}}]{Chen:2024vgh}%
  \BibitemOpen
  \bibfield  {author} {\bibinfo {author} {\bibfnamefont {J.}~\bibnamefont
  {Chen}}, \bibinfo {author} {\bibfnamefont {L.~H.}\ \bibnamefont {Nguyen}},\
  and\ \bibinfo {author} {\bibfnamefont {D.~J.~E.}\ \bibnamefont {Marsh}},\
  }\bibfield  {title} {\bibinfo {title} {{Vector Dark Matter Halo: From
  Polarization Dynamics to Direct Detection}},\ }\href@noop {} {\  (\bibinfo
  {year} {2024})},\ \Eprint {https://arxiv.org/abs/2407.17315}
  {arXiv:2407.17315 [astro-ph.CO]} \BibitemShut {NoStop}%
\bibitem [{\citenamefont {Zhang}(2023)}]{Zhang:2023ktk}%
  \BibitemOpen
  \bibfield  {author} {\bibinfo {author} {\bibfnamefont {H.-Y.}\ \bibnamefont
  {Zhang}},\ }\emph {\bibinfo {title} {{Probing ultralight dark fields in
  cosmological and astrophysical systems}}},\ \href@noop {} {Ph.D. thesis},\
  \bibinfo  {school} {Rice U.} (\bibinfo {year} {2023}),\ \Eprint
  {https://arxiv.org/abs/2401.00043} {arXiv:2401.00043 [hep-ph]} \BibitemShut
  {NoStop}%
\bibitem [{\citenamefont {Zhang}(2024)}]{Zhang:2024bjo}%
  \BibitemOpen
  \bibfield  {author} {\bibinfo {author} {\bibfnamefont {H.-Y.}\ \bibnamefont
  {Zhang}},\ }\bibfield  {title} {\bibinfo {title} {{Unified view of scalar and
  vector dark matter solitons}},\ }\href@noop {} {\  (\bibinfo {year}
  {2024})},\ \Eprint {https://arxiv.org/abs/2406.05031} {arXiv:2406.05031
  [hep-ph]} \BibitemShut {NoStop}%
\bibitem [{Movies showing the time evolution of the four Proca stars depicted
  in Fig~\ref{fig.polarizations}()}]{youtube}%
  \BibitemOpen
  Movies showing the time evolution of the four Proca stars depicted in
  Fig~\ref{fig.polarizations},\ \href@noop {} {}\bibinfo {howpublished}
  {\url{https://youtube.com/playlist?list=PLtSeM9Y95qPbmOVgqq1Gy0Kag8DJ7ZKZB&si=hg_tYzoFmDI6KwGN}}
  (\bibinfo {year} {2024})\BibitemShut {NoStop}%
\bibitem [{\citenamefont {Roque}\ \emph {et~al.}(2023)\citenamefont {Roque},
  \citenamefont {Ch\'avez~Nambo},\ and\ \citenamefont
  {Sarbach}}]{Roque:2023sjl}%
  \BibitemOpen
  \bibfield  {author} {\bibinfo {author} {\bibfnamefont {A.~A.}\ \bibnamefont
  {Roque}}, \bibinfo {author} {\bibfnamefont {E.}~\bibnamefont
  {Ch\'avez~Nambo}},\ and\ \bibinfo {author} {\bibfnamefont {O.}~\bibnamefont
  {Sarbach}},\ }\bibfield  {title} {\bibinfo {title} {{Radial linear stability
  of nonrelativistic \ensuremath{\ell}-boson stars}},\ }\href
  {https://doi.org/10.1103/PhysRevD.107.084001} {\bibfield  {journal} {\bibinfo
   {journal} {Phys. Rev. D}\ }\textbf {\bibinfo {volume} {107}},\ \bibinfo
  {pages} {084001} (\bibinfo {year} {2023})},\ \Eprint
  {https://arxiv.org/abs/2302.00717} {arXiv:2302.00717 [gr-qc]} \BibitemShut
  {NoStop}%
\bibitem [{\citenamefont {Ch\'avez~Nambo}(2021)}]{NamboThesis}%
  \BibitemOpen
  \bibfield  {author} {\bibinfo {author} {\bibfnamefont {E.}~\bibnamefont
  {Ch\'avez~Nambo}},\ }\emph {\bibinfo {title} {Sobre la existencia de
  estrellas de bosones newtonianas con momento angular en simetr\'ia
  esf\'erica}},\ \href@noop {} {Master's thesis},\ \bibinfo  {school}
  {Universidad Michoacana de San Nicol\'as de Hidalgo} (\bibinfo {year}
  {2021})\BibitemShut {NoStop}%
\bibitem [{\citenamefont {Matos}\ and\ \citenamefont
  {Urena-Lopez}(2007)}]{Matos:2007zza}%
  \BibitemOpen
  \bibfield  {author} {\bibinfo {author} {\bibfnamefont {T.}~\bibnamefont
  {Matos}}\ and\ \bibinfo {author} {\bibfnamefont {L.~A.}\ \bibnamefont
  {Urena-Lopez}},\ }\bibfield  {title} {\bibinfo {title} {{Flat rotation curves
  in scalar field galaxy halos}},\ }\href
  {https://doi.org/10.1007/s10714-007-0470-y} {\bibfield  {journal} {\bibinfo
  {journal} {Gen. Rel. Grav.}\ }\textbf {\bibinfo {volume} {39}},\ \bibinfo
  {pages} {1279} (\bibinfo {year} {2007})}\BibitemShut {NoStop}%
\bibitem [{\citenamefont {Ch\'avez~Nambo}\ \emph {et~al.}(2024)\citenamefont
  {Ch\'avez~Nambo}, \citenamefont {Diez-Tejedor}, \citenamefont {Roque},\ and\
  \citenamefont {Sarbach}}]{Nambo:2024gvs}%
  \BibitemOpen
  \bibfield  {author} {\bibinfo {author} {\bibfnamefont {E.}~\bibnamefont
  {Ch\'avez~Nambo}}, \bibinfo {author} {\bibfnamefont {A.}~\bibnamefont
  {Diez-Tejedor}}, \bibinfo {author} {\bibfnamefont {A.~A.}\ \bibnamefont
  {Roque}},\ and\ \bibinfo {author} {\bibfnamefont {O.}~\bibnamefont
  {Sarbach}},\ }\bibfield  {title} {\bibinfo {title} {{Linear stability of
  nonrelativistic self-interacting boson stars}},\ }\href
  {https://doi.org/10.1103/PhysRevD.109.104011} {\bibfield  {journal} {\bibinfo
   {journal} {Phys. Rev. D}\ }\textbf {\bibinfo {volume} {109}},\ \bibinfo
  {pages} {104011} (\bibinfo {year} {2024})},\ \Eprint
  {https://arxiv.org/abs/2402.07998} {arXiv:2402.07998 [gr-qc]} \BibitemShut
  {NoStop}%
\bibitem [{\citenamefont {Lieb}(1977)}]{Lieb1977}%
  \BibitemOpen
  \bibfield  {author} {\bibinfo {author} {\bibfnamefont {E.~H.}\ \bibnamefont
  {Lieb}},\ }\bibfield  {title} {\bibinfo {title} {{Existence and Uniqueness of
  the Minimizing Solution of Choquard's Nonlinear Equation}},\ }\href
  {https://doi.org/https://doi.org/10.1002/sapm197757293} {\bibfield  {journal}
  {\bibinfo  {journal} {Studies in Applied Mathematics}\ }\textbf {\bibinfo
  {volume} {57}},\ \bibinfo {pages} {93} (\bibinfo {year} {1977})}\BibitemShut
  {NoStop}%
\bibitem [{\citenamefont {Cazenave}\ and\ \citenamefont
  {Lions}(2017)}]{Cazenave1982}%
  \BibitemOpen
  \bibfield  {author} {\bibinfo {author} {\bibfnamefont {T.}~\bibnamefont
  {Cazenave}}\ and\ \bibinfo {author} {\bibfnamefont {P.~L.}\ \bibnamefont
  {Lions}},\ }\bibfield  {title} {\bibinfo {title} {{Orbital stability of
  standing waves for some nonlinear Schr\"odinger equations}},\ }\href
  {https://doi.org/10.1007/BF01403504} {\bibfield  {journal} {\bibinfo
  {journal} {Communications in Mathematical Physics}\ }\textbf {\bibinfo
  {volume} {85}},\ \bibinfo {pages} {549–561} (\bibinfo {year}
  {2017})}\BibitemShut {NoStop}%
\bibitem [{\citenamefont {Lions}(1984{\natexlab{a}})}]{LIONS1984109}%
  \BibitemOpen
  \bibfield  {author} {\bibinfo {author} {\bibfnamefont {P.}~\bibnamefont
  {Lions}},\ }\bibfield  {title} {\bibinfo {title} {The
  concentration-compactness principle in the calculus of variations. {T}he
  locally compact case, part 1},\ }\href
  {https://doi.org/https://doi.org/10.1016/S0294-1449(16)30428-0} {\bibfield
  {journal} {\bibinfo  {journal} {Annales de l'Institut Henri Poincaré C,
  Analyse non linéaire}\ }\textbf {\bibinfo {volume} {1}},\ \bibinfo {pages}
  {109} (\bibinfo {year} {1984}{\natexlab{a}})}\BibitemShut {NoStop}%
\bibitem [{\citenamefont {Lions}(1984{\natexlab{b}})}]{LIONS1984223}%
  \BibitemOpen
  \bibfield  {author} {\bibinfo {author} {\bibfnamefont {P.}~\bibnamefont
  {Lions}},\ }\bibfield  {title} {\bibinfo {title} {The
  concentration-compactness principle in the calculus of variations. {T}he
  locally compact case, part 2},\ }\href
  {https://doi.org/https://doi.org/10.1016/S0294-1449(16)30422-X} {\bibfield
  {journal} {\bibinfo  {journal} {Annales de l'Institut Henri Poincaré C,
  Analyse non linéaire}\ }\textbf {\bibinfo {volume} {1}},\ \bibinfo {pages}
  {223} (\bibinfo {year} {1984}{\natexlab{b}})}\BibitemShut {NoStop}%
\bibitem [{\citenamefont {Simon}(2005)}]{Simon2005}%
  \BibitemOpen
  \bibfield  {author} {\bibinfo {author} {\bibfnamefont {B.}~\bibnamefont
  {Simon}},\ }\bibinfo {title} {Sturm oscillation and comparison theorems},\
  in\ \href {https://doi.org/10.1007/3-7643-7359-8_2} {\emph {\bibinfo
  {booktitle} {Sturm-Liouville Theory: Past and Present}}},\ \bibinfo {editor}
  {edited by\ \bibinfo {editor} {\bibfnamefont {W.~O.}\ \bibnamefont {Amrein}},
  \bibinfo {editor} {\bibfnamefont {A.~M.}\ \bibnamefont {Hinz}},\ and\
  \bibinfo {editor} {\bibfnamefont {D.~P.}\ \bibnamefont {Pearson}}}\ (\bibinfo
   {publisher} {Birkh{\"a}user Basel},\ \bibinfo {address} {Basel},\ \bibinfo
  {year} {2005})\ pp.\ \bibinfo {pages} {29--43}\BibitemShut {NoStop}%
\bibitem [{\citenamefont {Chavanis}(2011)}]{Chavanis:2011zi}%
  \BibitemOpen
  \bibfield  {author} {\bibinfo {author} {\bibfnamefont {P.-H.}\ \bibnamefont
  {Chavanis}},\ }\bibfield  {title} {\bibinfo {title} {{Mass-radius relation of
  Newtonian self-gravitating Bose-Einstein condensates with short-range
  interactions: I. Analytical results}},\ }\href
  {https://doi.org/10.1103/PhysRevD.84.043531} {\bibfield  {journal} {\bibinfo
  {journal} {Phys. Rev. D}\ }\textbf {\bibinfo {volume} {84}},\ \bibinfo
  {pages} {043531} (\bibinfo {year} {2011})},\ \Eprint
  {https://arxiv.org/abs/1103.2050} {arXiv:1103.2050 [astro-ph.CO]}
  \BibitemShut {NoStop}%
\bibitem [{\citenamefont {Chavanis}\ and\ \citenamefont
  {Delfini}(2011)}]{Chavanis:2011zm}%
  \BibitemOpen
  \bibfield  {author} {\bibinfo {author} {\bibfnamefont {P.~H.}\ \bibnamefont
  {Chavanis}}\ and\ \bibinfo {author} {\bibfnamefont {L.}~\bibnamefont
  {Delfini}},\ }\bibfield  {title} {\bibinfo {title} {{Mass-radius relation of
  Newtonian self-gravitating Bose-Einstein condensates with short-range
  interactions: II. Numerical results}},\ }\href
  {https://doi.org/10.1103/PhysRevD.84.043532} {\bibfield  {journal} {\bibinfo
  {journal} {Phys. Rev. D}\ }\textbf {\bibinfo {volume} {84}},\ \bibinfo
  {pages} {043532} (\bibinfo {year} {2011})},\ \Eprint
  {https://arxiv.org/abs/1103.2054} {arXiv:1103.2054 [astro-ph.CO]}
  \BibitemShut {NoStop}%
\bibitem [{\citenamefont {Li}\ \emph {et~al.}(2021)\citenamefont {Li},
  \citenamefont {Zeng}, \citenamefont {Song},\ and\ \citenamefont
  {Wang}}]{Li:2020ffy}%
  \BibitemOpen
  \bibfield  {author} {\bibinfo {author} {\bibfnamefont {H.-B.}\ \bibnamefont
  {Li}}, \bibinfo {author} {\bibfnamefont {Y.-B.}\ \bibnamefont {Zeng}},
  \bibinfo {author} {\bibfnamefont {Y.}~\bibnamefont {Song}},\ and\ \bibinfo
  {author} {\bibfnamefont {Y.-Q.}\ \bibnamefont {Wang}},\ }\bibfield  {title}
  {\bibinfo {title} {{Self-interacting multistate boson stars}},\ }\href
  {https://doi.org/10.1007/JHEP04(2021)042} {\bibfield  {journal} {\bibinfo
  {journal} {JHEP}\ }\textbf {\bibinfo {volume} {04}},\ \bibinfo {pages}
  {042}},\ \Eprint {https://arxiv.org/abs/2006.11281} {arXiv:2006.11281
  [gr-qc]} \BibitemShut {NoStop}%
\bibitem [{\citenamefont {Virtanen}\ \emph {et~al.}(2020)\citenamefont
  {Virtanen}, \citenamefont {Gommers},\ and\ \citenamefont
  {et~al.}}]{2020SciPy-NMeth}%
  \BibitemOpen
  \bibfield  {author} {\bibinfo {author} {\bibfnamefont {P.}~\bibnamefont
  {Virtanen}}, \bibinfo {author} {\bibfnamefont {R.}~\bibnamefont {Gommers}},\
  and\ \bibinfo {author} {\bibnamefont {et~al.}},\ }\bibfield  {title}
  {\bibinfo {title} {{SciPy 1.0: Fundamental Algorithms for Scientific
  Computing in Python}},\ }\href {https://doi.org/10.1038/s41592-019-0686-2}
  {\bibfield  {journal} {\bibinfo  {journal} {Nature Methods}\ }\textbf
  {\bibinfo {volume} {17}},\ \bibinfo {pages} {261} (\bibinfo {year}
  {2020})}\BibitemShut {NoStop}%
\bibitem [{\citenamefont {Dormand}\ and\ \citenamefont
  {Prince}(1980)}]{DORMAND198019}%
  \BibitemOpen
  \bibfield  {author} {\bibinfo {author} {\bibfnamefont {J.}~\bibnamefont
  {Dormand}}\ and\ \bibinfo {author} {\bibfnamefont {P.}~\bibnamefont
  {Prince}},\ }\bibfield  {title} {\bibinfo {title} {{A family of embedded
  Runge-Kutta formulae}},\ }\href
  {https://doi.org/10.1016/0771-050X(80)90013-3} {\bibfield  {journal}
  {\bibinfo  {journal} {Journal of Computational and Applied Mathematics}\
  }\textbf {\bibinfo {volume} {6}},\ \bibinfo {pages} {19} (\bibinfo {year}
  {1980})}\BibitemShut {NoStop}%
\bibitem [{\citenamefont {Lawrence}(1986)}]{Lawrence1986SomePR}%
  \BibitemOpen
  \bibfield  {author} {\bibinfo {author} {\bibfnamefont {F.~S.}\ \bibnamefont
  {Lawrence}},\ }\bibfield  {title} {\bibinfo {title} {{Some practical
  Runge-Kutta formulas}},\ }\href@noop {} {\bibfield  {journal} {\bibinfo
  {journal} {Mathematics of Computation}\ }\textbf {\bibinfo {volume} {46}},\
  \bibinfo {pages} {135} (\bibinfo {year} {1986})}\BibitemShut {NoStop}%
\bibitem [{\citenamefont {Merzbacher}(1998)}]{Merzbacher}%
  \BibitemOpen
  \bibfield  {author} {\bibinfo {author} {\bibfnamefont {E.}~\bibnamefont
  {Merzbacher}},\ }\href@noop {} {\emph {\bibinfo {title} {{Quantum
  Mechanics}}}}\ (\bibinfo  {publisher} {Wiley},\ \bibinfo {address} {USA},\
  \bibinfo {year} {1998})\BibitemShut {NoStop}%
\bibitem [{\citenamefont {Clough}\ \emph {et~al.}(2022)\citenamefont {Clough},
  \citenamefont {Helfer}, \citenamefont {Witek},\ and\ \citenamefont
  {Berti}}]{Clough:2022ygm}%
  \BibitemOpen
  \bibfield  {author} {\bibinfo {author} {\bibfnamefont {K.}~\bibnamefont
  {Clough}}, \bibinfo {author} {\bibfnamefont {T.}~\bibnamefont {Helfer}},
  \bibinfo {author} {\bibfnamefont {H.}~\bibnamefont {Witek}},\ and\ \bibinfo
  {author} {\bibfnamefont {E.}~\bibnamefont {Berti}},\ }\bibfield  {title}
  {\bibinfo {title} {{Ghost Instabilities in Self-Interacting Vector Fields:
  The Problem with Proca Fields}},\ }\href
  {https://doi.org/10.1103/PhysRevLett.129.151102} {\bibfield  {journal}
  {\bibinfo  {journal} {Phys. Rev. Lett.}\ }\textbf {\bibinfo {volume} {129}},\
  \bibinfo {pages} {151102} (\bibinfo {year} {2022})},\ \Eprint
  {https://arxiv.org/abs/2204.10868} {arXiv:2204.10868 [gr-qc]} \BibitemShut
  {NoStop}%
\bibitem [{\citenamefont {Mou}\ and\ \citenamefont
  {Zhang}(2022)}]{Mou:2022hqb}%
  \BibitemOpen
  \bibfield  {author} {\bibinfo {author} {\bibfnamefont {Z.-G.}\ \bibnamefont
  {Mou}}\ and\ \bibinfo {author} {\bibfnamefont {H.-Y.}\ \bibnamefont
  {Zhang}},\ }\bibfield  {title} {\bibinfo {title} {{Singularity Problem for
  Interacting Massive Vectors}},\ }\href
  {https://doi.org/10.1103/PhysRevLett.129.151101} {\bibfield  {journal}
  {\bibinfo  {journal} {Phys. Rev. Lett.}\ }\textbf {\bibinfo {volume} {129}},\
  \bibinfo {pages} {151101} (\bibinfo {year} {2022})},\ \Eprint
  {https://arxiv.org/abs/2204.11324} {arXiv:2204.11324 [hep-th]} \BibitemShut
  {NoStop}%
\bibitem [{\citenamefont {Coates}\ and\ \citenamefont
  {Ramazano\u{g}lu}(2022)}]{Coates:2022qia}%
  \BibitemOpen
  \bibfield  {author} {\bibinfo {author} {\bibfnamefont {A.}~\bibnamefont
  {Coates}}\ and\ \bibinfo {author} {\bibfnamefont {F.~M.}\ \bibnamefont
  {Ramazano\u{g}lu}},\ }\bibfield  {title} {\bibinfo {title} {{Intrinsic
  Pathology of Self-Interacting Vector Fields}},\ }\href
  {https://doi.org/10.1103/PhysRevLett.129.151103} {\bibfield  {journal}
  {\bibinfo  {journal} {Phys. Rev. Lett.}\ }\textbf {\bibinfo {volume} {129}},\
  \bibinfo {pages} {151103} (\bibinfo {year} {2022})},\ \Eprint
  {https://arxiv.org/abs/2205.07784} {arXiv:2205.07784 [gr-qc]} \BibitemShut
  {NoStop}%
\bibitem [{\citenamefont {Barausse}\ \emph {et~al.}(2022)\citenamefont
  {Barausse}, \citenamefont {Bezares}, \citenamefont {Crisostomi},\ and\
  \citenamefont {Lara}}]{Barausse:2022rvg}%
  \BibitemOpen
  \bibfield  {author} {\bibinfo {author} {\bibfnamefont {E.}~\bibnamefont
  {Barausse}}, \bibinfo {author} {\bibfnamefont {M.}~\bibnamefont {Bezares}},
  \bibinfo {author} {\bibfnamefont {M.}~\bibnamefont {Crisostomi}},\ and\
  \bibinfo {author} {\bibfnamefont {G.}~\bibnamefont {Lara}},\ }\bibfield
  {title} {\bibinfo {title} {{The well-posedness of the Cauchy problem for
  self-interacting vector fields}},\ }\href
  {https://doi.org/10.1088/1475-7516/2022/11/050} {\bibfield  {journal}
  {\bibinfo  {journal} {JCAP}\ }\textbf {\bibinfo {volume} {11}},\ \bibinfo
  {pages} {050}},\ \Eprint {https://arxiv.org/abs/2207.00443} {arXiv:2207.00443
  [gr-qc]} \BibitemShut {NoStop}%
\bibitem [{\citenamefont {Aoki}\ and\ \citenamefont
  {Minamitsuji}(2022)}]{Aoki:2022woy}%
  \BibitemOpen
  \bibfield  {author} {\bibinfo {author} {\bibfnamefont {K.}~\bibnamefont
  {Aoki}}\ and\ \bibinfo {author} {\bibfnamefont {M.}~\bibnamefont
  {Minamitsuji}},\ }\bibfield  {title} {\bibinfo {title} {{Resolving the
  pathologies of self-interacting Proca fields: A case study of Proca stars}},\
  }\href {https://doi.org/10.1103/PhysRevD.106.084022} {\bibfield  {journal}
  {\bibinfo  {journal} {Phys. Rev. D}\ }\textbf {\bibinfo {volume} {106}},\
  \bibinfo {pages} {084022} (\bibinfo {year} {2022})},\ \Eprint
  {https://arxiv.org/abs/2206.14320} {arXiv:2206.14320 [gr-qc]} \BibitemShut
  {NoStop}%
\bibitem [{\citenamefont {Rubio}\ \emph {et~al.}(2024)\citenamefont {Rubio},
  \citenamefont {Lara}, \citenamefont {Bezares}, \citenamefont {Crisostomi},\
  and\ \citenamefont {Barausse}}]{Rubio:2024ryv}%
  \BibitemOpen
  \bibfield  {author} {\bibinfo {author} {\bibfnamefont {M.~E.}\ \bibnamefont
  {Rubio}}, \bibinfo {author} {\bibfnamefont {G.}~\bibnamefont {Lara}},
  \bibinfo {author} {\bibfnamefont {M.}~\bibnamefont {Bezares}}, \bibinfo
  {author} {\bibfnamefont {M.}~\bibnamefont {Crisostomi}},\ and\ \bibinfo
  {author} {\bibfnamefont {E.}~\bibnamefont {Barausse}},\ }\bibfield  {title}
  {\bibinfo {title} {{Fixing the dynamical evolution of self-interacting vector
  fields}},\ }\href {https://doi.org/10.1103/PhysRevD.110.063015} {\bibfield
  {journal} {\bibinfo  {journal} {Phys. Rev. D}\ }\textbf {\bibinfo {volume}
  {110}},\ \bibinfo {pages} {063015} (\bibinfo {year} {2024})},\ \Eprint
  {https://arxiv.org/abs/2407.08774} {arXiv:2407.08774 [gr-qc]} \BibitemShut
  {NoStop}%
\bibitem [{\citenamefont {Ch\'avez~Nambo}\ \emph {et~al.}(2023)\citenamefont
  {Ch\'avez~Nambo}, \citenamefont {Roque},\ and\ \citenamefont
  {Sarbach}}]{Nambo:2023yut}%
  \BibitemOpen
  \bibfield  {author} {\bibinfo {author} {\bibfnamefont {E.}~\bibnamefont
  {Ch\'avez~Nambo}}, \bibinfo {author} {\bibfnamefont {A.~A.}\ \bibnamefont
  {Roque}},\ and\ \bibinfo {author} {\bibfnamefont {O.}~\bibnamefont
  {Sarbach}},\ }\bibfield  {title} {\bibinfo {title} {{Are nonrelativistic
  ground state \ensuremath{\ell}-boson stars only stable for
  \ensuremath{\ell}=0 and \ensuremath{\ell}=1?}},\ }\href
  {https://doi.org/10.1103/PhysRevD.108.124065} {\bibfield  {journal} {\bibinfo
   {journal} {Phys. Rev. D}\ }\textbf {\bibinfo {volume} {108}},\ \bibinfo
  {pages} {124065} (\bibinfo {year} {2023})},\ \Eprint
  {https://arxiv.org/abs/2310.18405} {arXiv:2310.18405 [gr-qc]} \BibitemShut
  {NoStop}%
\end{thebibliography}%

\end{document}